\documentclass[a4paper,11pt]{article}
\usepackage{authblk}
\pdfoutput=1 

\usepackage{amsmath}
\usepackage{comment}
\usepackage{float}
\usepackage{caption}
\usepackage{subcaption}
\usepackage{preamble}
\usepackage[T1]{fontenc}
\usepackage[sorting=none,maxcitenames=50,maxnames=50]{biblatex}
\title{A String Theory for Two-Dimensional Yang-Mills Theory II}
\author{Ofer Aharony${}^{a}$ \footnote{ofer.aharony@weizmann.ac.il}, Suman Kundu${}^{a,b}$\footnote{sukundu@sissa.it}, and Tal Sheaffer${}^{a}$\footnote{sheaffertal@gmail.com}\\

\footnotesize{\it ${}^{a}$Department of Particle Physics and Astrophysics, Weizmann Institute of Science, Rehovot 7610001, Israel\\}
\footnotesize{\it ${}^{b}$SISSA, Via Bonomea 265, 34136 Trieste, Italy}
}
\date{}

\begin{document}

\maketitle

\begin{abstract}
In earlier work we proposed a string theory dual to two dimensional Yang-Mills theory at zero coupling (which can also be thought of as a $BF$ theory), given by a Polyakov-like generalization of Ho\v rava's topological rigid string theory, and we showed that it correctly reproduces (in the $1/N$ expansion) several partition functions of $SU(N)$ Yang-Mills theory. In the present paper, we generalise this to Wilson loop expectation values by adding boundaries with one Dirichlet and one Neumann boundary condition to our string worldsheets. We discuss in detail several examples, including examples where the worldsheet has branch points or orientation-reversing tubes, or where the Wilson loop has one or more self-intersections, and we show that in all of them the string theory reproduces the known Yang-Mills expectation values. We argue that examples with orientation-reversing tubes or self-intersecting Wilson loops cannot be brought to the conformal gauge, so we analyse them in a different gauge.
\end{abstract}

\newpage
\tableofcontents{}


\section{Introduction and summary}

More than 50 years ago, 't Hooft argued \cite{tHooft:1973alw} that any large $N$ gauge theory should have a dual description as a string theory with a string coupling $g_s \sim 1/N$. The AdS/CFT correspondence has led to many examples of this duality, but we still do not know how to find the string dual of general large $N$ gauge theories. 

A simple model where one may hope to derive a string dual is the case of two-dimensional Yang-Mills (YM) theory. This theory has no dynamical particles, and it is exactly solvable for any gauge group. In \cite{Gross:1993hu}, Gross and Taylor showed how the large $N$ expansion of the partition function of the $SU(N)$ YM theory can be rewritten as a string theory. The string theory involves a sum of mappings from two-dimensional worldsheets to space-time, where the mappings are allowed to have specific types of singularities that come with specific weights. It is interesting to construct a worldsheet action that would lead to the string theory of \cite{Gross:1993hu}, to put its definition on firmer ground and to allow its generalisation by adding additional matter fields. Following the early works \cite{Horava:1993aq,Cordes:1994fc,Cordes:1994sd} and in particular \cite{Horava:1995ic}, we presented in \cite{Aharony:2023tam} a conjecture for a worldsheet action that leads to the string theory of \cite{Gross:1993hu}, in the limit of vanishing Yang-Mills coupling\footnote{Note that large $N$ 2d YM theory on a sphere has a phase transition at finite 't Hooft coupling \cite{Douglas:1993iia,Minahan:1993tp,Caselle:1993gc,Caselle:1993mq,Billo_caselle:1998fb,Gross:1994mr,Taylor:1994zm}, so for the case of the sphere our theory describes the strong coupling phase and its continuation to small coupling, even though this is not the dominant contribution at small coupling.}. In the zero coupling limit the Yang-Mills theory is equivalent to a topological BF-theory containing the gauge field and a scalar $B$ in the adjoint representation, $S = \int d^2 x \, {\rm tr}(B(x) \epsilon_{\mu \nu} F^{\mu \nu}(x))$, but its partition function is still highly non-trivial, and we argued in \cite{Aharony:2023tam} that it is reproduced by our worldsheet action, to all orders in perturbation theory in $1/N$.\footnote{There are several other suggestions in the literature for worldsheet string theories that reproduce a chiral sector of the 2d Yang-Mills theory 
\cite{Cordes:1994sd,Cordes:1994fc,Rudd:1994ta,Vafa:2004qa,Benizri:2025xmz,Komatsu:2025sqo,GABERDIEL_forthcoming}, but as far as we know there are no other suggestions for worldsheet actions that reproduce the full Yang-Mils theory, including configurations with orientation-reversing tubes and twists. It would be interesting to understand the relations between these different worldsheet theories better.}

In this paper, we show that the worldsheet action in \cite{Aharony:2023tam} correctly reproduces also the expectation values of Wilson loops, whose large $N$ expansion was described in \cite{Gross:1993yt}. From the worldsheet point of view, one adds boundaries of the worldsheet ending on the Wilson loop (with exactly one worldsheet ending on any Wilson loop in the fundamental representation), and we show in this paper, in several examples, that our worldsheet reproduces the zero-coupling Yang-Mills results. Some issues arising in the computation, such as the appearance of branch points and of zero-size tubes, are similar to the issues that showed up for the partition function in \cite{Aharony:2023tam}, and we show that similar methods can be used to handle them. Wilson loops raise new issues, in particular when they have self-intersections which lead to degenerations (``twists'') of the worldsheet. Both for zero-size tubes and for twists, the conformal gauge is not convenient, and we introduce a different gauge called the ``induced gauge'', and use it to analyse these cases.

We begin in section \ref{sec:review_gross_taylor} by reviewing the Gross-Taylor formulation of the $1/N$ expansion of YM theory as a string theory, and Ho\v rava's suggestion for a worldsheet theory to describe it. In section \ref{sec: the string theory} we review the worldsheet theory we suggested in \cite{Aharony:2023tam}, and describe how to include in it boundary conditions corresponding to Wilson loops. We also discuss the two different gauges that we will use in our computations.

The rest of the paper is devoted to studying in detail several examples of Wilson loop computations that illustrate all the different issues and types of mappings that arise in these computations. In section \ref{sec:WLs in YM} we review the computation of five different Wilson loops in the large $N$ Yang-Mills theory, which were computed in section 7 of \cite{Gross:1993yt}. Then, in section \ref{sec:WLs in string theory}, we describe how our worldsheet computations of these Wilson loop expectation values precisely reproduce the Yang-Mills results. The computations of different Wilson loops illustrate different features of the mappings involved and their contributions.

So far, we do not have a general proof that our string theory reproduces the (zero-coupling) YM theory to all orders in $1/N$. We argue (but do not rigorously prove) that each worldsheet topology has a simple contribution \eqref{eq:general_result}. For holomorphic mappings (``chiral Yang-Mills theory''), it was argued in \cite{Cordes:1994fc,Cordes:1994sd} that this agrees with the Wilson loop computations, but we are not aware of an argument for this for general mappings, and it would be very interesting to show this.

In upcoming work 
we plan to generalize our results (both for closed strings and for open strings) to finite coupling (and possibly also to generalized Yang-Mills theories \cite{Ganor:1994bq}), by appropriate deformations of the worldsheet action. It would also be interesting to generalize our work to $O(N)$ and $USp(2N)$ gauge theories, and the corresponding non-orientable string theories \cite{Ramgoolam:1993hh,Naculich:1994kd}. In addition, we plan to study the string dual of $1+1$-dimensional QCD by adding matter fields in the fundamental or the adjoint representations.

\section{Review of results from the 1990s}\label{sec:review_gross_taylor}

\subsection{Gross Taylor: Partition function and Wilson loop}

The exact solvability of two-dimensional quantum Yang-Mills (YM) theory has enabled detailed investigations of its large \( N \) expansion. On a Riemann surface \( M \) with genus \( G \) and area \( A \), the partition function is given by \cite{Migdal:1975zg,Rusakov:1990rs,Fine:1990zz,Witten:1992xu,Blau:1991mp}:
\begin{equation} \label{YM_partition}
Z_M = \int \frac{[DA_\mu]}{{\rm gauge\  trans.}} e^{-\frac{1}{4\tilde{g}^2}\int_M d^2x \sqrt{g} \text{Tr} (F_{\mu\nu}F^{\mu\nu})} = \sum_R (\text{dim } (R))^{2-2G} e^{-\frac{{\tilde{g}}^2 A}{2} C_2(R)},
\end{equation}
where the sum runs over irreducible representations \( R \) of the gauge group. Here, \( \text{dim } (R) \) and \( C_2(R) \) are the dimension and quadratic Casimir of \( R \), respectively. For $SU(N)$ gauge groups, we define the 't Hooft coupling \( \lambda \equiv \tilde{g}^2 N \).

Gross and Taylor conjectured that the \( SU(N) \) Yang-Mills theory admits an equivalent description as a string theory in the large-\( N \) limit \cite{Gross:1993hu}. In this dual picture, the string coupling is \( g_s = 1/N \), and the string tension is \( \lambda/2 \). The key to verifying this conjecture lies in rewriting the large $N$ limit of the partition function, where only special representations survive, as a sum over maps of a two-dimensional worldsheet onto the target space \( M \). These maps encode the string dual of the gauge theory.

We consider maps from a two-dimensional worldsheet to $M$, 
that locally cover \( M \) at all but a finite set of points. Such maps are characterized by having (at a generic point) \( n \) sheets with an orientation matching that of \( M \), and \( \tilde{n} \) sheets with the opposite orientation. The partition function for a target space of genus $G$ can be expanded as:
%
\begin{align} \label{eq: zymnew}
    Z_{YM} = \sum_{n,{\tilde n}=0}^{\infty} & \frac{1}{n! {\tilde n}!} e^{-(n+{\tilde n}) \lambda A / 2} \sum_{s,t=0}^{\infty} (-1)^s \frac{(\lambda A)^{s+t}}{s!t!} N^{(n+{\tilde n})(2-2G)-s-2t} \frac{(n-{\tilde n})^{2t}}{2^t} \times \\
    & \sum_{k=0}^{\infty} (-1)^k \binom{2G+k-3}{k} \sum_{p_1,\cdots,p_s} \sum_{a_1,\cdots,a_G,b_1,\cdots,b_G} \delta(p_1 \cdots p_s {\tilde \Omega}_{n,\tilde n}^k \prod_{j=1}^G (a_j b_j a_j^{-1} b_j^{-1})). \nonumber 
\end{align}
Here, $s$ is the number of branch points such that two sheets of the same orientation are permuted when going around them, and $p_i$ denotes the specific $Z_2$ permutation in $S_n\times S_{\tilde n}$ at the $i$'th branch point; $a_j$ and $b_j$ denote possible permutations in $S_n\times S_{\tilde n}$ when going around non-contractible cycles of the target space; and $t$ is the number of additional allowed degeneration points of the manifold, which can be
contracted (zero-size) tubes between sheets of the same or opposite orientation, or contracted (zero-size) handles\footnote{If we consider the $U(N)$ gauge theory instead of the $SU(N)$ gauge theory, then in the large $N$ limit the same representations appear in \eqref{YM_partition}, since baryons are exponentially suppressed in $N$. However, the quadratic Casimir is slightly different in this case, so the sum over $t$ (and the associated degeneration points) does not appear in \eqref{eq: zymnew}.}. The exponential of $N$ is the Euler number of the corresponding worldsheet. Finally, $k$ is the number of so-called $\Omega$-points, which are special degeneration points of the mapping, where sheets undergo arbitrary permutations, and/or are connected by orientation-reversing tubes (including combinations of branch points and orientation-reversing tubes to give ``branched orientation-reversing tubes''). We will not need here the general expression for the contribution ${\tilde\Omega}_{n,\tilde n}$ of these points (that depends on the specific details of the permutations and orientation-reversing tubes involved), which can be found in \cite{Gross:1993yt,Cordes:1994fc,Cordes:1994sd,Horava:1995ic,Aharony:2023tam}.
A significant simplification of the Gross-Taylor string theory is the absence of mappings containing folds of nonzero length, related to the absence of propagating particles in two-dimensional YM.

The string theory exhibits a bi-sectoral structure with two distinct chiral sectors. These correspond to covering maps with purely one orientation or the opposite one and, in the gauge theory, to the contributions of products of a finite number of either fundamental or anti-fundamental representations of \( SU(N) \). The sectors are coupled only through orientation-reversing tubes.

In the zero coupling limit that we will discuss here, only mappings with $s=t=0$ appear. However, we can still have $\Omega$-points; for a sphere target space $G=0$, their number can be $k=0,1,2$.

In \cite{Gross:1993yt}, Gross and Taylor extended their framework to compute the vacuum expectation values (VEVs) of Wilson loops, providing a complete formulation for calculating the VEVs of arbitrary products of Wilson loops (to all orders in $1/N$). 
A Wilson loop, defined in the gauge theory framework as  
\begin{equation}
W(R, \gamma) = \text{Tr}_R \left[\, \mathcal{P} \exp\left(i \oint_\gamma A_\mu dx^\mu\right) \right] \equiv \chi_R(U_\gamma),
\end{equation}
associates each loop \( \gamma \) with a representation \( R \) of the gauge group. The path-ordered exponential is traced in the representation \( R \), and \( U_\gamma \) represents the holonomy of the gauge field along the contour \( \gamma \). While this representation is natural in the conventional gauge theory framework, Gross and Taylor demonstrated that for the string-theoretic interpretation, it is more appropriate to express the \( 1/N \) expansion in terms of observables labelled by elements of the symmetric group \( S_k \), rather than by gauge group representations.

Gross and Taylor \cite{Gross:1993yt} gave an algorithm for computing the Wilson loop expectation values, using their formulation of the sum over mappings, which is a bit different from \eqref{eq: zymnew} and involves also so-called $\Omega^{-1}$-points; they showed that the topology of $M$ constrains the number of singular points of each type in every region of the target space, and used this to compute the expectation values. A simpler algorithm for writing the partition function (leading to \eqref{eq: zymnew}) and for computing the contributions from only one orientation to Wilson loop VEVs was described in \cite{Cordes:1994fc,Cordes:1994sd}.
 We will not give the details of these computations here, which can be found in the original papers; for the chiral theory, there is a closed-form expression for the Wilson loop VEVs \cite{Gross:1993yt,Cordes:1994fc}, but we do not know of such a formula for the full theory. In this paper, we will review the results of \cite{Gross:1993yt} for the VEVs of five different Wilson loops, and compare them (in the zero coupling limit) to what we find using our worldsheet theory.

\subsection{Ho\v rava's Harmonic Sigma model}

Ho\v rava in \cite{Horava:1993aq,Horava:1995ic,Horava:1998wf} proposed a string which could reproduce the $1/N$ expansion of the $2$D YM partition function; however, some subtleties in his formalism are addressed in the alternative Polyakov-like formalism that we review below. Ho\v rava's formalism is effectively a harmonic sigma model with ${\cal N}=(1,1)$ supersymmetry. In the superspace formulation of this theory, worldsheet fields are combined into a superfield \( X^\mu(\sigma^a;\theta,\bar{\theta}) \), which depends on bosonic worldsheet coordinates \( \sigma^a \) and on Grassmann (fermionic) coordinates \( \theta \) and \( \bar{\theta} \). The superfield is expanded as 
\begin{equation}
    X^\mu = x^\mu + \theta \psi^\mu + \bar{\theta} \chi^\mu + \theta \bar{\theta} B^\mu,
\end{equation}
where \( x^\mu \) represents the bosonic fields of the mapping to space-time, \( \psi^\mu \) and \( \chi^\mu \) are fermionic components, and \( B^\mu \) is an auxiliary bosonic field. Objects that depend on \( \theta \) and \( \bar{\theta} \) are referred to as bi-graded, including superfields. The induced metric is also extended to a superfield, defined as \( H^I_{ab} \equiv \partial_a X \cdot \partial_b X \), with the dot product contracting indices using the spacetime metric \( g_{\mu\nu} \).

The worldsheet gauge symmetries in this formalism generalize the diffeomorphisms into bi-graded diffeomorphisms, parameterized by bi-graded vectors \( \epsilon^a(\sigma^a; \theta, \bar{\theta}) \). These transformations act on the superfield \( X^{\mu} \) as \( \delta X^{\mu} = \epsilon^a \partial_a X^{\mu} \), and the action should be invariant. Superfields constructed with bosonic derivatives transform tensorially under these transformations, while those involving super-derivatives may transform non-tensorially. 

The simplest action in this superspace formalism is written compactly as 
\begin{equation} \label{horava}
    S_H = t \int d^2 \sigma \, d^2 \theta \, \sqrt{H^I},
\end{equation} 
where \( d^2 \theta = d\theta d\bar{\theta} \). This action is invariant under the fermionic BRST symmetries \( Q = \partial_\theta \) and \( \bar{Q} = \partial_{\bar{\theta}} \), as well as 
an \( SU(2) \) R-symmetry group rotating $\theta$ and $\bar{\theta}$, under which \( Q \) and \( \bar{Q} \) rotate in the fundamental representation. The theory is topological in the target space, meaning that changes to the spacetime metric lead to \( Q \)-exact terms, leaving the physical content of the theory unchanged.

At the formal level, the path integral with the action \eqref{horava} is independent of $t$, and by taking $t\to \infty$ one can show that it localizes to configurations where the bottom component of $X$ is a solution to the Nambu-Goto equations of motion (which would follow from an action involving the bottom component rather than the top component of the superfield in \eqref{horava}), namely it extremizes the area. However, computing this path integral is subtle. For instance, any smooth mapping to a $1+1$ dimensional target space automatically satisfies the Nambu-Goto equations, and locally any change in the mapping is equivalent to a diffeomorphism; the content of the Nambu-Goto equations in 2d is just the statement that the worldsheet cannot fold back on itself (at least for folds of finite size). Mappings including zero-size tubes, collapsed handles or branch points are allowed. The non-trivial solutions are all singular mappings, and it is not clear how to regularize the singularities; in addition, the existence of a moduli space of mappings leads to zero modes, and it is not clear how to swallow the fermionic zero modes.

In order to avoid these subtleties, we introduce a different Polyakov-like version of this string theory, which we review in the next section, explaining how to use it also in the presence of boundaries.

\section{Ho\v rava-Polyakov formalism with boundaries}\label{sec: the string theory}


In the Polyakov-like formalism, we add on the worldsheet an extra worldsheet metric superfield, $H_{ab}= h_{ab} +\theta \bar h_{\theta,ab} +\thetabar h_{\theta,ab} +\theta\thetabar \bar h_{ab}$.
The action is \cite{Aharony:2023tam}:
\begin{equation}\label{eq: super-Polyakov action}
    S=-it S_{0} + S_1 = -i\frac{t}{2} \intop d^2\sigma d^2\theta \sqrt{H}H^{ab}\partial_a X \cdot \partial_b X+\intop d^2\sigma d^2\theta \sqrt{h^I}\partial_\theta X^\mu K_{\mu\nu} \partial_{\bar\theta} X^\nu.
\end{equation}
The term \( S_0 \), a superspace version of the Polyakov action, plays a distinguished role in localization, as \( t \) is taken to infinity, while \( S_1 \) (which involves the extrinsic curvature tensor $K_{\mu \nu}$ constructed from the induced worldsheet metric superfield $h^I_{ab} = \partial_a X^{\mu} \partial_b X_{\mu}$) is only important in this context when discussing zero modes of $S_0$. Integrating out \( H \) in this action reduces the formulation to one resembling the super-space Nambu-Goto (NG) action described above. Specifically, one finds that (in the $t\to \infty$ limit)
\( h_{ab} \) is conformally equivalent to the induced metric, while the fluctuations of the fermionic components cancel the one-loop determinant of the bosonic components. The net effect is that \( H \) is replaced by the induced metric \( \partial X \cdot \partial X \), recovering the superspace NG action.

The action $S_0$ in \eqref{eq: super-Polyakov action} is (super)-Weyl-invariant, both classically and in the quantum theory.
Due to Weyl invariance, the determinant part of \( H \) decouples, rendering its integration ill-defined. To handle this mode, a gauge-fixing procedure for the (super)-Weyl symmetry using the Faddeev-Popov (FP) method is necessary, in addition to the gauge-fixing needed for the (super)-diffeomorphisms. 
In this paper, we use two types of gauge conditions depending on the case. For smooth worldsheet mappings (including possible branch points), we use the conformal gauge. However, the mappings that appear when we have zero-size tubes or ``twists'' (self-intersections of Wilson loops) involve lines on the worldsheet that are mapped to points, such that the induced metric degenerates along these lines (one of its eigenvalues vanishes), and the metric cannot be brought to the conformal gauge. For such worldsheet mappings, we use an ``induced gauge'', which we will describe in detail below. 

The corresponding path integral is formally expressed as:  
\begin{equation}
Z = \int \frac{DXDH}{{\rm Vol}({\rm diff}^{2|2} \ltimes {\rm Weyl}^{2|2})} \exp(-S[X, H]).
\end{equation}

{\bf Boundary Conditions}\\
In this paper, we consider non-dynamical boundaries corresponding to Wilson loops in the gauge theory, such that the worldsheet boundary is fixed to lie on a specific curve in space-time (and is identified with it up to reparameterizations). In the Nambu-Goto formalism, this implies that on the boundary
$\delta X \propto \dot X$ (or $\epsilon_{\mu \nu} \dot{X}^{\mu} \delta X^{\nu} =0$), where dots denote derivatives along the boundary.
Upon introducing the dynamical auxiliary metric $H_{ab}$, the condition for vanishing of the boundary variation of $S_0$ becomes (if $\sigma$ is the coordinate transverse to the boundary, and $\tau$ is the direction along it):
\begin{equation}
    \sqrt{H}H^{\sigma a}\partial_a X \cdot \delta X=0.
\end{equation}

We can expand any vector on the boundary in terms of $t^\mu$, the unit normalized tangent vector to the Wilson loop, and $n^\mu (= \epsilon^{\mu\nu}t_\nu)$, the outward normal to the loop. In particular, we can write
\begin{equation}
    \delta X^\mu = \delta X_\perp n^\mu + \delta X_{||} t^\mu.
\end{equation}
Then, the boundary condition becomes
\begin{equation}
\begin{split}
    \sqrt{H}H^{\sigma a}\partial_a X_\mu (\delta X_\perp n^\mu + \delta X_{||} t^\mu) & = 0,\\
     \sqrt{H}H^{\sigma a} n\cdot \partial_a X ~ \delta X_\perp + \sqrt{H}H^{\sigma a} t\cdot \partial_a X ~ \delta X_{||} & = 0.
\end{split}
\end{equation}
For Wilson loops, the transverse position of the boundary should remain fixed, whereas the string can move along the boundary due to re-parametrization symmetry, so we choose the following boundary conditions (using $t^\mu \propto \partial_\tau X^\mu$):
\begin{equation}
    \begin{split}
        \delta X_\perp = 0, \quad\quad \sqrt{H}H^{\sigma a} \partial_a X \cdot \partial_\tau X = 0:
    \end{split}
\end{equation}
The boundary of the worldsheet should not change during the variation, but the end of the string can move along the boundary, which is equivalent to the re-parametrization symmetry of the boundary. The second condition is equivalent to
\begin{equation} \label{general_bc}
    \begin{split}
        H^{\sigma \sigma} \partial_\sigma X\cdot \partial_\tau X + H^{\sigma \tau} \partial_\tau X\cdot \partial_\tau X & = 0, \\
        H^{\sigma \sigma} h_{\sigma\tau}^I + H^{\sigma \tau} h_{\tau\tau}^I & = 0,
    \end{split}
\end{equation}
which puts a constraint on the metric at the boundary: $H^{\sigma \sigma}/H^{\sigma \tau}$  must be the same for the auxiliary metric as for the induced one. 

\subsection{Branch points, folds and Wilson loop boundaries}
\label{folds}

As we mentioned above, the singularities allowed in mappings appearing in the zero-coupling theory involve only $\Omega$-points, so they can be branch points, orientation-reversing tubes, or both.

Branch points are points such that when we go around them $(n+1)$ sheets are permuted by a cyclic permutation; locally, the mapping of the worldsheet (labelled by $z$) to the target space (labelled by $Z$) at these points looks like $Z = z^{n+1} / (n+1)$. If we have a boundary corresponding to a Wilson loop in the fundamental representation, only one sheet can end on this boundary, and thus, we cannot have a branch point on the boundary; it has to sit in the bulk.

Orientation-reversing tubes are the limits of circular folds where the fold goes to zero size in space-time. General mappings of two-dimensional world sheets to two-dimensional target space can have lines of folds where the world sheet bends back on itself; locally, the mapping near a fold looks like
\begin{equation} \label{simple_fold}
    X = x, \qquad Y = y^2,
\end{equation}
where we map a worldsheet labeled by $(x,y)$ to a target space labeled by $(X, Y)$, with a fold at $Y=0$ on the target space (and at $y=0$ on the worldsheet), such that the region $Y > 0$ of the target space is covered twice (once with each worldsheet orientation), and the region $Y < 0$ is not covered at all. Configurations with folds are generally not solutions to the Nambu-Goto equations of motion, since such solutions extremize the induced area, and when we have a fold, the area can decrease by moving the fold in the positive $Y$ direction. The only exception is a zero-size circular fold. We can write a circular fold in radial coordinates for the worldsheet and target space as
\begin{equation}
    R = (r - 1)^2 + R_0, \qquad \Theta = \theta,
\end{equation}
with a circular fold located at $r=1$ in the worldsheet, and at $R=R_0$ in the target space. In general this is not a solution to the NG equations of motion, except for the special case $R_0=0$ where the fold goes to zero size in the target space; in that case the configuration maximizes the worldsheet area (for the given topology), so it is a solution (though there is a negative-action fluctuation around this solution, corresponding to increasing the fold to finite size). Note that the fold has zero size in the target space, but a finite size on the worldsheet. In general, when we have a zero-size fold and several sheets of the worldsheet with each orientation, we can also have permutations of these sheets when we go around the fold, and the generalization to this case is straightforward.

In the presence of boundaries corresponding to WLs in the fundamental representation, a fold like \eqref{simple_fold} can end on a Wilson loop (located, say, at $X=0$). When we have a Wilson loop, we must have either a worldsheet with one orientation ending on it from the right, or a worldsheet with the opposite orientation ending on it from the left. So if we have a configuration like \eqref{simple_fold} at $X>0$, we must have one of its sheets ending on the boundary from the positive $X$ direction (at $Y > 0$), while the other sheet must continue to the other side, and then end on the boundary (from its other side) from the negative $X$ direction (at $Y < 0$). An example of a mapping which does this, taking the worldsheet to be the upper half-plane $y \geq 0$, is
\begin{equation}
    X = - 2 x y, \qquad Y = (y-x) |x+y|,
\end{equation}
where the fold sits on the worldsheet at $x+y=0$, and the worldsheet ends on the boundary with different orientations for $x>0$ compared to $x<0$.

\subsection{Conformal gauge}\label{subsec:Conformal Gauge}

Smooth mappings, or mappings with branch points, can be brought by a diffeomorphism+Weyl transformation to the conformal gauge $H_{ab} = \delta_{ab}$. To implement this
we add a Lagrange multiplier $\Lambda^{ab}(\sigma)$ to fix the auxiliary metric to the identity matrix, deforming the action by
\begin{equation} \label{conformal_gauge}
    \begin{split}
        -i\int d^2\sigma d^2\theta \sqrt{H} \Lambda^{ab}(H_{ab}-\delta_{ab}).
    \end{split}
\end{equation}

We can integrate out $\Lambda_{ab}$ and $H_{ab}$, without worrying about $1$-loop determinants, since those are cancelled between fermions and bosons. This indicates that the FP determinant should also be trivial, but we should carefully include it to show the action with \eqref{conformal_gauge} is truly equivalent to the original one. Any residual gauge transformations after this gauge-fixing (namely, conformal transformations) are automatically extended to superspace. The FP determinant gives rise to two extra superfields:
\begin{itemize}
    \item An extended diffeomorphism ghost $C^a = c^a +\theta \bar c_\theta ^a +\thetabar c_\theta^a + \theta\thetabar \bar c^a$.
    \item A symmetric-matrix antighost $B_{ab}=b_{ab} +\theta\, \bar b_{\bar\theta,ab} +\thetabar\, b_{\theta,ab} + \theta\thetabar \, \bar b_{ab}$, which will be made traceless by the integral over the Weyl ghost.
\end{itemize}

{\bf Equation of motion}

The action $S_0$ is $Q$-exact. So, we can formally take $t\rightarrow \infty $ without affecting the path integral, such that the variations of all fields, including $X^\mu$, are only sensitive to the first part of the action $S_0$. As in \cite{Horava:1995ic,Aharony:2023tam}, the path integral localizes to solutions of the equation of motion coming from the bottom component of the superfield in $S_0$; in the conformal gauge, we get
\begin{equation}
    \partial_a \partial^a X^\mu = 0.
\end{equation}
In $(z,\bar z)$ coordinates the equation has the form
\begin{equation}
    \partial_z \partial_{\bar z} X^\mu = 0.
\end{equation}
In this gauge, a generic solution in complex coordinates ($Z, \bar Z$ in space-time, $z, \bar z$ on the worldsheet) has the form
\begin{equation}
    Z = f(z)+\tilde f(\bar z),
\end{equation}
where $f$ and $\tilde f$ are two arbitrary single-variable functions.

In addition, the equation of motion of the metric imposes the Virasoro condition $T_{ab}=0$, 
where
\begin{equation}
    \begin{split}
        T_{ab} &= \left.\frac{1}{\sqrt{H}}\frac{\delta S}{\delta H_{ab}}\right|_{H_{ab} = \delta_{ab}} \\
        &= \left.\partial_a X\cdot \partial_b X - \frac{1}{2}H^{cd}\partial_c X\cdot \partial_d X ~ H_{ab}\right|_{H_{ab} = \delta_{ab}}  \\
        &= h_{ab}^I - \frac{1}{2}Tr(h^I) ~ \delta_{ab}.
    \end{split}
\end{equation}
Since $T$ is traceless, this is equivalent to $T \equiv T_{zz} = 0$, ${\bar T} \equiv T_{{\bar z}{\bar z}} = 0$. This implies that the solution should be either holomorphic or anti-holomorphic, and not a combination of both.

{\bf Boundary conditions}

In conformal gauge the condition \eqref{general_bc} becomes
\begin{equation}
    \partial_{\sigma} X \cdot \partial_{\tau} X = 0.
\end{equation}
This boundary condition can also be written in terms of the stress tensor:
\be
T_{\sigma \tau} = 0;
\ee
Note that this follows from the equation of motion of the auxiliary metric, but we impose it also off-shell and not just on-shell. For the disk, this can be written as $\left(z^{2}T-\bar{z}^{2}\bar{T}\right)\Big|_{\partial D}=0$.   
If we consider solutions where $X$ is holomorphic,
then the solutions to the boundary condition will be holomorphic quadratic differentials \cite{Polchinski_textbook:1998rq} -- in one-to-one correspondence with complex structure deformations (and 0-modes of $B_{ab}$).

{\bf Ghosts}

The super-ghost action in components is
\begin{equation}
        \frac{1}{4 \pi }\int_{WS}d^2\theta~d^2\sigma~B_{ab}\partial^a{C}^b =\frac{1}{4 \pi }\int_{WS}\left(b_{ab}\partial^a{\bar c}^b+\bar b_{ab}\partial^a{c}^b+b_{\theta,ab}\partial^a{\bar c}_\theta^b-\bar b_{\theta,ab}\partial^a{c}_\theta^b\right).
\end{equation}

As in the closed string case discussed in \cite{Aharony:2023tam}, and as in bosonic string theory, we have a $C$ zero mode for every conformal Killing vector of the worldsheet, and a $B$ zero mode for every modulus of the worldsheet. The $C$ zero modes are lifted when we perform extra gauge-fixings of the (super-)diffeomorphisms associated with the corresponding conformal Killing vectors, and the $B$ zero modes are lifted by the extra integrations over the (super-)moduli (see equation (4.6) of \cite{Aharony:2023tam} and the subsequent discussion, which generalizes straightforwardly to the case with boundaries). The non-zero modes all cancel between the bosonic and fermionic modes of the ghosts.

\subsection{Induced gauge}\label{subsec:Induced Gauge}

It will sometimes be convenient to use a different gauge, where we fix the worldsheet metric $H_{ab}$ not to the identity matrix, but instead to the induced metric of some specific configuration $x_0(\sigma)$, $h^0_{ab} \equiv \partial_a x_0^{\mu} \partial_b x_{0,\mu}$. The advantage of this gauge is that this gauge-fixing can be done even when the induced metric degenerates on some points or lines; the disadvantage is that (for each topological class of mappings) we need to find and choose some specific mapping (in that topological class). To impose this gauge, we similarly add:
\begin{equation}
    \begin{split}
        -i\int d^2\theta d^2\sigma \sqrt{H} \Lambda^{ab}(H_{ab}-h_{0,ab}).
    \end{split}
\end{equation}

Most of the analysis follows the one above for the conformal gauge. In particular, again we get a ghost $C$ and an anti-ghost $B$, whose action we write down below.

{\bf Equation of motion}

As above, the action we are considering is $Q$-exact, so we can formally take $t\rightarrow \infty $, and the variations of all fields, including $X^\mu$, are then only sensitive to the first part of the action $S_0$. The equation of motion we get is
\begin{equation}
    \nabla^2 X^\mu = 0,
\end{equation}
where the covariantization is with respect to the metric $h^0_{ab}$. In the induced gauge, it is easy to check that one solution to this equation is the specific configuration $x_0$ that we used to define the `induced gauge'. This is because, for this case, the equation of motion becomes the same one as in the Nambu-Goto string (where we have a Laplacian with respect to the induced metric), and as discussed for instance in \cite{Aharony:2023tam}, this equation holds automatically for any mapping that is not degenerate. The equation does not hold only when there are folds of finite extent in the target space. It does allow for zero-size folds \cite{Aharony:2023tam}, so we will allow such folds in our mappings.

The equation of motion of the metric (the Virasoro condition) also holds automatically for the mapping $x_0$, assuming it is smooth.
This is because
\begin{equation} \label{virasoro_induced}
    \begin{split}
        T_{ab} &= \left.\frac{1}{\sqrt{H}}\frac{\delta S}{\delta H_{ab}}\right|_{H_{ab} = h^0_{ab},X^\mu=x_0^\mu} \\
        &= \left.\partial_a x_0 \cdot \partial_b x_0 - \frac{1}{2} H_{ab} ~ H^{cd}\partial_c x_0\cdot \partial_d x_0 \right|_{H_{ab} = h^0_{ab}}  = 0.
    \end{split}
\end{equation}

{\bf Boundary conditions}

As discussed above, the boundary condition is satisfied when \eqref{general_bc} holds.
In the induced gauge, it is automatically satisfied for the mapping $x_0$, since this obeys $H^{\sigma a} = h_0^{\sigma a}$ on the boundary. For other configurations which solve the equations of motion, the Virasoro condition \eqref{virasoro_induced} implies that the induced metric is related by a Weyl transformation to the induced metric $h_0$, and then the boundary condition will also hold.

{\bf Uniqueness of the classical solution}

Next, we want to ask if there
are other solutions to the equations of motion (and the boundary conditions) in a particular induced gauge?
We start from a specific configuration $x_0^{\mu}$ and we look for nearby solutions, where we can expand $X^{\mu} = x_0^{\mu} + V^{\mu}$ for small $V^{\mu}$. At linear order in $V$, we find that we need to satisfy
\begin{equation} \label{linear_NG}
    \nabla^2 V^{\mu} = 0,
\end{equation}
\begin{equation} \label{linear_virasoro}
    \nabla_{(a} V^{\mu} \nabla_{b)} x_{0,\mu} = h_{ab}^0 h^{0,cd} \nabla_c V^{\mu} \nabla_d x_{0,\mu},
\end{equation}
where all covariant derivatives are with respect to the metric $h^0_{ab} = \partial_a x_0 \cdot \partial_b x_0$.

At any regular point on the worldsheet, where the metric is non-degenerate such that the matrix $\nabla_a x_{0,\mu}$ is invertible, we can write
\begin{equation} \label{diff_decomp}
    V^{\mu} = \xi^a \nabla_a x_0^{\mu},
\end{equation}
namely, the deformation is a worldsheet diffeomorphism with parameter $\xi^a$. The Virasoro condition \eqref{linear_virasoro} then implies:
\begin{equation}
    \nabla_{(a} (\xi^c \nabla_c x_0^{\mu}) \nabla_{b)} x_{0,\mu} = h^0_{ab} h^{0,cd} \nabla_c (\xi^e \nabla_e x_0^{\mu}) \nabla_d x_{0,\mu}.
\end{equation}
We can write this as
\begin{equation}
\nabla_{(a} \xi^c \nabla_c x_0^{\mu} \nabla_{b)} x_{0,\mu} 
+ \xi^c \nabla_{(a} \nabla_c x_0^{\mu} \nabla_{b)} x_{0,\mu} 
    = h^0_{ab} h^{0,cd} (\nabla_c (\xi^e \nabla_e x_0^{\mu} \nabla_d x_{0,\mu}) - \xi^e \nabla_e x_0^{\mu} \nabla_c \nabla_d x_{0,\mu}).
\end{equation}
Using the fact that the metric is covariantly constant, this is equivalent to:
\begin{equation}
    \nabla_{(a} \xi_{b)} = h^0_{ab} \nabla_c \xi^c.
\end{equation}
Thus, the Virasoro constraints force $V^{\mu}$ to be of the form $\xi^a \nabla_a x_0^{\mu}$, where $\xi_a$ are conformal Killing vectors of the induced metric $h^0_{ab}$. This is the same as the statement that $V^{\mu}$ should be a conformal Killing vector of the target space metric; in the special case when the target space metric is proportional to the unit matrix, and we use complex coordinates, this means it should be an (infinitesimal) holomorphic function.
So, we find that the dimension of the space of nearby solutions is the same as the number of conformal Killing vectors that are consistent with the boundary conditions. All the solutions are worldsheet diffeomorphisms, so our original solution is unique up to such diffeomorphisms.

If the mapping has no degenerate points, then this is the final answer. However, in many cases, we are interested in mappings with degenerate points, such as branch points (that degenerate at a single point), or tubes or twists (that degenerate along a line). Away from the degeneration, we can still write \eqref{diff_decomp}, but now $\xi$ may be singular at the degeneration (even when $V^{\mu}$ is non-singular), such that it no longer corresponds to a diffeomorphism.

Let us begin with the case of a branch point. A mapping with a degree $n$ branch point behaves as $Z(z) = \frac{z^{n+1}}{n+1}$. The induced metric is in this case proportional to the identity matrix, so the Virasoro condition tells us that $V^z$ and thus $\xi^z$ have to be purely functions of $z$ and not $\bar z$. However, since $V^z = z^n \xi^z$, $\xi^z$ does not have to be holomorphic, but it is allowed to have a pole up to degree $n$. The coefficients of the terms in $\xi^z$ going as $z^{-k}$ ($k=1,\cdots,n$) give us $n$ deformations of the mapping that are not diffeomorphisms (since $\xi$ is singular), corresponding to the $n$-dimensional manifold that we get by splitting the branch point into $n$ simple branch points that can move around in the target space. This is the same as what we find also in the conformal gauge (where we simply get holomorphic mappings, and the holomorphic deformations of $Z(z)$ are the same as the ones we found above); indeed, in this case, the two gauges are related just by a Weyl transformation.

The cases of zero-size tubes or twists are different; in these cases a line on the worldsheet maps to a point in target space, where for a tube the line is closed and does not intersect the boundary, while for a twist it ends (on both sides) on the boundary (at the point in space-time where the Wilson loop has a self-intersection). In these cases, one eigenvalue of the induced metric (and of the matrix $\nabla_a x_0^{\mu}$) goes to zero while the other one does not, and again we get extra solutions because the appropriate component of $\xi$ is allowed to be singular on the singular line. Without loss of generality we can assume that the eigenvalue of $\nabla_a x_0^{\mu}$ goes to zero linearly as we approach the line, and then we get one extra deformation that is not a diffeomorphism, in which $V^{\mu}$ is a constant (this obviously satisfies \eqref{linear_NG} and \eqref{linear_virasoro}). For the case of a zero-size tube, this extra deformation corresponds to moving the tube in the target space, and it gives us the extra modulus associated with the tube (in general, to satisfy the boundary conditions, we will need to perform an extra diffeomorphism together with shifting the tube). We will exhibit this explicitly in an example in section \ref{sec: disk with ORT} below. For a twist, this deformation is not allowed by the boundary conditions, since the two segments of the Wilson loop that intersect go in different directions, so a constant shift of the degeneration line (that ends on different segments, at the same point in target space) is not possible. 

Note that when the metric degenerates on a line, one can use different conformal Killing vectors on the two sides of this line (as long as they vanish on the line).

At zero 't Hooft coupling, collapsed handles doesn't contribute to the partition function or the Wilson loop expectation values. We show explicitly in appendix \ref{appendix:coll_handle} that this is true also in our string theory, because of zero modes related to internal deformations of the handle that have a vanishing metric on their moduli space, leading the path integral to vanish.

{\bf Ghosts}

The super-ghost action for the induced gauge in components is
\begin{equation}
        \frac{1}{4 \pi }\int_{WS}d^2\theta\sqrt{H}~B_{ab}\nabla^a{C}^b \\=\frac{1}{4 \pi }\int_{WS}\sqrt{h^0}\left(b_{ab}\nabla^a{\bar c}^b+\bar b_{ab}\nabla^a{c}^b+b_{\theta,ab}\nabla^a{\bar c}_\theta^b-\bar b_{\theta,ab}\nabla^a{c}_\theta^b\right),
\end{equation}
where we now have covariantization with respect to the induced metric $h^0_{ab}$.

Since the kinetic operator is exactly the same between bosonic and fermionic ghost fields, contributions coming from all non-zero modes will cancel each other. The only non-trivial contribution can come from the zero modes. 
Zero modes for $c^a$ satisfy 
\begin{equation}
\nabla_a c_b+\nabla_b c_a-h_{ab}\nabla\cdot c = 0,
\end{equation}
and the $c^a$ boundary condition is $c^an_a = 0$, where $n$ is a vector orthogonal to the boundary. They correspond to conformal Killing vectors of the induced metric $h^0_{ab}$, and they will be lifted when we fix the corresponding diffeomorphisms. Similarly, zero modes of $B$ will be lifted by the integration over worldsheet moduli, as before.

\subsection{The worldsheet path integral}

In the absence of boundaries, we argued in \cite{Aharony:2023tam} (following \cite{Horava:1995ic}) that the worldsheet path integral for some topological class of mappings computes the Euler number of the moduli space of mappings (the space of classical solutions to the EOM modulo diffeomorphisms and Weyl transformations). The way this works is that the fermionic symmetry implies that the path integral is independent of $t$, and in the $t\to \infty$ limit, the path integral localizes to the extrema of the bottom component of the integrand in $S_0$, which are extremal area metrics. In general, these metrics come in families labelled by moduli (positions of branch points and orientation-reversing tubes), which have fermionic partner zero modes. To get a non-zero answer, one has to saturate these zero modes by bringing down factors of $S_1$ from the action. These give a metric on the moduli space, and saturating the fermion zero modes gives factors of the Riemann tensor on the moduli space, which combine together to give precisely the Euler number of the moduli space of mappings (independently of the precise form of the metric), see appendix A of \cite{Aharony:2023tam}. According to \cite{Cordes:1994fc,Cordes:1994sd,Horava:1995ic}, the appearance of this Euler number (together with the standard contribution of $N$ to the power of the worldsheet genus) in the sum over mappings reproduces the Yang-Mills partition function. (We did not write explicitly in our worldsheet action the Einstein-Hilbert term that gives the worldsheet genus, but it should also be present, together with the appropriate boundary term when boundaries are present.) The integrals over the non-zero bosonic and fermionic modes cancel each other exactly, up to an overall sign; each mode with a negative eigenvalue (that reduces the area rather than increasing it) gives a factor of $(-1)$.

The same arguments can also be made in the presence of boundaries, so we conjecture that in the presence of boundaries :

{\bf Conjecture:} Consider worldsheet mappings with $s$ branch points and ${\tilde t}$ orientation reversing tubes, then their contribution to the zero-coupling Wilson loop expectation value that we obtain from our string theory is:
    \begin{equation} \label{eq:general_result}
        (-1)^{\tilde t}\, \chi({\rm Moduli~space}) \, N^{\chi({\rm world~sheet})},
    \end{equation}
where $\chi({\rm world~sheet})$ is the Euler number of the worldsheet (now with boundaries), and $\chi({\rm Moduli~space})$ is the Euler number of the moduli space of inequivalent mappings in this topological class. This moduli space arises from the positions of the branch-points and the orientation-reversing tubes (ORTs), and it also now has boundaries when these points approach the Wilson loop (on a sheet that ends on it). The factor of $(-1)^{\tilde t}$ comes from the single negative mode that each ORT has.

In addition, we conjecture that the sum over all world sheets with the weight \eqref{eq:general_result} reproduces the $1/N$ expansion of the Wilson loop expectation value at zero coupling, derived in \cite{Gross:1993yt}. For holomorphic mappings (the ``chiral YM theory'') at zero coupling, arguments for this were given in \cite{Cordes:1994fc,Cordes:1994sd}, and we conjecture that it is true also more generally. In the next section, we show that this conjecture agrees with the YM computations of \cite{Gross:1993yt}, and in the following section, we show that our worldsheet computations indeed give rise to \eqref{eq:general_result}.

\section{Wilson loop expectation values in 2D YM} \label{sec:WLs in YM}

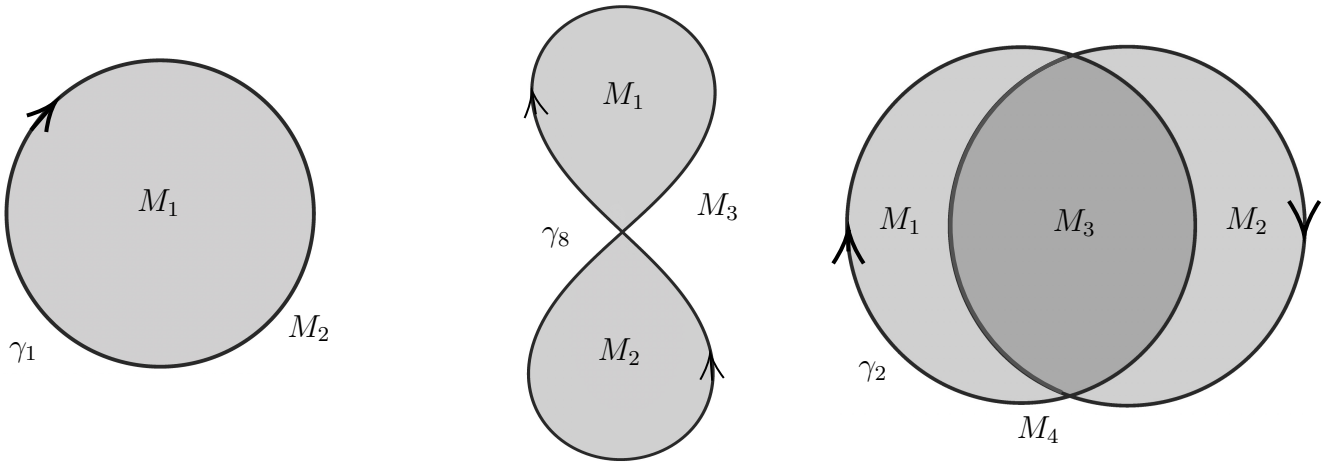
\begin{figure}[h]
    \centering

\tikzset{every picture/.style={line width=0.75pt}} 

\begin{tikzpicture}[x=0.75pt,y=0.75pt,yscale=-1,xscale=1]

\draw (610,184) node  {\includegraphics[width=177.75pt,height=141.75pt]{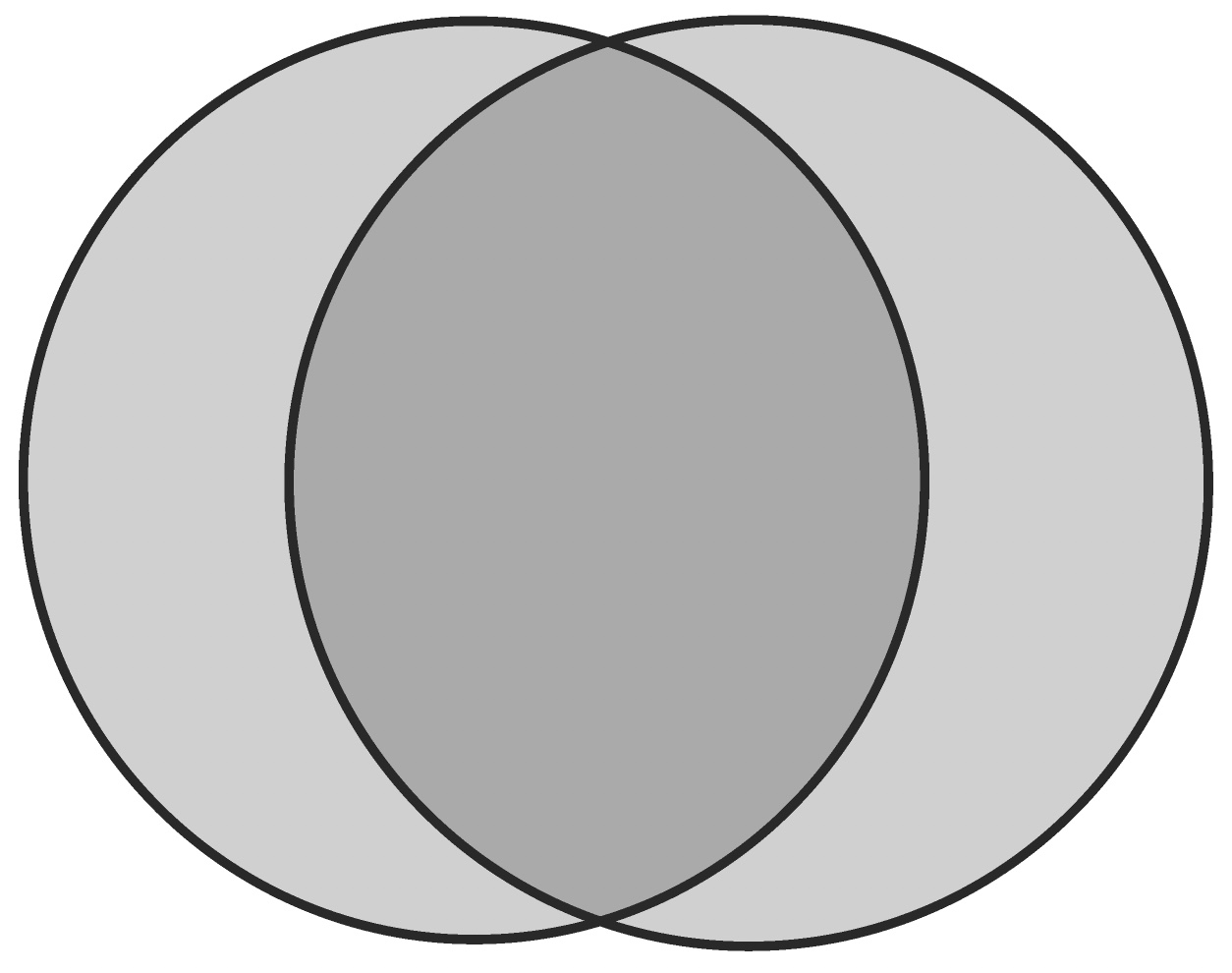}};
\draw (154,177) node  {\includegraphics[width=120.75pt,height=120.75pt]{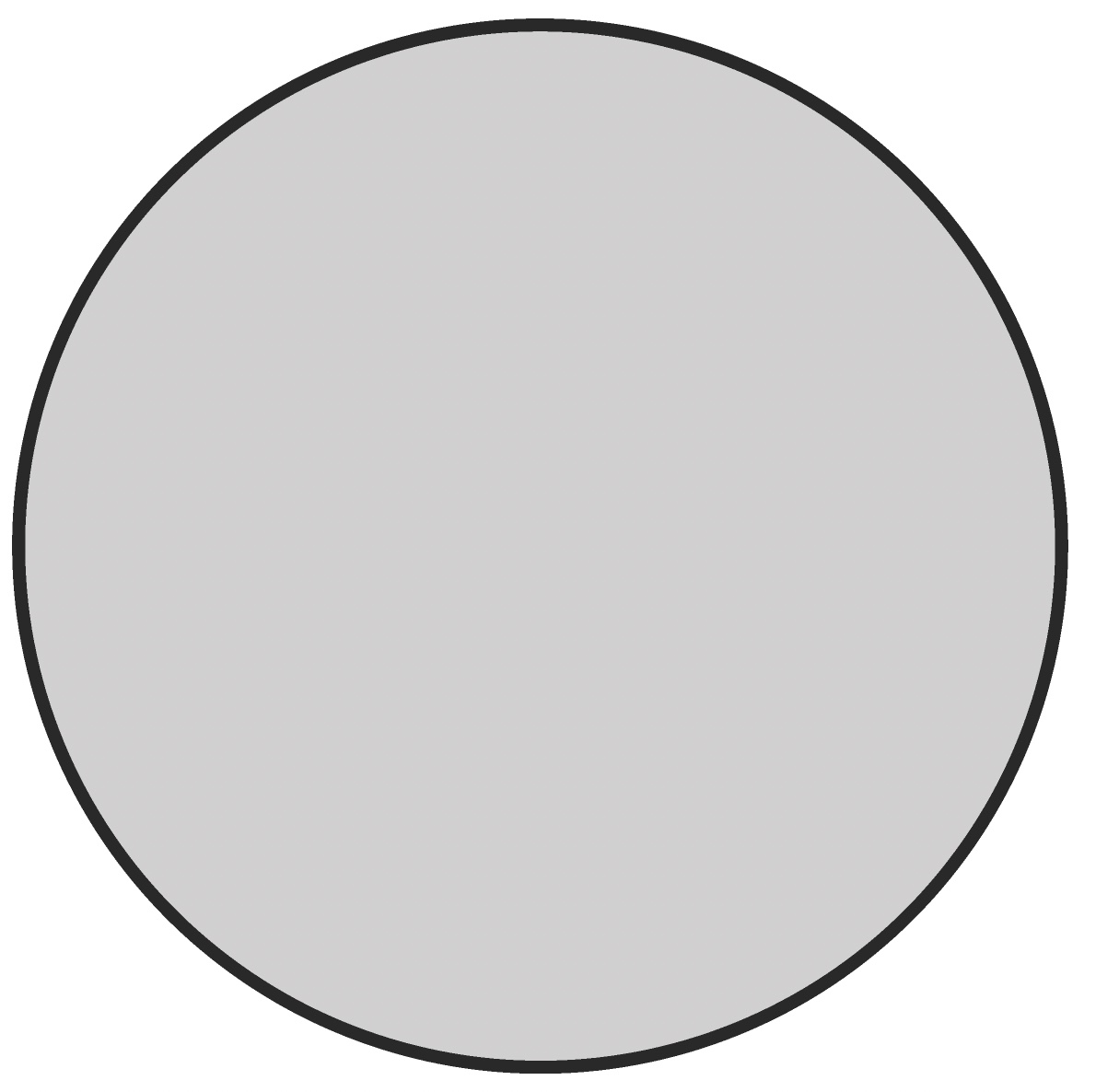}};
\draw (383.5,187.25) node  {\includegraphics[width=72pt,height=172.88pt]{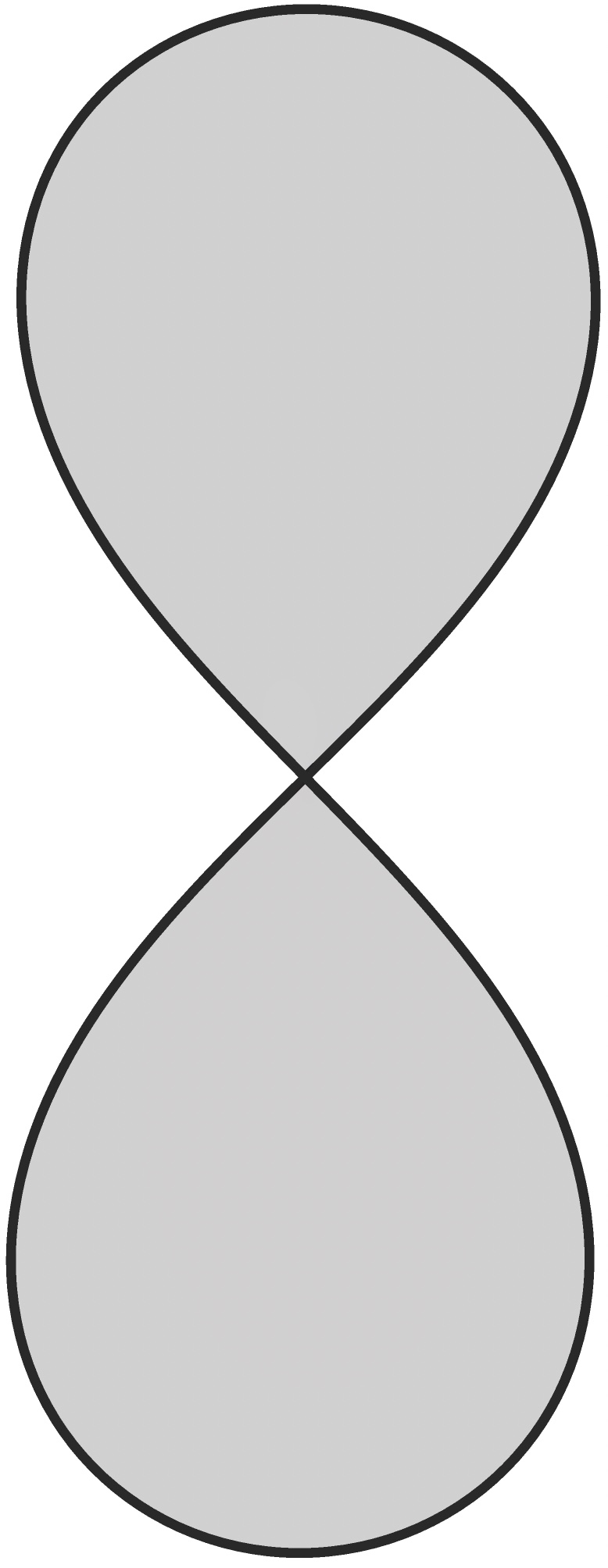}};
\draw [line width=1.5]    (88.5,135) -- (98.55,123.28) ;
\draw [shift={(100.5,121)}, rotate = 130.6] [color={rgb, 255:red, 0; green, 0; blue, 0 }  ][line width=1.5]    (14.21,-6.37) .. controls (9.04,-2.99) and (4.3,-0.87) .. (0,0) .. controls (4.3,0.87) and (9.04,2.99) .. (14.21,6.37)   ;
\draw [line width=0.75]    (340.5,127) -- (338.83,116.97) ;
\draw [shift={(338.5,115)}, rotate = 80.54] [color={rgb, 255:red, 0; green, 0; blue, 0 }  ][line width=0.75]    (13.12,-5.88) .. controls (8.34,-2.76) and (3.97,-0.8) .. (0,0) .. controls (3.97,0.8) and (8.34,2.76) .. (13.12,5.88)   ;
\draw [line width=0.75]    (428.5,261) -- (427.64,248.99) ;
\draw [shift={(427.5,247)}, rotate = 85.91] [color={rgb, 255:red, 0; green, 0; blue, 0 }  ][line width=0.75]    (13.12,-5.88) .. controls (8.34,-2.76) and (3.97,-0.8) .. (0,0) .. controls (3.97,0.8) and (8.34,2.76) .. (13.12,5.88)   ;
\draw [line width=1.5]    (496.5,196) -- (496.5,188) ;
\draw [shift={(496.5,185)}, rotate = 90] [color={rgb, 255:red, 0; green, 0; blue, 0 }  ][line width=1.5]    (17.05,-7.64) .. controls (10.84,-3.59) and (5.16,-1.04) .. (0,0) .. controls (5.16,1.04) and (10.84,3.59) .. (17.05,7.64)   ;
\draw [line width=1.5]    (724,182.75) -- (724,184.75) ;
\draw [shift={(724,187.75)}, rotate = 270] [color={rgb, 255:red, 0; green, 0; blue, 0 }  ][line width=1.5]    (17.05,-7.64) .. controls (10.84,-3.59) and (5.16,-1.04) .. (0,0) .. controls (5.16,1.04) and (10.84,3.59) .. (17.05,7.64)   ;
\draw  [draw opacity=0][line width=1.5]  (603.04,266.83) .. controls (570.4,253.01) and (547.5,220.68) .. (547.5,183) .. controls (547.5,144.31) and (571.65,111.25) .. (605.7,98.09) -- (638.5,183) -- cycle ; \draw  [color={rgb, 255:red, 82; green, 82; blue, 82 }  ,draw opacity=1 ][line width=1.5]  (603.04,266.83) .. controls (570.4,253.01) and (547.5,220.68) .. (547.5,183) .. controls (547.5,144.31) and (571.65,111.25) .. (605.7,98.09) ;  

\draw (342,182.4) node [anchor=north west][inner sep=0.75pt]  [font=\large]  {$\gamma _{8}$};
\draw (500,250) node [anchor=north west][inner sep=0.75pt]  [font=\large]  {$\gamma _{2}$};
\draw (372,109.4) node [anchor=north west][inner sep=0.75pt]  [font=\large]  {$M_{1}$};
\draw (370,238.4) node [anchor=north west][inner sep=0.75pt]  [font=\large]  {$M_{2}$};
\draw (419,165.4) node [anchor=north west][inner sep=0.75pt]  [font=\large]  {$M_{3}$};
\draw (76,240) node [anchor=north west][inner sep=0.75pt]  [font=\large]  {$\gamma _{1}$};
\draw (140,163.4) node [anchor=north west][inner sep=0.75pt]  [font=\large]  {$M_{1}$};
\draw (510,172.4) node [anchor=north west][inner sep=0.75pt]  [font=\large]  {$M_{1}$};
\draw (215,226.4) node [anchor=north west][inner sep=0.75pt]  [font=\large]  {$M_{2}$};
\draw (683,172.4) node [anchor=north west][inner sep=0.75pt]  [font=\large]  {$M_{2}$};
\draw (597,173.4) node [anchor=north west][inner sep=0.75pt]  [font=\large]  {$M_{3}$};
\draw (579,277.4) node [anchor=north west][inner sep=0.75pt]  [font=\large]  {$M_{4}$};

\end{tikzpicture}

    \caption{Three Wilson loops on the sphere, with numbers denoting the different regions.}
    \label{fig:All WL}
\end{figure}

Our goal is to match the Wilson loop expectation values computed by Gross and Taylor with our string theory results. In this section, we list the results calculated by Gross and Taylor in section 7 of \cite{Gross:1993yt} for the large $N$ expansion of the VEVs of five specific Wilson loops (the first three drawn in figure \ref{fig:All WL}) that have $k = 1$, namely they involve the trace of the holonomy of the gauge field in the fundamental representation. For simplicity, we consider Wilson loops on the sphere or on the plane; if we start from a Wilson loop $\gamma$ on the sphere, taking the area $A_i$ of the region $M_i$ outside the loop to infinity reduces $\langle W_\gamma \rangle$ to its value on the infinite plane\footnote{Note that there is an order of limits involved here; this relation between the sphere and the plane holds for any finite coupling, and we should take the limit first and only then take the coupling to zero. Alternatively, we can directly work in the zero-coupling theory, and compute there separately the expectation values for the loops on the sphere and on the plane.}. In this limit, only a finite number of covering maps contribute, as the number of sheets in the infinite region must vanish.

Note that on the plane, since there are no non-trivial gauge bundles, the expectation value in the free gauge theory of any Wilson loop in the fundamental representation is simply equal to $N$. Similarly, the expectation value of a product of $k$ such Wilson loops is equal to $N^k$. However, as we will review, in the stringy formalism of Gross and Taylor, these answers sometimes arise in a non-trivial fashion.

Gross and Taylor computed the number of $\Omega$ 
points in different regions, and used this to calculate various Wilson loop expectation values and to interpret them in string theory. These points generally lead to mappings with
orientation-reversing tubes and branch points. In this section, we compare the YM results to our string theory results given by \eqref{eq:general_result}.

\subsection{$G=0$, $\gamma_1$, $k=1$ : Disk}
\label{sec:disk}

Let us consider the target space to be a sphere, \( G=0 \), with a circular Wilson loop \( \gamma_1 \) of winding number \( k=1 \) (the left-most figure in figure \ref{fig:All WL}), such that the simplest worldsheet topology is a disk. We denote the number of sheets in each region $i$ by $n_i$ with one orientation (holomorphic), and ${\tilde n}_i$ with the opposite orientation (anti-holomorphic). The orientation of the Wilson loop implies that the relation between the numbers of holomorphic and anti-holomorphic covering sheets in the two regions is given by \( n_1 - \tilde{n}_1 = n_2 - \tilde{n}_2 + 1 \). Additionally, since only one sheet can end on the WL, the number of sheets must satisfy the conditions \( n_1 \geq n_2 \) and \( \tilde{n}_2 \geq \tilde{n}_1 \). For the case of the WL on the plane, where region 2 is infinite, we need to have $n_2 = \tilde{n}_2 = 0$ and then the only possibility is $n_1=1$, $\tilde{n}_1 = 0$, while on the sphere there are also additional possibilities, and we review the lowest ones below.

\subsubsection{$n_1=1$, $n_2=0$}

This case is a purely chiral mapping, and from the Yang-Mills (YM) chiral sector calculation, Gross-Taylor obtain in equation (7.2) of \cite{Gross:1993yt} (for $\lambda=0$)
\begin{equation}  \label{simple disk}
   \boxed{\langle W_1^+ \rangle = N.}  
\end{equation}  

Since this calculation is restricted to the chiral sector of YM theory, only holomorphic mappings are allowed. As a result, after fixing (in conformal gauge) the remaining $SL(2,\mathbb R)$ conformal transformations, the only possible worldsheet is a single disk with no moduli, $Z=z$. Since there is no moduli space and no ORT, and since \( N^{\chi(D)} = N \), our general result \eqref{eq:general_result}
matches with the YM result \eqref{simple disk}. We will explain how this comes from the string theory in detail in section \ref{sec: the disk}. On the sphere, there is also a similar anti-holomorphic contribution from the mapping $Z = 1 / {\bar z}$, with only $\tilde{n}_2=1$.

\subsubsection{$n_1=2$, $n_2=1$}  

For this configuration, with the worldsheet covering numbers \( n_1 = 2 \), \( n_2 = 1 \), ${\tilde n}_1 = {\tilde n}_2 = 0$, from the YM chiral sector calculation, Gross-Taylor derive the result (equation (7.5) of \cite{Gross:1993yt}):  
\begin{equation}  
   \boxed{\langle W_2^+ \rangle = N^3.}  
\end{equation}  

Since this calculation also remains within the chiral sector of YM theory, only holomorphic mappings are allowed. In this case, the only possible worldsheet is a disconnected combination of a sphere (\( S \)) and a disk (\( D \)), with no moduli. After fixing the \( SL(2,\mathbb{R}) \) symmetries, the mapping simplifies to \( Z = z \) for both spaces. The Euler characteristic of this configuration as a worldsheet is given by \( \chi(D) + \chi(S) = 1 + 2 = 3 \).  

Using the result from \cite{Aharony:2023tam} for the sphere, and the result mentioned above for the disk,
the contribution of this configuration is \( N^{\chi(D) + \chi(S)} = N^3 \), consistent with the YM result. Again, on the sphere there is also a similar anti-holomorphic contribution, with $\tilde{n}_1=1, \tilde{n}_2=2$.

\subsubsection{$n_1 = \tilde{n}_1 = 1$, $\tilde{n}_2 = 1$ or $n_2 = \tilde{n}_2 = 1$, $n_1 = 1$}  \label{subsubsec: disk with ORT}

Next, we consider a setup where \( n_1 = \tilde{n}_1 = 1 \), \( \tilde{n}_2 = 1 \), \( n_2 = 0 \), or \( n_2 = \tilde{n}_2 = 1 \), \( n_1 = 1 \), \( \tilde{n}_1 = 0 \). The second line of equation (7.7) of \cite{Gross:1993yt} gives the contribution from these configurations (the first line there gives the contributions listed above), up to an overall factor of $N^2$ that is missing there. For $\lambda=0$ it gives
\begin{equation}  
    \boxed{\langle W_{mixed} \rangle = 2N^3 - 2N}.
\end{equation}  

The first term here ($2N^3$) arises from disconnected but mixed-orientation configurations, specifically \( S \otimes \bar{D} \) and \( \bar{S} \otimes D \), when we have a sphere wrapped holomorphically and a disk anti-holomorphically or vice versa. Similar to the discussion above, the Euler characteristic for these configurations is \( \chi(S \otimes \bar{D}) = \chi(\bar{S} \otimes D) = 3 \), resulting in a total contribution of \( 2N^3 \).  

The second term is a contribution from connected worldsheets with an orientation-reversing tube (ORT) (for $\lambda=0$ only one such tube is allowed in the discussion of \cite{Gross:1993yt}, coming from an $\Omega$-point). A worldsheet disk (\( D \)) with an ORT (such that in target space we have a disk and a sphere connected by the ORT), and an anti-disk (\( \bar{D} \)) with an ORT, each contribute negatively, due the presence of a negative mode in the one loop fluctuation spectrum around the ORT. These mappings have a moduli space labelled by the target-space position of the ORT, which is a disk. The Euler characteristic of the moduli space for these configurations is thus \( \chi(D \text{ with ORT}) = \chi(\bar{D} \text{ with ORT}) = 1 \), leading to a total negative contribution of \( -2N \) (since the worldsheet is a disk with $\chi=1$). 
The details of these contributions and their topological interpretations will be further elaborated in Section \ref{sec: disk with ORT}.  

By summing all the contributions described above (with all possible orientations), we arrive at the full result (up to these winding numbers):  
\begin{equation}
    \boxed{\langle W_1 \rangle = 2N^3 + 2N + 2N^3 - 2N  = 4N^3,}
\end{equation}
which agrees with \cite{Gross:1993yt}.

\subsection{$G=0$, $\gamma_8$, $k=1$  : Twist}

Let us next consider the target space to be a sphere, \( G=0 \), with the self-intersecting ``figure $8$'' Wilson loop \( \gamma_8 \) (in the middle of figure \ref{fig:All WL}), with winding number \( k=1 \). The mapping of the worldsheet in this case can include a twist. The relationship between the covering numbers is given by \( n_1 - \tilde{n}_1 = n_3 - \tilde{n}_3 + 1 = n_2 - \tilde{n}_2 + 2 \). In addition, we have the constraints on the number of worldsheets $n_1\geq n_3\geq n_2$ and $\tilde n_2 \geq\tilde n_3 \geq\tilde n_1$. 

\subsubsection{$\tilde n_i=0$, $n_1 = n_3 + 1 = n_2 + 2$,~$n_2=0$} \label{twice_winding_disk}

\begin{figure}[h]
    \centering

\tikzset{every picture/.style={line width=0.75pt}} 

\begin{tikzpicture}[x=0.75pt,y=0.75pt,yscale=-1,xscale=1]

\draw (463.75,203.25) node  {\includegraphics[width=202.13pt,height=204.38pt]{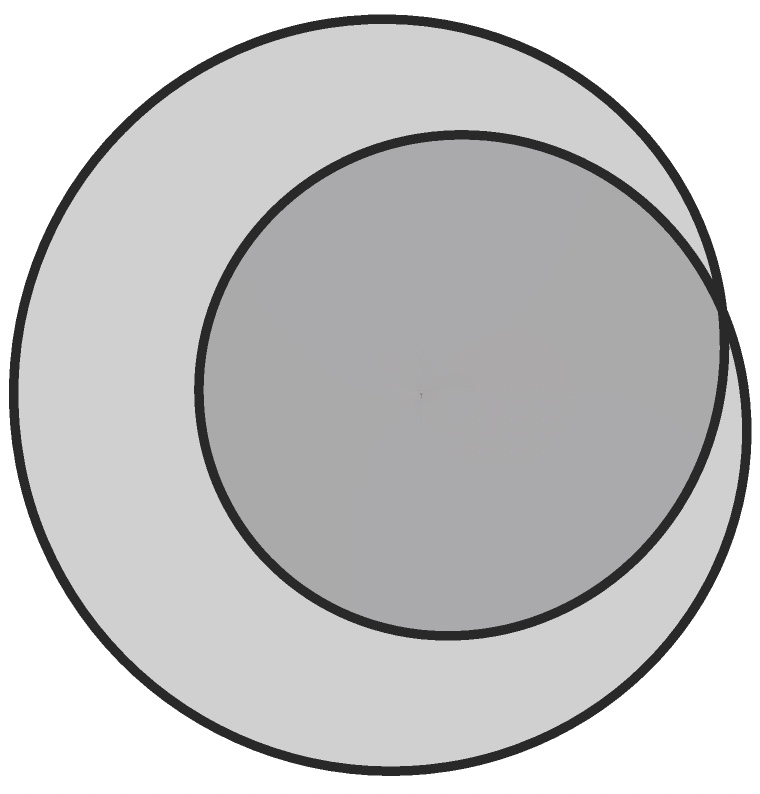}};
\draw [line width=1.5]    (364.5,120) -- (374.55,108.28) ;
\draw [shift={(376.5,106)}, rotate = 130.6] [color={rgb, 255:red, 0; green, 0; blue, 0 }  ][line width=1.5]    (14.21,-6.37) .. controls (9.04,-2.99) and (4.3,-0.87) .. (0,0) .. controls (4.3,0.87) and (9.04,2.99) .. (14.21,6.37)   ;
\draw [line width=1.5]    (413.5,154) -- (423.55,142.28) ;
\draw [shift={(425.5,140)}, rotate = 130.6] [color={rgb, 255:red, 0; green, 0; blue, 0 }  ][line width=1.5]    (14.21,-6.37) .. controls (9.04,-2.99) and (4.3,-0.87) .. (0,0) .. controls (4.3,0.87) and (9.04,2.99) .. (14.21,6.37)   ;

\draw (327,90.4) node [anchor=north west][inner sep=0.75pt]  [font=\Large]  {$\gamma _{2}$};
\draw (480,180.4) node [anchor=north west][inner sep=0.75pt]  [font=\Large]  {$M_{1}$};
\draw (282,260.4) node [anchor=north west][inner sep=0.75pt]  [font=\large]  {$M_{2}$};
\draw (353,192.4) node [anchor=north west][inner sep=0.75pt]  [font=\Large]  {$M_{3}$};

\end{tikzpicture}

    \caption{A different drawing of the ``figure 8'' Wilson loop, which is more suitable when the area of the second region goes to infinity.}
    \label{fig8:2}
\end{figure}
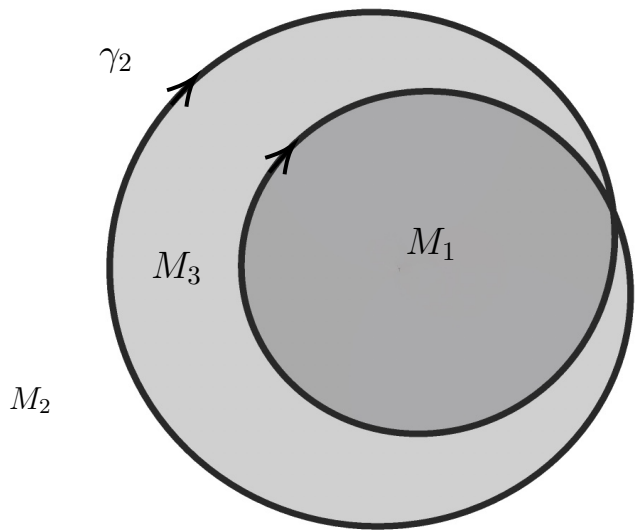

Since $n_2=\tilde n_2=0$, an interesting limiting case where this is the only contribution is if we take the $2^{\rm nd}$ region to have infinite area. Then we get a Wilson loop on the plane that looks like figure \ref{fig8:2}.

The exact answer for this Wilson loop (at $\lambda=0$) is (equation (7.13) of \cite{Gross:1993yt})
\begin{equation} \label{disk_two}
    \boxed{\lim_{A_2\rightarrow \infty} \langle W_\gamma \rangle  = N.}
\end{equation}
This configuration is equivalent (by taking $A_3 \to 0$) to a circular Wilson loop with winding number $k=2$. The only modulus is a branch point, and we have a holomorphic mapping. The moduli space in this case (corresponding to the position of the branch point) is a Disk, so we get from \eqref{eq:general_result}
$\chi(D) N^{\chi(D)} = \bf N$, in agreement with \eqref{disk_two}. 
We will explore this in detail in section \ref{sec: branch point}. 

Similarly, the expectation value on the plane of any other Wilson loop winding $k$ times around the disk is also equal to $N$ (see \cite{Bralic:1980ra} for an explicit computation in the $k=3$ case for any coupling).

\subsubsection{$n_1=1$, $\tilde{n}_2=1$}  \label{figure_eight}

Alternatively, we can take the third region to have infinite size so that we have a ``figure 8'' on the plane. Then, the only configuration that contributes has winding numbers \( n_1 = 1 \) and \( \tilde{n}_2 = 1 \), defining a mapping from a disk to a figure 8 configuration with a twist (see figure \ref{fig:All WL}).
The geometry has no moduli, simplifying the calculation of the contribution.  The resulting contribution to the Wilson loop from YM is (from equation (7.16) of \cite{Gross:1993yt})  
\begin{equation}  
   \boxed{\langle W \rangle = N.}  
\end{equation}  

The Euler characteristic of the disk is \( \chi(D) = 1 \), hence we obtain from our string theory \( N^{\chi(D)} = \mathbf{N} \). We will explore this in detail in section \ref{sec: twist}.

On the sphere, there are also additional configurations with higher winding numbers, whose analysis we leave to future work.

\subsection{$G=0$, $\gamma_2$, $k=1$ : two intersecting disks with the same orientation}  

The target space is again a sphere, \( G=0 \), with a Wilson loop \( \gamma_2 \) (on the right in figure \ref{fig:All WL}) with winding number \( k=1 \). The geometry in this case corresponds to two intersecting circles that share the same orientation. The relationship between the quantum numbers is given by \( n_3-\tilde{n}_3 = n_1-\tilde{n}_1 + 1 = n_2-\tilde{n}_2 + 1 = n_4-\tilde{n}_4 + 2 \). This configuration captures the interplay between the disks ending on the two circles. 

\subsubsection{$\tilde n_i=0$, $n_3 = n_1 + 1 = n_2 + 1 = n_4 + 2$, $n_4 = 0$ : Holomorphic}

In the limit $A_4\rightarrow \infty$, we cannot have any worldsheet on the 4th region. The Wilson loop expectation value for this configuration is (equation (7.21) of \cite{Gross:1993yt})
\begin{equation}
    \boxed{\lim_{A_4\rightarrow \infty} \langle W_\gamma \rangle  = N^2.}
\end{equation}

Looking at the orientation of the Wilson loop, we can easily see that with $n_3=2,~n_1=n_2=1$, the only option is to have two disconnected oriented disks without any moduli, giving $N^{\chi(D)+\chi(D)} = \bf N^2$.

\subsubsection{$A_2\rightarrow \infty$, $n_3=\tilde n_4=1$ : Coupled theory}

\begin{figure}[h]
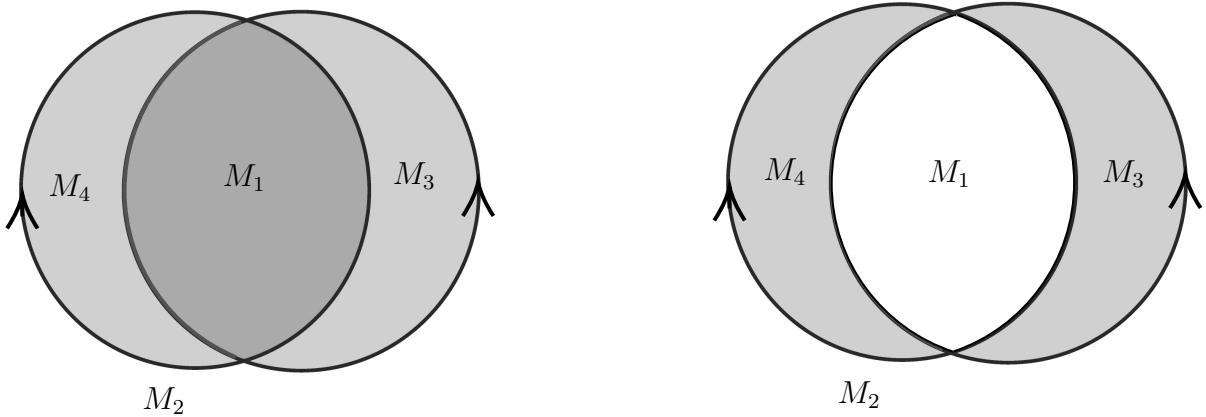

    \centering

\tikzset{every picture/.style={line width=0.75pt}} 

\begin{tikzpicture}[x=0.75pt,y=0.75pt,yscale=-1,xscale=1]

\draw (196,175) node  {\includegraphics[width=177.75pt,height=141.75pt]{Image6.jpeg}};
\draw [line width=1.5]    (82.5,187) -- (82.5,179) ;
\draw [shift={(82.5,176)}, rotate = 90] [color={rgb, 255:red, 0; green, 0; blue, 0 }  ][line width=1.5]    (17.05,-7.64) .. controls (10.84,-3.59) and (5.16,-1.04) .. (0,0) .. controls (5.16,1.04) and (10.84,3.59) .. (17.05,7.64)   ;
\draw [line width=1.5]    (310,173.75) -- (310,172.75) ;
\draw [shift={(310,169.75)}, rotate = 90] [color={rgb, 255:red, 0; green, 0; blue, 0 }  ][line width=1.5]    (17.05,-7.64) .. controls (10.84,-3.59) and (5.16,-1.04) .. (0,0) .. controls (5.16,1.04) and (10.84,3.59) .. (17.05,7.64)   ;
\draw  [draw opacity=0][line width=1.5]  (189.04,257.83) .. controls (156.4,244.01) and (133.5,211.68) .. (133.5,174) .. controls (133.5,135.31) and (157.65,102.25) .. (191.7,89.09) -- (224.5,174) -- cycle ; \draw  [color={rgb, 255:red, 82; green, 82; blue, 82 }  ,draw opacity=1 ][line width=1.5]  (189.04,257.83) .. controls (156.4,244.01) and (133.5,211.68) .. (133.5,174) .. controls (133.5,135.31) and (157.65,102.25) .. (191.7,89.09) ;  
\draw (549,171) node  {\includegraphics[width=177.75pt,height=141.75pt]{Image6.jpeg}};
\draw [line width=1.5]    (435.5,183) -- (435.5,175) ;
\draw [shift={(435.5,172)}, rotate = 90] [color={rgb, 255:red, 0; green, 0; blue, 0 }  ][line width=1.5]    (17.05,-7.64) .. controls (10.84,-3.59) and (5.16,-1.04) .. (0,0) .. controls (5.16,1.04) and (10.84,3.59) .. (17.05,7.64)   ;
\draw [line width=1.5]    (663,169.75) -- (663,168.75) ;
\draw [shift={(663,165.75)}, rotate = 90] [color={rgb, 255:red, 0; green, 0; blue, 0 }  ][line width=1.5]    (17.05,-7.64) .. controls (10.84,-3.59) and (5.16,-1.04) .. (0,0) .. controls (5.16,1.04) and (10.84,3.59) .. (17.05,7.64)   ;
\draw  [draw opacity=0][line width=1.5]  (542.04,253.83) .. controls (509.4,240.01) and (486.5,207.68) .. (486.5,170) .. controls (486.5,131.31) and (510.65,98.25) .. (544.7,85.09) -- (577.5,170) -- cycle ; \draw  [color={rgb, 255:red, 82; green, 82; blue, 82 }  ,draw opacity=1 ][line width=1.5]  (542.04,253.83) .. controls (509.4,240.01) and (486.5,207.68) .. (486.5,170) .. controls (486.5,131.31) and (510.65,98.25) .. (544.7,85.09) ;  
\draw  [draw opacity=0][fill={rgb, 255:red, 255; green, 255; blue, 255 }  ,fill opacity=1 ] (547.59,254.99) .. controls (512.68,243.24) and (487.5,209.85) .. (487.5,170.5) .. controls (487.5,130.84) and (513.08,97.24) .. (548.41,85.74) -- (575.25,170.5) -- cycle ; \draw   (547.59,254.99) .. controls (512.68,243.24) and (487.5,209.85) .. (487.5,170.5) .. controls (487.5,130.84) and (513.08,97.24) .. (548.41,85.74) ;  
\draw  [draw opacity=0][fill={rgb, 255:red, 255; green, 255; blue, 255 }  ,fill opacity=1 ] (547.07,255.03) .. controls (511.89,243.33) and (486.5,209.9) .. (486.5,170.5) .. controls (486.5,130.79) and (512.29,97.15) .. (547.9,85.7) -- (574.75,170.5) -- cycle ; \draw   (547.07,255.03) .. controls (511.89,243.33) and (486.5,209.9) .. (486.5,170.5) .. controls (486.5,130.79) and (512.29,97.15) .. (547.9,85.7) ;  
\draw  [draw opacity=0][fill={rgb, 255:red, 255; green, 255; blue, 255 }  ,fill opacity=1 ] (549.29,85.75) .. controls (583.23,98.16) and (607.5,131.2) .. (607.5,170) .. controls (607.5,209.49) and (582.36,243.01) .. (547.48,254.89) -- (519.54,170) -- cycle ; \draw   (549.29,85.75) .. controls (583.23,98.16) and (607.5,131.2) .. (607.5,170) .. controls (607.5,209.49) and (582.36,243.01) .. (547.48,254.89) ;  

\draw (181,159.4) node [anchor=north west][inner sep=0.75pt]  [font=\large]  {$M_{1}$};
\draw (141,271.4) node [anchor=north west][inner sep=0.75pt]  [font=\large]  {$M_{2}$};
\draw (266,158.4) node [anchor=north west][inner sep=0.75pt]  [font=\large]  {$M_{3}$};
\draw (94,164.4) node [anchor=north west][inner sep=0.75pt]  [font=\large]  {$M_{4}$};
\draw (533,157.4) node [anchor=north west][inner sep=0.75pt]  [font=\large]  {$M_{1}$};
\draw (488,268.4) node [anchor=north west][inner sep=0.75pt]  [font=\large]  {$M_{2}$};
\draw (620,157.4) node [anchor=north west][inner sep=0.75pt]  [font=\large]  {$M_{3}$};
\draw (451,156.4) node [anchor=north west][inner sep=0.75pt]  [font=\large]  {$M_{4}$};

\end{tikzpicture}

    \caption{A different drawing of the two-circle Wilson loop from figure \ref{fig:All WL}, which is more suitable when the second area goes to infinity.}
    \label{fig: two opposite loop}
\end{figure}

In the limit $A_2\to \infty$, it is more natural to draw the Wilson loop as in
figure \ref{fig: two opposite loop} on the plane, where we have two intersecting circles with opposite orientations. 
The simplest configurations have covering numbers \( n_3 = \tilde{n}_4 = 1 \).

{\bf Case 1: \( n_1 = \tilde{n}_1 = 0 \)}

In this case, the worldsheet is confined to regions \( 3 \) and \( 4 \) of the Wilson loop diagram, with opposite orientations, as on the right-hand side of figure \ref{fig: two opposite loop}. The YM answer associated with this configuration is (equation (7.22) of \cite{Gross:1993yt})
\begin{equation}  
    \boxed{\langle W \rangle = 1.}  
\end{equation}  
The mapping corresponds to an annulus (two boundaries) with two twists (which locally look like the figure $8$ diagram discussed above), and there are no moduli associated with this geometry. The Euler characteristic of the annulus is \( \chi(A) = 0 \). 
So the annulus configuration, with opposite orientations in regions \( 3 \) and \( 4 \), yields \( N^{\chi(A)} = \mathbf{1} \).

{\bf Case 2 : $n_1=\tilde n_1 = 1$}

In this case, the Wilson loop contribution is given by (equation (7.24) of \cite{Gross:1993yt}) 
\begin{equation}  
    \boxed{\langle W \rangle = N^2 - 1.}  
\end{equation}

The allowed configurations now look like the left-hand side of figure \ref{fig: two opposite loop}. They include the disconnected contributions from \( D \otimes \bar{D} \) (a disk and an oppositely oriented disk), or a connected annulus \( A \) with an orientation-reversing tube (ORT) connecting the two sheets. 

$N^2$: This contribution comes from the product of disconnected worldsheets \( D \otimes \bar{D} \), where the Euler characteristics satisfy \( \chi(D) + \chi(\bar{D}) = 2 \). With no moduli, this gives \( N^{\chi(D) + \chi(\bar{D})} = \mathbf{N^2} \).

$N^0$: This arises from the annulus \( A \) with an ORT, where the Euler characteristic of the worldsheet is \( \chi(A) = 0 \), contributing \( -N^{\chi(A)} = -\mathbf{1} \). The ORT introduces a negative sign, explaining the \(-1\) term due to its negative mode responsible for expanding tube size, and its moduli space has the topology of a disk (since the ORT can be anywhere in region 1), with $\chi(D)=1$. We will explain this in detail in section \ref{sec: two disk with ORT}.

\subsection{$G=0$, $\gamma_x$, $k=1$ : Special twist}
\label{sec:special_twist_YM}

\begin{figure}[h]
    \centering

\tikzset{every picture/.style={line width=0.75pt}} 

\begin{tikzpicture}[x=0.75pt,y=0.75pt,yscale=-1,xscale=1]

\draw (619.14,366.65) node  {\includegraphics[width=140.04pt,height=116.48pt]{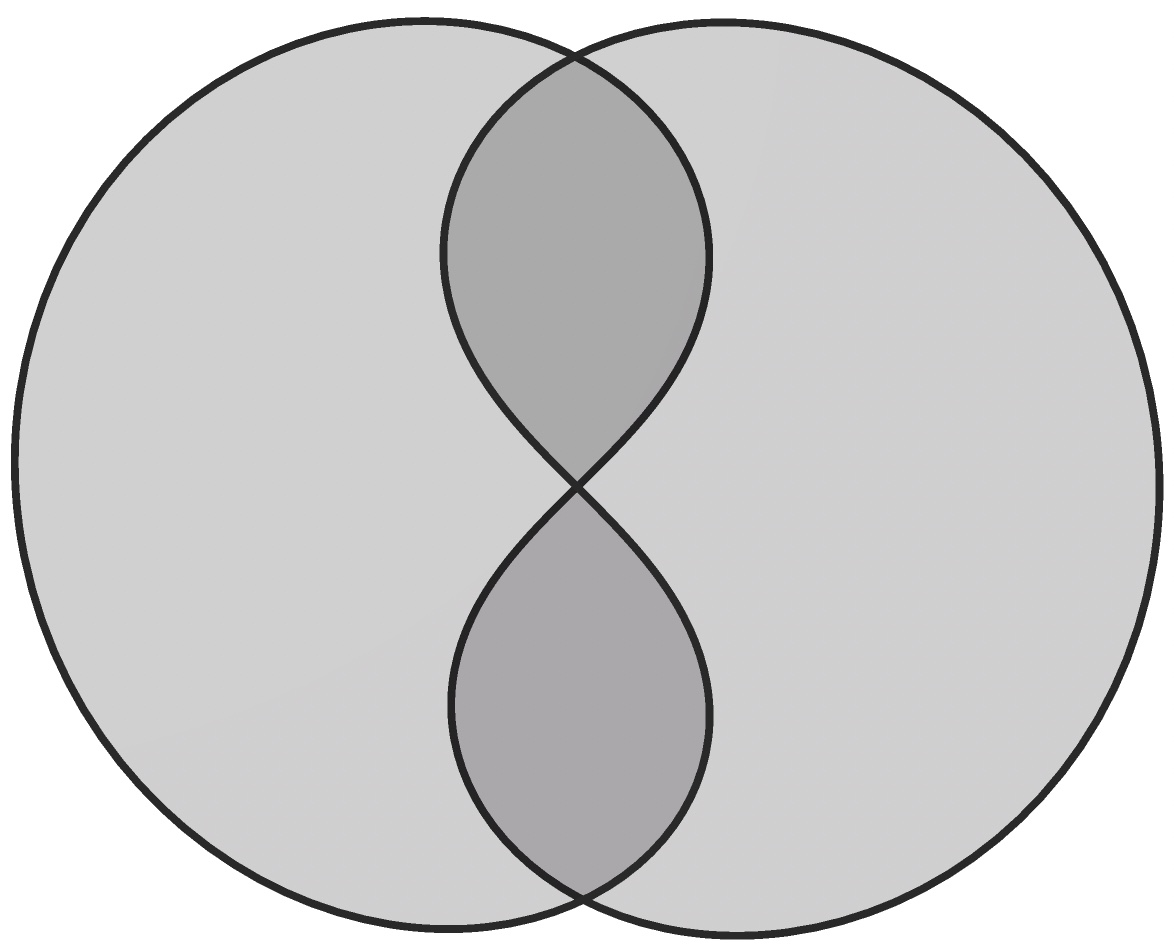}};
\draw [line width=0.75]    (527.86,364.56) -- (528.36,359.24) ;
\draw [shift={(528.55,357.25)}, rotate = 95.42] [color={rgb, 255:red, 0; green, 0; blue, 0 }  ][line width=0.75]    (9.84,-4.41) .. controls (6.25,-2.07) and (2.97,-0.6) .. (0,0) .. controls (2.97,0.6) and (6.25,2.07) .. (9.84,4.41)   ;
\draw [line width=0.75]    (709.72,360.04) -- (710.14,362.93) ;
\draw [shift={(710.42,364.91)}, rotate = 261.9] [color={rgb, 255:red, 0; green, 0; blue, 0 }  ][line width=0.75]    (9.84,-4.41) .. controls (6.25,-2.07) and (2.97,-0.6) .. (0,0) .. controls (2.97,0.6) and (6.25,2.07) .. (9.84,4.41)   ;
\draw  [color={rgb, 255:red, 255; green, 255; blue, 255 }  ,draw opacity=1 ][fill={rgb, 255:red, 255; green, 255; blue, 255 }  ,fill opacity=1 ] (598.56,404.26) .. controls (598.56,393.46) and (607.29,384.71) .. (618.05,384.71) .. controls (628.81,384.71) and (637.53,393.46) .. (637.53,404.26) .. controls (637.53,415.06) and (628.81,423.81) .. (618.05,423.81) .. controls (607.29,423.81) and (598.56,415.06) .. (598.56,404.26) -- cycle ;
\draw  [color={rgb, 255:red, 255; green, 255; blue, 255 }  ,draw opacity=1 ][fill={rgb, 255:red, 255; green, 255; blue, 255 }  ,fill opacity=1 ] (628.12,381.63) -- (632.67,388.98) -- (635.61,394.93) -- (636.15,396.25) -- (598.97,399.5) -- (600.74,393.12) -- (602.44,389.17) -- (605.91,383.48) -- (611.54,376.9) -- (617.71,370.37) -- (622.71,375.48) -- cycle ;
\draw  [color={rgb, 255:red, 255; green, 255; blue, 255 }  ,draw opacity=1 ][fill={rgb, 255:red, 255; green, 255; blue, 255 }  ,fill opacity=1 ] (632.67,423.06) -- (636.49,414.71) -- (637.53,406.35) -- (637.96,400.08) -- (598.7,402.09) -- (599.24,407.74) -- (600.05,414.01) -- (604.02,422.29) -- (611.16,430.03) -- (618.48,435.25) -- (626.43,430.03) -- cycle ;
\draw [color={rgb, 255:red, 55; green, 55; blue, 55 }  ,draw opacity=0.89 ][line width=0.75]    (597.97,400.08) -- (597.47,405.05) ;
\draw [shift={(597.27,407.04)}, rotate = 275.69] [color={rgb, 255:red, 55; green, 55; blue, 55 }  ,draw opacity=0.89 ][line width=0.75]    (9.84,-4.41) .. controls (6.25,-2.07) and (2.97,-0.6) .. (0,0) .. controls (2.97,0.6) and (6.25,2.07) .. (9.84,4.41)   ;
\draw [line width=0.75]    (638.92,404.96) -- (637.27,397.51) ;
\draw [shift={(636.84,395.55)}, rotate = 77.51] [color={rgb, 255:red, 0; green, 0; blue, 0 }  ][line width=0.75]    (9.84,-4.41) .. controls (6.25,-2.07) and (2.97,-0.6) .. (0,0) .. controls (2.97,0.6) and (6.25,2.07) .. (9.84,4.41)   ;
\draw  [color={rgb, 255:red, 255; green, 255; blue, 255 }  ,draw opacity=1 ][fill={rgb, 255:red, 255; green, 255; blue, 255 }  ,fill opacity=1 ] (633.02,348.27) -- (636.11,341.23) -- (637.15,334.96) -- (636.88,332.18) -- (597.62,334.19) -- (598.85,340.54) -- (601.17,345.95) -- (605.02,352.99) -- (609.77,358.99) -- (617.4,367.35) -- (627.93,355.93) -- cycle ;
\draw  [color={rgb, 255:red, 255; green, 255; blue, 255 }  ,draw opacity=1 ][fill={rgb, 255:red, 255; green, 255; blue, 255 }  ,fill opacity=1 ] (627.82,307.57) -- (632.37,312.83) -- (635.99,320.18) -- (637.22,326.38) -- (597.97,325.45) -- (599.51,319.88) -- (602.13,313.72) -- (605.6,308.73) -- (611.23,303.55) -- (617.4,299.1) -- (622.4,302.82) -- cycle ;
\draw  [color={rgb, 255:red, 255; green, 255; blue, 255 }  ,draw opacity=1 ][fill={rgb, 255:red, 255; green, 255; blue, 255 }  ,fill opacity=1 ] (597.62,329.6) .. controls (597.62,318.53) and (606.56,309.56) .. (617.6,309.56) .. controls (628.63,309.56) and (637.57,318.53) .. (637.57,329.6) .. controls (637.57,340.67) and (628.63,349.64) .. (617.6,349.64) .. controls (606.56,349.64) and (597.62,340.67) .. (597.62,329.6) -- cycle ;
\draw [color={rgb, 255:red, 64; green, 64; blue, 64 }  ,draw opacity=0.89 ][line width=0.75]    (637.92,326.38) -- (638.41,331.35) ;
\draw [shift={(638.61,333.34)}, rotate = 264.31] [color={rgb, 255:red, 64; green, 64; blue, 64 }  ,draw opacity=0.89 ][line width=0.75]    (9.84,-4.41) .. controls (6.25,-2.07) and (2.97,-0.6) .. (0,0) .. controls (2.97,0.6) and (6.25,2.07) .. (9.84,4.41)   ;
\draw [line width=0.75]    (596.23,334.19) -- (597.04,327.43) ;
\draw [shift={(597.27,325.45)}, rotate = 96.79] [color={rgb, 255:red, 0; green, 0; blue, 0 }  ][line width=0.75]    (9.84,-4.41) .. controls (6.25,-2.07) and (2.97,-0.6) .. (0,0) .. controls (2.97,0.6) and (6.25,2.07) .. (9.84,4.41)   ;
\draw (612.33,168.2) node  {\includegraphics[width=141.25pt,height=115.8pt]{Image4.jpeg}};
\draw [line width=0.75]    (520.26,166.12) -- (520.77,160.84) ;
\draw [shift={(520.96,158.85)}, rotate = 95.5] [color={rgb, 255:red, 0; green, 0; blue, 0 }  ][line width=0.75]    (9.84,-4.41) .. controls (6.25,-2.07) and (2.97,-0.6) .. (0,0) .. controls (2.97,0.6) and (6.25,2.07) .. (9.84,4.41)   ;
\draw [line width=0.75]    (703.7,161.62) -- (704.11,164.49) ;
\draw [shift={(704.4,166.47)}, rotate = 261.78] [color={rgb, 255:red, 0; green, 0; blue, 0 }  ][line width=0.75]    (9.84,-4.41) .. controls (6.25,-2.07) and (2.97,-0.6) .. (0,0) .. controls (2.97,0.6) and (6.25,2.07) .. (9.84,4.41)   ;
\draw  [color={rgb, 255:red, 255; green, 255; blue, 255 }  ,draw opacity=1 ][fill={rgb, 255:red, 255; green, 255; blue, 255 }  ,fill opacity=1 ] (591.58,205.59) .. controls (591.58,194.86) and (600.38,186.15) .. (611.23,186.15) .. controls (622.09,186.15) and (630.89,194.86) .. (630.89,205.59) .. controls (630.89,216.32) and (622.09,225.03) .. (611.23,225.03) .. controls (600.38,225.03) and (591.58,216.32) .. (591.58,205.59) -- cycle ;
\draw  [color={rgb, 255:red, 255; green, 255; blue, 255 }  ,draw opacity=1 ][fill={rgb, 255:red, 255; green, 255; blue, 255 }  ,fill opacity=1 ] (621.39,183.09) -- (625.98,190.4) -- (628.94,196.32) -- (629.48,197.63) -- (591.99,200.86) -- (593.78,194.51) -- (595.49,190.59) -- (598.99,184.93) -- (604.67,178.39) -- (610.89,171.89) -- (615.93,176.98) -- cycle ;
\draw [color={rgb, 255:red, 64; green, 64; blue, 64 }  ,draw opacity=0.89 ][line width=0.75]    (631.27,128.16) -- (631.77,133.09) ;
\draw [shift={(631.97,135.08)}, rotate = 264.23] [color={rgb, 255:red, 64; green, 64; blue, 64 }  ,draw opacity=0.89 ][line width=0.75]    (9.84,-4.41) .. controls (6.25,-2.07) and (2.97,-0.6) .. (0,0) .. controls (2.97,0.6) and (6.25,2.07) .. (9.84,4.41)   ;
\draw [line width=0.75]    (589.23,135.93) -- (590.04,129.22) ;
\draw [shift={(590.28,127.23)}, rotate = 96.89] [color={rgb, 255:red, 0; green, 0; blue, 0 }  ][line width=0.75]    (9.84,-4.41) .. controls (6.25,-2.07) and (2.97,-0.6) .. (0,0) .. controls (2.97,0.6) and (6.25,2.07) .. (9.84,4.41)   ;
\draw  [color={rgb, 255:red, 255; green, 255; blue, 255 }  ,draw opacity=1 ][fill={rgb, 255:red, 255; green, 255; blue, 255 }  ,fill opacity=1 ] (625.98,224.28) -- (629.83,215.98) -- (630.89,207.67) -- (631.31,201.44) -- (591.72,203.44) -- (592.26,209.05) -- (593.08,215.28) -- (597.08,223.52) -- (604.28,231.21) -- (611.67,236.4) -- (619.68,231.21) -- cycle ;
\draw [color={rgb, 255:red, 55; green, 55; blue, 55 }  ,draw opacity=0.89 ][line width=0.75]    (590.98,201.44) -- (590.48,206.37) ;
\draw [shift={(590.28,208.36)}, rotate = 275.77] [color={rgb, 255:red, 55; green, 55; blue, 55 }  ,draw opacity=0.89 ][line width=0.75]    (9.84,-4.41) .. controls (6.25,-2.07) and (2.97,-0.6) .. (0,0) .. controls (2.97,0.6) and (6.25,2.07) .. (9.84,4.41)   ;
\draw [line width=0.75]    (632.29,206.28) -- (630.62,198.89) ;
\draw [shift={(630.19,196.94)}, rotate = 77.34] [color={rgb, 255:red, 0; green, 0; blue, 0 }  ][line width=0.75]    (9.84,-4.41) .. controls (6.25,-2.07) and (2.97,-0.6) .. (0,0) .. controls (2.97,0.6) and (6.25,2.07) .. (9.84,4.41)   ;
\draw (313.61,378.22) node  {\includegraphics[width=140.83pt,height=120.32pt]{Image4.jpeg}};
\draw [line width=0.75]    (221.82,376.06) -- (222.33,370.49) ;
\draw [shift={(222.52,368.5)}, rotate = 95.28] [color={rgb, 255:red, 0; green, 0; blue, 0 }  ][line width=0.75]    (9.84,-4.41) .. controls (6.25,-2.07) and (2.97,-0.6) .. (0,0) .. controls (2.97,0.6) and (6.25,2.07) .. (9.84,4.41)   ;
\draw [color={rgb, 255:red, 55; green, 55; blue, 55 }  ,draw opacity=0.89 ][line width=0.75]    (292.32,412.75) -- (291.82,417.95) ;
\draw [shift={(291.62,419.94)}, rotate = 275.54] [color={rgb, 255:red, 55; green, 55; blue, 55 }  ,draw opacity=0.89 ][line width=0.75]    (9.84,-4.41) .. controls (6.25,-2.07) and (2.97,-0.6) .. (0,0) .. controls (2.97,0.6) and (6.25,2.07) .. (9.84,4.41)   ;
\draw [line width=0.75]    (333.51,417.78) -- (331.83,410.03) ;
\draw [shift={(331.41,408.07)}, rotate = 77.83] [color={rgb, 255:red, 0; green, 0; blue, 0 }  ][line width=0.75]    (9.84,-4.41) .. controls (6.25,-2.07) and (2.97,-0.6) .. (0,0) .. controls (2.97,0.6) and (6.25,2.07) .. (9.84,4.41)   ;
\draw [line width=0.75]    (404.71,371.38) -- (405.13,374.44) ;
\draw [shift={(405.41,376.42)}, rotate = 262.11] [color={rgb, 255:red, 0; green, 0; blue, 0 }  ][line width=0.75]    (9.84,-4.41) .. controls (6.25,-2.07) and (2.97,-0.6) .. (0,0) .. controls (2.97,0.6) and (6.25,2.07) .. (9.84,4.41)   ;
\draw  [color={rgb, 255:red, 255; green, 255; blue, 255 }  ,draw opacity=1 ][fill={rgb, 255:red, 255; green, 255; blue, 255 }  ,fill opacity=1 ] (292.26,342.6) .. controls (292.26,331.45) and (301.03,322.41) .. (311.86,322.41) .. controls (322.68,322.41) and (331.45,331.45) .. (331.45,342.6) .. controls (331.45,353.76) and (322.68,362.8) .. (311.86,362.8) .. controls (301.03,362.8) and (292.26,353.76) .. (292.26,342.6) -- cycle ;
\draw  [color={rgb, 255:red, 255; green, 255; blue, 255 }  ,draw opacity=1 ][fill={rgb, 255:red, 255; green, 255; blue, 255 }  ,fill opacity=1 ] (322.34,317.18) -- (326.91,322.62) -- (330.56,330.21) -- (331.8,336.61) -- (292.32,335.65) -- (293.87,329.89) -- (296.51,323.54) -- (300,318.38) -- (305.66,313.03) -- (311.87,308.43) -- (316.89,312.28) -- cycle ;
\draw  [color={rgb, 255:red, 255; green, 255; blue, 255 }  ,draw opacity=1 ][fill={rgb, 255:red, 255; green, 255; blue, 255 }  ,fill opacity=1 ] (327.57,359.23) -- (330.68,351.96) -- (331.72,345.48) -- (331.45,342.6) -- (291.97,344.68) -- (293.21,351.24) -- (295.54,356.83) -- (299.42,364.11) -- (304.19,370.3) -- (311.87,378.93) -- (322.45,367.14) -- cycle ;
\draw [color={rgb, 255:red, 64; green, 64; blue, 64 }  ,draw opacity=0.89 ][line width=0.75]    (332.5,336.61) -- (333,341.81) ;
\draw [shift={(333.2,343.8)}, rotate = 264.46] [color={rgb, 255:red, 64; green, 64; blue, 64 }  ,draw opacity=0.89 ][line width=0.75]    (9.84,-4.41) .. controls (6.25,-2.07) and (2.97,-0.6) .. (0,0) .. controls (2.97,0.6) and (6.25,2.07) .. (9.84,4.41)   ;
\draw [line width=0.75]    (290.58,344.68) -- (291.39,337.64) ;
\draw [shift={(291.62,335.65)}, rotate = 96.61] [color={rgb, 255:red, 0; green, 0; blue, 0 }  ][line width=0.75]    (9.84,-4.41) .. controls (6.25,-2.07) and (2.97,-0.6) .. (0,0) .. controls (2.97,0.6) and (6.25,2.07) .. (9.84,4.41)   ;
\draw (310.13,171.63) node  {\includegraphics[width=140.06pt,height=119.44pt]{Image4.jpeg}};
\draw [line width=0.75]    (218.82,171.23) -- (219.38,162.88) ;
\draw [shift={(219.51,160.88)}, rotate = 93.81] [color={rgb, 255:red, 0; green, 0; blue, 0 }  ][line width=0.75]    (9.84,-4.41) .. controls (6.25,-2.07) and (2.97,-0.6) .. (0,0) .. controls (2.97,0.6) and (6.25,2.07) .. (9.84,4.41)   ;
\draw [line width=0.75]    (329.77,215.82) -- (329.21,207.47) ;
\draw [shift={(329.08,205.47)}, rotate = 86.19] [color={rgb, 255:red, 0; green, 0; blue, 0 }  ][line width=0.75]    (9.84,-4.41) .. controls (6.25,-2.07) and (2.97,-0.6) .. (0,0) .. controls (2.97,0.6) and (6.25,2.07) .. (9.84,4.41)   ;
\draw [line width=0.75]    (287.73,140.18) -- (287.73,131.03) ;
\draw [shift={(287.73,129.03)}, rotate = 90] [color={rgb, 255:red, 0; green, 0; blue, 0 }  ][line width=0.75]    (9.84,-4.41) .. controls (6.25,-2.07) and (2.97,-0.6) .. (0,0) .. controls (2.97,0.6) and (6.25,2.07) .. (9.84,4.41)   ;
\draw [line width=0.75]    (400.74,164.86) -- (401.19,168.45) ;
\draw [shift={(401.43,170.44)}, rotate = 262.95] [color={rgb, 255:red, 0; green, 0; blue, 0 }  ][line width=0.75]    (9.84,-4.41) .. controls (6.25,-2.07) and (2.97,-0.6) .. (0,0) .. controls (2.97,0.6) and (6.25,2.07) .. (9.84,4.41)   ;
\draw [color={rgb, 255:red, 42; green, 42; blue, 42 }  ,draw opacity=0.89 ][line width=0.75]    (329.08,129.03) -- (329.6,135) ;
\draw [shift={(329.77,136.99)}, rotate = 265.05] [color={rgb, 255:red, 42; green, 42; blue, 42 }  ,draw opacity=0.89 ][line width=0.75]    (9.84,-4.41) .. controls (6.25,-2.07) and (2.97,-0.6) .. (0,0) .. controls (2.97,0.6) and (6.25,2.07) .. (9.84,4.41)   ;
\draw [color={rgb, 255:red, 42; green, 42; blue, 42 }  ,draw opacity=0.89 ][line width=0.75]    (288.42,207.06) -- (288.94,213.04) ;
\draw [shift={(289.11,215.03)}, rotate = 265.05] [color={rgb, 255:red, 42; green, 42; blue, 42 }  ,draw opacity=0.89 ][line width=0.75]    (9.84,-4.41) .. controls (6.25,-2.07) and (2.97,-0.6) .. (0,0) .. controls (2.97,0.6) and (6.25,2.07) .. (9.84,4.41)   ;

\draw (539.96,424.5) node [anchor=north west][inner sep=0.75pt]  [font=\large]  {$\gamma _{x}$};
\draw (489.61,401.52) node [anchor=north west][inner sep=0.75pt]  [font=\normalsize]  {$M_{5}$};
\draw (552.53,357.86) node [anchor=north west][inner sep=0.75pt]  [font=\normalsize]  {$M_{1}$};
\draw (656.2,356.25) node [anchor=north west][inner sep=0.75pt]  [font=\normalsize]  {$M_{4}$};
\draw (605.4,319.18) node [anchor=north west][inner sep=0.75pt]  [font=\normalsize]  {$M_{2}$};
\draw (606.44,393.57) node [anchor=north west][inner sep=0.75pt]  [font=\normalsize]  {$M_{3}$};
\draw (532.57,225.65) node [anchor=north west][inner sep=0.75pt]  [font=\large]  {$\gamma _{x}$};
\draw (481.81,202.8) node [anchor=north west][inner sep=0.75pt]  [font=\normalsize]  {$M_{5}$};
\draw (544.22,160.41) node [anchor=north west][inner sep=0.75pt]  [font=\normalsize]  {$M_{1}$};
\draw (649.84,157.79) node [anchor=north west][inner sep=0.75pt]  [font=\normalsize]  {$M_{4}$};
\draw (595.58,119.94) node [anchor=north west][inner sep=0.75pt]  [font=\normalsize]  {$M_{2}$};
\draw (597.33,193.11) node [anchor=north west][inner sep=0.75pt]  [font=\normalsize]  {$M_{3}$};
\draw (234.05,438.34) node [anchor=north west][inner sep=0.75pt]  [font=\large]  {$\gamma _{x}$};
\draw (183.43,414.6) node [anchor=north west][inner sep=0.75pt]  [font=\normalsize]  {$M_{5}$};
\draw (246.67,366.4) node [anchor=north west][inner sep=0.75pt]  [font=\normalsize]  {$M_{1}$};
\draw (352.97,367.83) node [anchor=north west][inner sep=0.75pt]  [font=\normalsize]  {$M_{4}$};
\draw (298.87,328.51) node [anchor=north west][inner sep=0.75pt]  [font=\normalsize]  {$M_{2}$};
\draw (299.22,408.12) node [anchor=north west][inner sep=0.75pt]  [font=\normalsize]  {$M_{3}$};
\draw (244.54,240.16) node [anchor=north west][inner sep=0.75pt]  [font=\large]  {$\gamma _{x}$};
\draw (195.9,214.68) node [anchor=north west][inner sep=0.75pt]  [font=\normalsize]  {$M_{5}$};
\draw (238.62,160.53) node [anchor=north west][inner sep=0.75pt]  [font=\normalsize]  {$M_{1}$};
\draw (348.19,163.71) node [anchor=north west][inner sep=0.75pt]  [font=\normalsize]  {$M_{4}$};
\draw (295.13,121.71) node [anchor=north west][inner sep=0.75pt]  [font=\normalsize]  {$M_{2}$};
\draw (294.44,199.55) node [anchor=north west][inner sep=0.75pt]  [font=\normalsize]  {$M_{3}$};

\end{tikzpicture}

    \caption{A Wilson loop with 3 self-intersections, and four different contributions to its expectation value.}
    \label{fig:special twist}
\end{figure}

Next, we consider the self-intersecting Wilson loop $\gamma_x$ shown in figure \ref{fig:special twist}, which intersects itself at three points on the plane. The allowed configurations \cite{Gross:1993yt} all have $n_1=1, {\tilde n}_1=0, n_4=0, {\tilde n}_4=1$, while the middle regions have $n_2={\tilde n}_2$ and $n_3={\tilde n}_3$, which can each take the values $0$ or $1$.

When $n_2=n_3=0$ (the bottom right of figure \ref{fig:special twist}), we have 3 twist points at the 3 self-intersection points, each looks like the twist of the figure 8 diagram discussed in section \ref{figure_eight}. The worldsheet topology of this configuration is a disk with a handle (a cycle going through two twist points is incontractible). There is no moduli space here, and the genus of the worldsheet is $(-1)$, so this configuration contributes
\begin{equation}
    \boxed{\langle W_{0,0} \rangle = N^{-1},}
\end{equation}
consistent with (7.27) of \cite{Gross:1993yt}.

The cases $n_2=1,n_3=0$ and $n_3=1,n_2=0$ are similar, so let us just discuss the first one, on the top right of figure \ref{fig:special twist}. In this case, the top two self-intersection points in the target space are just regular points on the worldsheet, where we have two sheets of opposite orientation, each ends on one of the intersecting lines. The bottom self-intersection point is a twist point, locally similar to the one discussed in section \ref{figure_eight}. The simplest worldsheet topology here is a disk, and there are no moduli, so it contributes $N$. However, in this case, we can also have an ORT in region two connecting the two sheets, such that the worldsheet topology is that of a disk with a handle. The ORT has a negative mode, and its moduli space has the topology of a disk, so this gives a contribution of $-N^{-1}$, and the overall contribution of this configuration,
\begin{equation}
    \boxed{\langle W_{1,0} \rangle = N - N^{-1},}
\end{equation}
is consistent with equation (7.29) of \cite{Gross:1993yt}.

The final case is $n_2=n_3=1$ (the top left of figure \ref{fig:special twist}). In this case, the top and bottom self-intersection points are smooth points on the worldsheet with two sheets sitting on top of each other. However, the middle point is now a different type of twist point than we had in the figure 8, where each worldsheet that ends there covers 3 quadrants instead of just 1. We will call this a ``special twist'' point, and discuss mappings that realize this in section \ref{sec:special_twist} below. 

A ``special twist'' point also has no moduli. 
However, it has an instability towards folding back the covering, either in region 2 or in region 3 (by creating a finite-size fold inside these regions that ends on the two boundaries, as discussed in section \ref{folds} above), so it gives a negative contribution. Naively, one may think that these are two separate negative modes, but in fact, they are not, since once you fold into region 2, the self-intersection point just becomes a regular point, so you cannot also fold it into region 3, and vice versa. Note that the limit where the fold (say in region 2) becomes as large as possible, eating up the whole region, is also a solution, but it precisely coincides with the $n_2=0,n_3=1$ configuration that we discussed above.

There is actually another thing you can naively do to the mapping, which is to fold it into both region 2 and region 3 at the same time, such that the self-intersection point becomes a twist point similar to that in the figure 8. However, this is not really possible if we start with the mapping from the disk, since the resulting topology has a higher genus (indeed, when the folds go to their maximal size, this goes over to the $n_2=n_3=0$ configuration discussed above). This is related to the fact that we can also add an ORT to this $n_2=n_3=1$ configuration, which sits either in region 2 or in region 3, with a worldsheet topology that is a disk with a handle. Naively, the moduli space of the ORT is comprised of two separate regions. However, we can in fact take the limit where the ORT goes to the ``special twist'' point, and this limit gives precisely the configurations discussed above, namely the zero-size limit of folds in regions 2 and 3 together. Thus, the actual moduli space of the ORT is the union of regions 2, 3 and the self-intersection point, and its topology has Euler number $1$. Thus, the total contribution from worldsheet topologies of a disk with a handle is $(-1)\times(-1)\times N^{-1}$, with separate factors of $(-1)$ from the ORT and from the instability of the ``special twist'' point. The overall contribution from these mappings is thus
\begin{equation}
    \boxed{\langle W_{1,1} \rangle = -N + N^{-1},}
\end{equation}
consistent with (7.30) of \cite{Gross:1993yt}.

\subsection{$G=1$, $\gamma_a \cup \gamma_{a'}$, $k=1$: Torus}
\label{torus_YM}

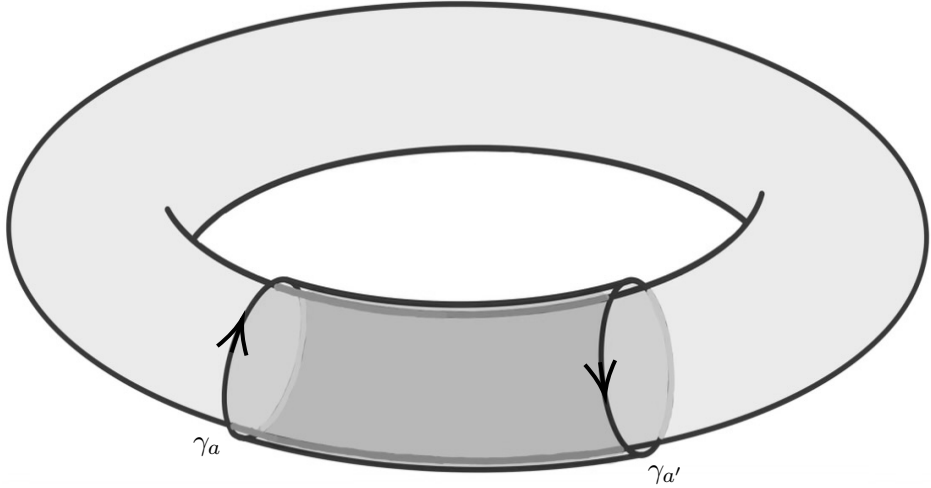
\begin{figure}[h]
    \centering

\tikzset{every picture/.style={line width=0.5pt}} 

\begin{tikzpicture}[x=0.75pt,y=0.75pt,yscale=-1,xscale=1]

\draw (414.25,197) node  {\includegraphics[width=351.38pt,height=186pt]{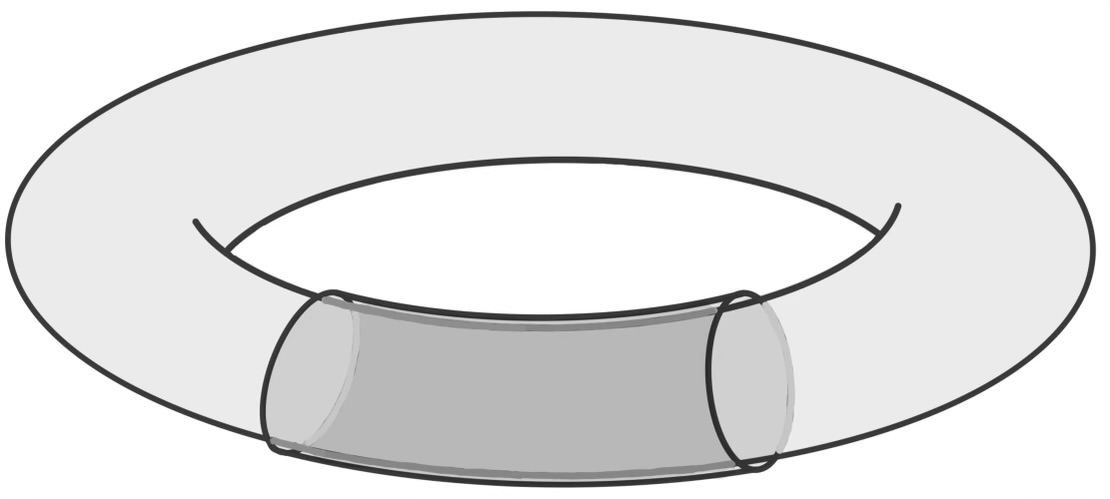}};
\draw [line width=1.5]    (294.5,255) -- (299.08,240.85) ;
\draw [shift={(300,238)}, rotate = 107.93] [color={rgb, 255:red, 0; green, 0; blue, 0 }  ][line width=1.5]    (17.05,-7.64) .. controls (10.84,-3.59) and (5.16,-1.04) .. (0,0) .. controls (5.16,1.04) and (10.84,3.59) .. (17.05,7.64)   ;
\draw [line width=1.5]    (478.5,260) -- (480.56,274.03) ;
\draw [shift={(481,277)}, rotate = 261.63] [color={rgb, 255:red, 0; green, 0; blue, 0 }  ][line width=1.5]    (17.05,-7.64) .. controls (10.84,-3.59) and (5.16,-1.04) .. (0,0) .. controls (5.16,1.04) and (10.84,3.59) .. (17.05,7.64)   ;

\draw (274,296.4) node [anchor=north west][inner sep=0.75pt]    {$\gamma _{a}$};
\draw (501,310.4) node [anchor=north west][inner sep=0.75pt]    {$\gamma _{a'}$};

\end{tikzpicture}

    \caption{Two Wilson loops on the torus, depicting a specific worldsheet contributing to their expectation value.}
    \label{fig:Torus}
\end{figure}

Finally, let us discuss a Wilson loop on a torus. If we just have a single $k=1$ Wilson loop wrapping a cycle of the torus, there are obviously no corresponding world sheets, so its expectation value vanishes; the simplest topologically non-trivial non-zero configuration involves two WLs wrapping a cycle with two different orientations, as in figure \ref{fig:Torus}.

The leading order contributions involve an annulus worldsheet, that fills the region between the WLs on one or the other side of the torus. In both cases, there are no moduli, so one gets a contribution equal to $N^0=1$, and the total contribution is 
\begin{equation}
    \boxed{\langle W \rangle = 2,}
\end{equation}
consistent with (7.33) of \cite{Gross:1993yt}. 

At the same order in $1/N$, there are also more complicated configurations where the annulus wraps around the torus any number of times.
The existence of an infinite series of possible mappings means that this Wilson loop actually diverges in the zero-coupling theory, with a similar divergence to the one appearing in the partition function of the torus, arising from multiple coverings; these divergences disappear once we turn on a finite coupling (they are related to the fact that at zero coupling, the partition function on the torus \eqref{YM_partition} is simply the sum over representations, with no suppression). In any case, we can still compare our results for given covering numbers to the results from the YM theory with the same covering numbers, and we find agreement.

\section{String theory analysis of various examples} \label{sec:WLs in string theory}

\subsection{An example without any branch points or tubes -- the Disk}
%
\label{sec: the disk}

Let us begin with the simplest case of a circular Wilson loop of radius one on the plane (see section \ref{sec:disk}). In this case, the only worldsheet configuration is a disk, with boundary conditions specified by the unit circle in the target space, $|Z| = 1$.

Since there are no complex structure deformations, the on-shell value of the stress tensor must vanish. Given that $T = \partial Z \partial \bar{Z}$, this implies that the solutions must either be holomorphic or antiholomorphic, but not a combination of both. 
In principle, solutions that are piecewise-holomorphic and piecewise-antiholomorphic could exist; however, there are no such solutions in this case.

The moduli space of classical solutions is precisely $SL^{2|2}(2,\mathbb{R})$, with all solutions equivalent by conformal transformations preserving the boundary (which are the gauge transformations coming from unfixed diffeomorphisms in the conformal gauge). Explicitly, choosing the worldsheet to be a disk of radius one, the general classical solution winding once around the boundary is:
\begin{equation}
    Z_{\rm cl} = e^{i\varphi_0} \frac{z - z_0}{1 - z_0^* z},
\end{equation}
where $z_0$ is some point inside the disk. This indicates that there is essentially one solution, with its simplest representative being $Z = z$. Importantly, all parameters ($z_0$ and $\varphi_0$) are functions of $\theta$ and $\thetabar$.  

To verify the absence of additional solutions, we analyze the zero modes of the kinetic operator for small deformations around a solution, $Z = z + \delta Z$. The linearized Neumann condition on $\delta Z$ becomes:
\begin{equation}
    0 = \left(z^{2}\partial\delta\bar{Z} + \bar{z}^{2}\bar{\partial}\delta Z\right)\Big|_{\partial D}.
\end{equation}
For solutions to the equations of motion, $\partial \delta \bar{Z}$ is holomorphic, and the above condition forces it to vanish (each Fourier coefficient must vanish). This implies that $\delta Z$ is holomorphic. The linearized Dirichlet condition is:
\begin{equation}
    0 = \left(z\delta\bar{Z} + \bar{z}\delta Z\right)\Big|_{\partial D}.
\end{equation}
From this, we obtain the general solution:
\begin{equation}
    \delta Z = a\left(1 - z^{2}\right) + ib\left(1 + z^{2}\right) + icz.
\end{equation}
This corresponds to the variation under an infinitesimal $SL(2,\mathbb{R})$ transformation, $\delta Z = v(z)\partial Z_{\rm cl} = v(z)$, where $v(z)$ is a conformal Killing vector.

What does the path integral evaluate to? Intuitively, the integration over solutions cancels the factor ${\rm Vol}\left(SL^{2|2}(2,\mathbb{R})\right)^{-1}$ from the unfixed gauge transformations. Over each solution, the fluctuations of the non-zero modes contribute a factor of 1. In total, we obtain:
\begin{equation}
    Z_{\rm disk}\left(X(\partial D) = \left\{|X| = 1\right\}\right) = N,
\end{equation}
Consistent with the expectations from Yang-Mills theory described in section \ref{sec:disk} (the factor of $N$ is obtained in the standard way, since the Euler number of the worldsheet is $1$).

To elaborate, we require three real gauge-fixing conditions to reduce the integration over the $X$-field to one that intersects the solution space only at $Z_{\rm cl} = z$. These conditions can be chosen, for instance, as:
\begin{equation}
    0 = \intop_{D}dzd\bar{z}Z = \intop_{D}dzd\bar{z}\frac{\bar{z}Z - z\bar{Z}}{2i}.
\end{equation}
We then argue that the Faddeev-Popov determinant evaluates to 1. Finally, the integration over the non-zero modes also yields a factor of 1. Thus, the final result is as expected.

When we have a circular Wilson loop on the sphere instead of the plane, there are also additional contributions that we will discuss in section \ref{sec: disk with ORT} below.

\subsection{An example with a branch point -- the doubly-covered disk}\label{sec: branch point}

Next, let's consider the case that the boundary of the disk double-covers a circle in the target space, exemplified by the solution $Z=z^2$; this corresponds to a Wilson loop winding twice over the unit circle in the plane. This can be viewed as a limit of the Wilson loop shown in figure \ref{fig8:2} and discussed in section \ref{twice_winding_disk} above, where we take the $1^{\text{st}}$ region as a circular region with unit radius, the $3^{\text{rd}}$ region as having zero size, and the $2^{\text{nd}}$ region to infinite size. The general (not gauge-fixed) mapping involves a double-cover of the disk with both covers having the same (holomorphic) orientation, given by:
\be
e^{i\theta_{0}}\frac{\left(z-z_{0}\right)\left(z-z_{1}\right)}{\left(1-z_{0}^{*}z\right)\left(1-z_{1}^{*}z\right)}.
\ee
The moduli space is now 5 dimensional; 3 of the directions are accounted for by the $SL(2,\mathbb{R})$  conformal symmetry of the disk, and the 2 remaining modes correspond to the position of the branch point in the target space (its worldsheet position can be fixed by $SL(2,\mathbb{R})$).
Expanding around the specific solution $Z = z^2$, the zero-modes can again be written as $\delta Z=v(z)\partial Z_{\rm cl}=2 zv(z)$, but now $v(z)$ is permitted to have a simple pole where $Z_{\rm cl}$ has a branch point ($z=0$), in which case it is not a legitimate generator of $SL(2,\mathbb{R})$ (but rather moves the branch point). 

We can use the $SL(2,\mathbb{R})$ symmetry to choose the worldsheet branch point to be at $z=0$, and fix also the rotations around it to have just one representative for each physical solution, for instance
\bea
    Z&=\frac{z^{2}+Z_{0}}{1+z^{2}Z_{0}^{*}},
\eea

where the parameter $Z_0$ represents the target space position to which the branch-point is mapped, $Z(0)=Z_0$.

The two zero-modes in excess of the gauge redundancy make the path integral ill-defined, unless we add some exact term to saturate them; this is similar to what we found for closed strings with branch points in \cite{Aharony:2023tam}. As discussed in \cite{Aharony:2023tam}, one natural candidate is $S_1$ involving the extrinsic curvature, which reduces to a delta function on branch points (and on folds, if they are present).

Generally, if we have a solution with a degree $n$ branch point $Z=\frac{1}{n+1} z^{n+1}$, there are $(2n+3)$  zero-modes, 3 of which correspond to $SL(2,\mathbb{R})$ transformations and the rest to displacements of the $n$ branch points. 
The invariant we use to couple to branch points, which localizes to branch points and measures the excess angle there, is:
\begin{equation}
    -24P_{\mu\rho}P_{\nu\sigma}h^{ab}h^{cd}\partial_{a}\partial_{[c}X^{\rho}\partial_{b]}\partial_{d}X^{\sigma}=2\pi ng_{\mu\nu}\frac{1}{\sqrt{h}}\delta^{2}\left(\sigma\right),
\end{equation}
where
\begin{equation}
    P_{\mu\nu}=g_{\mu\nu}-g_{\mu\rho}\partial_{a}X^{\rho}h^{ab}\partial_{b}X^{\sigma}g_{\sigma\nu}
\end{equation}
is the projector on directions transverse to the worldsheet (which do not exist at generic points). One way to compute it, as described in Appendix B of \cite{Aharony:2023tam},  is to 
regulate the inverse metric $h^{ab}\to \frac{\det (h)}{\det (h) + \epsilon} h^{ab}$. 

We can then add to the worldsheet action (as in \cite{Aharony:2023tam}) the term $S_1$ from \eqref{eq: super-Polyakov action}; plugging in a classical solution of a branch point, it evaluates to

\begin{align}
   & -24\intop d^{2}\sigma d^2\theta \sqrt{h}\partial_{\theta}X^{\mu}\partial_{\bar{\theta}}X^{\nu}P_{\mu\rho}P_{\nu\sigma}h^{ab}h^{cd}\partial_{a}\partial_{[c}X^{\rho}\partial_{b]}\partial_{d}X^{\sigma} \nonumber \\ & \qquad\qquad\qquad =  2\pi n \int d^2\theta g_{\mu\nu}\left(X\left(0,\theta,\bar{\theta}\right)\right)\partial_{\theta}X^{\mu}\left(0,\theta,\bar{\theta}\right)\partial_{\bar{\theta}}X^{\nu}\left(0,\theta,\bar{\theta}\right),
\end{align}
giving an expression that includes 4-fermion terms involving the fermion zero modes. As in the closed string case \cite{Aharony:2023tam}, we then obtain (by integrating over the zero modes of the fermions and the auxiliary fields) an integral over the moduli space (the position of the branch point) with an integrand which is the Chern class, giving the Euler number of the moduli space, which in our case is $\chi(D)=1$. This assumes that in this integral we have the appropriate boundary terms that make it a topological invariant; as long as our formalism preserves the diffeomorphism invariance of the target space (which may require adding some specific worldsheet boundary terms related to $S_1$), this is guaranteed.

Thus, the bottom line is that these configurations give us the expected result of $N$, in agreement with the YM result discussed in section \ref{twice_winding_disk}.

We can also get contributions to this Wilson line with these covering numbers from configurations with a higher odd number of branch points (corresponding to a higher genus worldsheet); however, in this case, the Euler number of the moduli space (the positions of the branch points, at separate points on the disk) vanishes, so they do not contribute.

\subsubsection{Multiple branch points - $k$-winding Wilson loops}

Another example, which was not analyzed in \cite{Gross:1993yt}, is the case of a Wilson line winding $k$ times around the disk. While the answer for this at general coupling is quite complicated, in the free theory it is simply equal to $N$.

The worldsheet analysis is very similar to the one above. The leading order contribution comes from mappings from the disk to the disk which wind around it $k$ times, of the general form
\be
Z(z) = e^{i\theta_{0}}\frac{\left(z-z_{1}\right)\left(z-z_{2}\right)\cdots \left(z-z_k\right)}{\left(1-z_{1}^{*}z\right)\left(1-z_{2}^{*}z\right)\cdots \left(1-z_k^* z\right)}.
\ee
After gauge-fixing 3 of the parameters, we are left with $(k-1)$ complex parameters, which at generic points on the moduli space can be identified with the positions of $(k-1)$ simple branch points (points where $dZ/dz=0$).

The mapping has $k$ sheets, two of which are permuted at each branch point, and the sheets are connected with a specific cyclic ordering as we go around the boundary. When we arrange the sheets in this specific cyclic ordering, the sheets that are permuted at the $(k-1)$ branch points are $(12), (23), \cdots, ((k-1) k)$, up to an overall cyclic permutation. Thus, there is a natural ordering for the branch points, and they are distinguishable by their position in this ordering. The topology of the moduli space is thus $D^{k-1}$. Note that there are no singularities when different branch points approach each other; if they are adjacent in the cyclic ordering, then we simply get a higher order branch point, and otherwise they do not feel each other. So, the integration over the moduli space gives a factor of $\chi(D^{k-1})=1$, and the total answer from the string theory is $N^{\chi(D)}=N$, in agreement with the gauge theory computation.

As in the previous case, we can add additional branch points and obtain a mapping from a higher genus Riemann surface to the disk, but with additional branch points, we do obtain singularities when some branch points approach each other, and the Euler number of the resulting moduli space vanishes.

\subsection{Examples with Orientation-Reversing Tubes (ORTs)}

\subsubsection{Two Disks connected with an ORT}\label{sec: two disk with ORT}

Consider two separate Wilson loops in the plane, with unit radius but with opposite orientations, as in figure 
\ref{fig: two opposite loop}. Instead of looking at a generic configuration, we can take the two disks to be centred at the origin, such that the $3^{\text{rd}}$ and $4^{\text{th}}$ regions in the figure reduce to zero size.

There is obviously a disconnected configuration with separate disks filling the two Wilson loops, and the next simplest configuration has the topology of an annulus/cylinder, with the two disks connected by an orientation-reversing tube.

The worldsheet cylinder has a single real modulus, and we choose coordinates where the worldsheet is an annulus
\begin{equation} \label{eq:annulus}
    z = r e^{i \theta}, \quad \frac{1}{T} \leq r \leq T.
\end{equation}
We begin by choosing the conformal gauge for the worldsheet metric.
In these coordinates, the modulus $T \in \left(1,\infty\right)$ is related to the existence of the holomorphic quadratic differential:
\begin{equation}
    b_{zz} = \frac{A}{z^2}, \quad A \in \mathbb{R},
\end{equation}
which obeys the boundary conditions:
\begin{equation}
    0 = b_{r\theta} \propto z^2 b_{zz} - \bar{z}^2 b_{\bar{z}\bar{z}}.
\end{equation}
There is one remaining worldsheet symmetry, $z \to z e^{i c}$, related to the conformal Killing vector (CKV):
\begin{equation}
    c^z = i C z, \quad C \in \mathbb{R},
\end{equation}
which obeys the boundary conditions:
\begin{equation}
    0 = c^r \propto \bar{z} c^z + z c^{\bar{z}}.
\end{equation}

Now, let us look at the mapping of this worldsheet to the target space. As discussed in \cite{Aharony:2023tam}, in the presence of worldsheet moduli but before we integrate over them, not all the equations of motion of the worldsheet metric are obeyed; in particular the Virasoro condition needs to be obeyed only up to the holomorphic quadratic differential, namely we need to find a solution to the EOM obeying $T_{zz} = A / z^2$. An appropriate solution, if we gauge-fix the rotations by choosing $Z(T)=1$, is given by
\begin{equation} \label{annulus_sol}
    Z = \frac{T}{1+T^2} \left(z + \frac{1}{\bar z} \right).
\end{equation}
This solution has a fold at $|z|=1$, which maps in the target space to the circle $|Z| = 2 T / (1 + T^2)$, centred at the origin. In the $T \to \infty$ limit, the fold shrinks to zero size and we get an ORT at $|Z|=0$. The mapping we found is singular in this limit, as expected for ORTs in the conformal gauge; we will discuss ORTs in other gauges below, and we will see that they can be written in a non-singular way.

If we first perform the integral over all modes except $T$, we localize to the solution \eqref{annulus_sol}. The lowest component of the integrand over superspace in $S_0$, whose extrema determine where the path integral localizes, is given by
\begin{equation}
        S^0_{0}[X_{cl}] = 2\pi \left( 1-\frac{2}{T^2+1} \right).
\end{equation}

Considering now the integration over $T$, we see that $S_0^0$ has a minimum at $T=1$, where the worldsheet degenerates to the Wilson loops (and does not cover the disk), and a maximum at $T \rightarrow \infty$, corresponding to the case where the size of the tube is zero and it is localized at the origin. The degenerate worldsheet is not allowed, so the worldsheet path integral localizes to the singular configuration with the zero-size tube.

Once the tube has zero size, we expect to obtain an extra modulus corresponding to the position of the zero-size tube (we will exhibit this extra mode explicitly in a different gauge below). The remaining task is to integrate over the position of this ORT. The moduli space for this position is topologically a disk. Consequently, the contribution of the moduli space integral to this saddle point is $-\chi(\text{Disk}) = -1$, and this is also its total contribution, since the Euler number of the annulus is zero. The minus sign 
arises from the negative mode associated with this saddle, since it is an area-maximizing configuration (the mode responsible for deforming the tube size has a negative eigenvalue).

In the conformal gauge, since the solution for the ORT is singular, it is difficult to explicitly see and analyze the moduli space (we encountered similar difficulties for closed strings in \cite{Aharony:2023tam}). Thus, it is worthwhile to discuss the same configurations in an induced gauge corresponding to some specific solution.
Consider the mapping from the annulus \eqref{eq:annulus} to the disk given by
\begin{equation}
    Z_0(z) = f_T(|z|) / {\bar z},
\end{equation}
where $f_T(x)$ is some specific function obeying that $f_T(T)=T$, $f_T(1/T)=1/T$, and $f_T(x)$ is positive for all $1/T < x < T$ except for a single value $x_0$ where $f_T(x_0)=0$, where $f_T(x)/x$ is monotonically decreasing for $1/T < x < x_0$, and monotonically increasing for $x_0 < x < T$. For example, we can choose $f_T(x) = (x-1)^2 T/(T-1)^2$. Such a configuration describes a mapping of the annulus to the disk which covers it once with each orientation and has a zero-size tube, where now the zero-size tube sits on the worldsheet at $|z|=x_0$ (and in space-time at $Z=0$), allowing the mapping to be non-singular (and non-degenerate away from the zero-size tube). We can now gauge-fix the diffeomorphisms and Weyl transformations, for each value of $T$, to the induced metric associated with this specific solution $Z_0(z)$. The general arguments of section \ref{subsec:Induced Gauge} tell us that the solution will be unique up to the possiblity of moving the position of the ORT in the target space. In particular, we can exhibit explicitly the solutions (in the induced gauge of $Z_0$) for other ORT positions; they take the form 
\begin{equation}
    Z(z) = \frac{Z_0(z) + Z_{ORT}}{1 + Z_{ORT}^* Z_0(z)},
\end{equation}
with the ORT now at $Z=Z_{ORT}$.

In this induced gauge, we have for every $T$ a moduli space of solutions, parameterized by $Z_{ORT}$, which is a disk. Assuming that $K_{\mu \nu}$ appearing in $S_1$ of \eqref{eq: super-Polyakov action} is non-zero on the zero-size fold (as we show in appendix \ref{appendix:folds}) , the integral over the zero modes coming from this moduli space gives the Euler number of the moduli space, as in our previous discussions. This justifies the result mentioned above. Note that in this gauge, unlike the conformal gauge, we do not have a saddle point in the integral over the (super) modulus $T$, but we get contributions from all values. However, the zero modes of the moduli are still lifted together with the $B$-ghost zero modes, so the joint integration over the moduli and the $B$ ghosts just gives a factor of $1$ (as shown in \cite{Aharony:2023tam}).

Since we found that a single ORT is a solution, we may expect to also have solutions with any number of ORTs connecting the two sheets (and with a higher genus worldsheet). However, similar to computations done in \cite{Horava:1995ic,Aharony:2023tam}, one can show that the Euler number of the moduli space of the positions of these multiple ORTs (which cannot be on top of each other) vanishes, so they do not contribute to the partition function. For instance, when we have two ORTs (which cannot be at the same position), the result is proportional to $\chi(D\times D) - \chi(D) = 0$.

\subsubsection{Disk with an ORT} \label{sec: disk with ORT}

Next, let us return to the case where the Wilson loop is a circle of unit radius centered around the origin in the target space, and now take the target space to be a sphere. The leading terms in this case are discussed in section \ref{sec: the disk}, and they come from a disk with one or the other orientation, covering the interior or the exterior of the circle. In order to get the additional contributions discussed in section \ref{subsubsec: disk with ORT}, we need to consider mappings with ORTs. In this case, the world sheet is just a disk, so it does not have any moduli, and we cannot construct in the conformal gauge non-singular mappings whose limit will give an ORT as in the previous subsection. 

One way to proceed to obtain non-singular mappings in the conformal gauge, as in \cite{Aharony:2023tam}, is to add extra variables in the path integrals as in ``constrained instanton'' computations. This can allow us to obtain the desired configurations as a singular limit of non-singular configurations, which can be used to compute their action and some of their other properties, similarly to what we did in the conformal gauge in the previous subsection.

However, as in the previous subsection, it is simpler to use the induced gauge instead. For instance, we can consider the mapping from the disk ($|z| \leq 1$) to the sphere
\begin{equation} \label{zero_fold_sol}
    Z_0(z) = \frac{2}{\bar z} (|z| - \frac{1}{2}) ,
\end{equation}
which covers the interior of the target-space disk $|Z|=1$ twice and its exterior once, and describes a zero-size tube at $Z=0$ in space-time, sitting on the world sheet at $|z|=\frac{1}{2}$ (this is an arbitrary choice, by a world sheet diffeomorphism we can move the fold to be on any other curve inside the disk). We can now choose the induced gauge corresponding to the solution $Z_0$, and ask if there are any nearby solutions to the equations of motion that will contribute to the path integral. Analyzing the linear deviations away from the solution \eqref{zero_fold_sol}, which solves the equations of motion in the induced gauge, as in section \ref{subsec:Induced Gauge}, we find that before taking into account the boundary conditions, they are given in radial coordinates $z = r e^{i\theta}$ by
\begin{equation}
    \delta Z = \sum_{k=-\infty}^{\infty} c_k \left|\frac{1-2r}{r}\right|^{|k|} e^{ik\theta},
\end{equation}
for arbitrary constants $c_k$, where the $k=0$ mode describes the shift in the position of the tube in the target space, and the other modes are diffeomorphisms that are conformal Killing vectors of the induced metric from \eqref{zero_fold_sol}. Imposing the boundary condition $|Z(r=1)|=1$ leaves us with only two deformations, $\delta Z = c_1 |\frac{1-2r}{r}| e^{i\theta}$ which corresponds to a worldsheet rotation ($\delta Z \propto dZ/d\theta$), and $\delta Z = c_0 - c_0^* (\frac{1-2r}{r})^2 e^{2i\theta}$, which corresponds (at linearized order) to moving the tube in the target space (by a complex parameter $c_0$).

In fact, it is easy to write down the most general solution (up to diffeomorphisms) in the induced gauge corresponding to $Z_0$; it takes the form
\begin{equation}
    Z(z) = \frac{Z_0(z) + Z_{ORT}}{1 + Z_{ORT}^* Z_0(z)},
\end{equation}
with an orientation-reversing tube at $Z=Z_{ORT}$ (which needs to lie within the unit disk in the target space).
Up to diffeomorphisms, we thus find a moduli space of solutions labeled by $Z_{ORT}$, whose topology is a disk.

The solutions we described with the ORT have action $S^0_0=6\pi$ (the area of the sphere and the disk together), and since they are a maximum of the action in the direction of making the fold have finite size, they have a negative mode and contribute to the path integral with a minus sign (the mode responsible for deforming the tube size has a negative eigenvalue). The moduli space is a disk, so we obtain that the contribution of these configurations is
$-\chi(D) N^{\chi(D)} = -N$, in agreement with the YM analysis.

\begin{figure}
    \centering
    \includegraphics[width=0.5\linewidth]{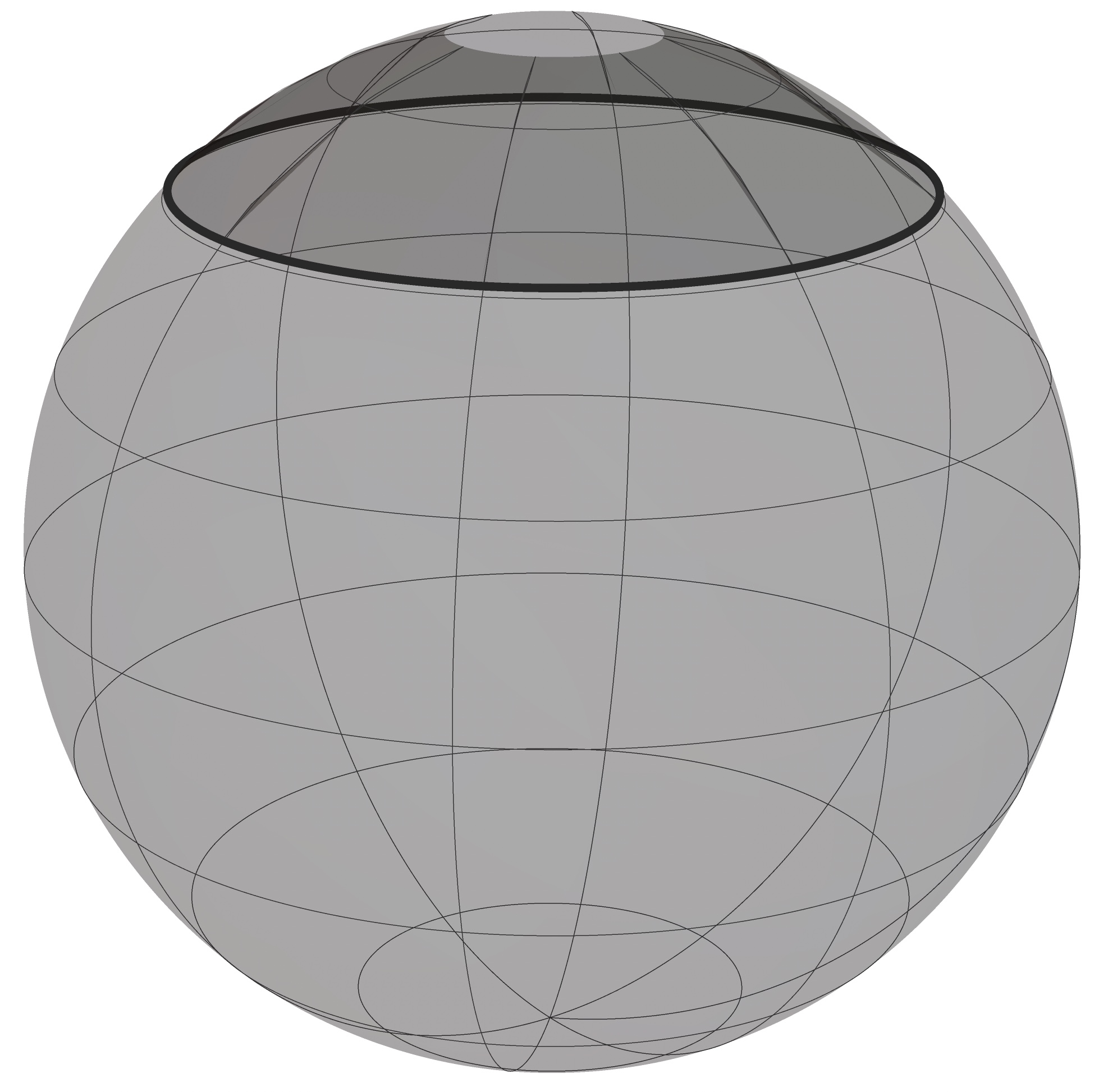}
    \caption{The mapping of a disk with an orientation-reversing tube to the sphere, depicting the tube as a finite-size fold (which we would get using the constrained instanton method). The actual solution is the limit of this mapping where the tube goes to zero size.}
    \label{fig:enter-label}
\end{figure}

There are also solutions to the EOM with additional ORTs, either with higher genus worldsheets or with more sheets covering the sphere.
When we have more than one sheet of each orientation, we can also have branched orientation-reversing tubes connecting several sheets at the same point, as discussed in \cite{Aharony:2023tam}. We expect that the sum over all these configurations will vanish so that we match with the field theory result, but we leave a detailed analysis of this to future work.

\subsection{Examples with twists}\label{sec: twist}

Next, consider the ``figure $8$'' Wilson loop shown in the middle of figure \ref{fig:All WL} on the plane, which is a different limit of the Wilson loop drawn in figure \ref{fig8:2} above, in which region 3 is taken to have infinite size. In this case, two separate points on the boundary of the worldsheet map to the same point in space-time (the intersection point of the WL), and in configurations that extremize the area, there will actually be a whole line on the worldsheet connecting these points that maps to the intersection point. Along this line, the induced metric is degenerate, so we cannot use the conformal gauge to analyze such mappings, and we will use the induced gauge instead.

We consider a unit disk as the worldsheet, with boundary $x^2+y^2=1$. One simple mapping that maps the boundary to the figure 8 WL is given by
\begin{equation} \label{regular_twist}
Z = x e^{\pi i y / 4}
\end{equation}
with a degeneration (mapping to the twist point $Z=0$) along the equator of the disk $x=0$. The corresponding Wilson loop sits on the curve
\begin{equation}
Z = x e^{\pm 2\pi i \sqrt{1-x^2} / 4},\qquad -1 \leq x \leq 1,
\end{equation}
which has the same form as the middle figure in figure \ref{fig:All WL} (rotated by 90 degrees). If we choose the induced gauge associated with this mapping, then both the equations of motion of $X$ and the Virasoro condition hold automatically.

To study the path integral we need to see whether in the induced gauge associated with this specific solution, there are any other solutions to the equations of motion nearby. It is easy to check that there are no such solutions, as implied by the general arguments of section \ref{subsec:Induced Gauge}. Thus, we have no moduli space, and since the worldsheet is a disk, the worldsheet path integral simply gives a factor of $N$.

The mapping we described above gives a specific geometry for the self-intersecting Wilson loop; similar mappings can be found for any other Wilson loop with the same topology.

\subsection{An example with a ``special twist''}
\label{sec:special_twist}

Next, consider a Wilson loop with 3 self-intersection points, as in figure \ref{fig:special twist}. A simple mapping from the worldsheet (taken to be a disk $x^2+y^2 \leq 1$) to the target space that maps the boundary to this Wilson loop is given by
\begin{equation} \label{special_twist_WL}
    Z = x e^{3\pi i y / 4}.
\end{equation}
It is not holomorphic or anti-holomorphic, so (as for other twist configurations) we cannot discuss it in the conformal gauge, but we can use the induced gauge.
The mapping \eqref{special_twist_WL} looks very similar to the regular-twist configuration \eqref{regular_twist} discussed above, with a twist at $x=0$, but the larger coefficient in the exponent means that the worldsheet now covers 3 quadrants on each side of the twist instead of just 1, so we obtain a mapping that looks like figure \ref{fig:special twist} with covering numbers $n_2=n_3=1$. The middle self-intersection point comes from the twist at $x=0$ on the worldsheet, while the other self-intersection points are just regular points on the worldsheet.

As discussed in section \ref{sec:special_twist_YM}, this mapping has an instability towards a fold moving into either region 2 or region 3 but not both, so it contributes with a minus sign. There is no moduli space, so the full contribution is $(-N)$.

Constructing mappings with $n_2=0,n_3=1$ or $n_2=1,n_3=0$ raises no new issues, so we will not discuss them explicitly. In all these cases, we can also have an ORT, as discussed in section \ref{sec:special_twist_YM}, which does not raise any new issues compared to the ORT configurations we discussed above.

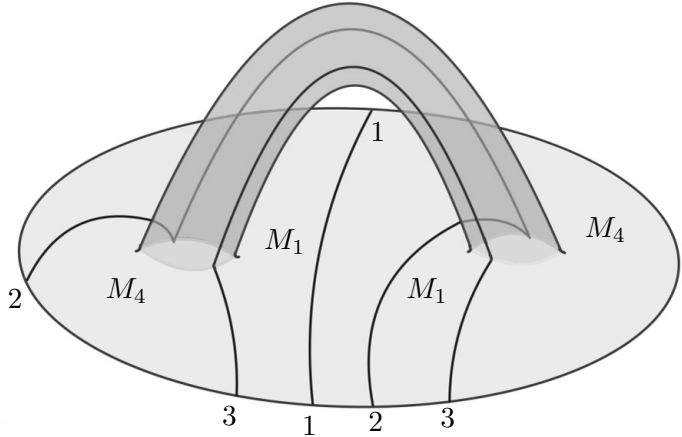
\begin{figure}[h]
    \centering

\tikzset{every picture/.style={line width=0.75pt}} 

\begin{tikzpicture}[x=0.75pt,y=0.75pt,yscale=-1,xscale=1]

\draw (381.5,175.5) node  {\includegraphics[width=258pt,height=163.5pt]{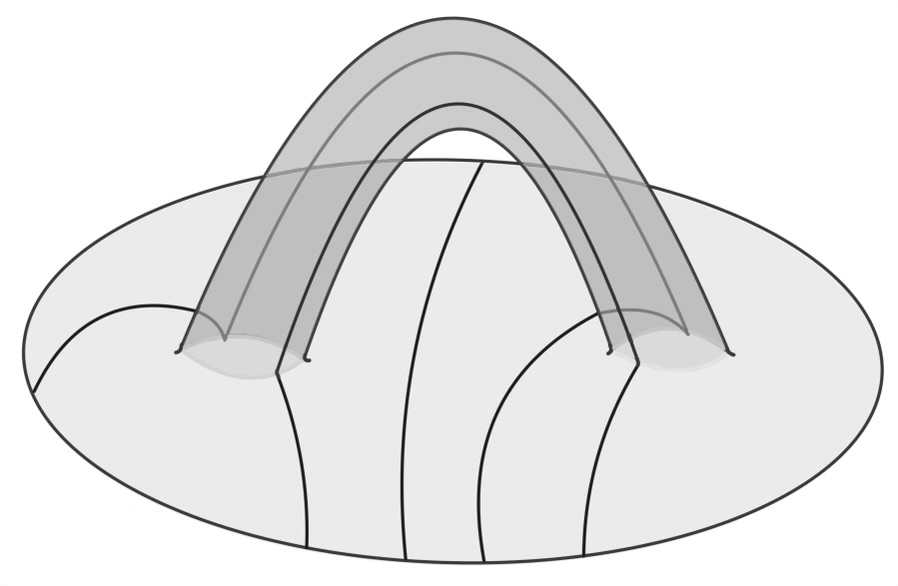}};

\draw (358,279.4) node [anchor=north west][inner sep=0.75pt]    {$1$};
\draw (392,131.4) node [anchor=north west][inner sep=0.75pt]    {$1$};
\draw (391,277.4) node [anchor=north west][inner sep=0.75pt]    {$2$};
\draw (211,215.4) node [anchor=north west][inner sep=0.75pt]    {$2$};
\draw (318,274.4) node [anchor=north west][inner sep=0.75pt]    {$3$};
\draw (427,275.4) node [anchor=north west][inner sep=0.75pt]    {$3$};
\draw (339,185.4) node [anchor=north west][inner sep=0.75pt]    {$M_{1}$};
\draw (499,179.4) node [anchor=north west][inner sep=0.75pt]    {$M_{4}$};
\draw (410,210) node [anchor=north west][inner sep=0.75pt]    {$M_{1}$};
\draw (260,210) node [anchor=north west][inner sep=0.75pt]    {$M_{4}$};

\end{tikzpicture}

    \caption{The worldsheet corresponding to the case of $n_2 = n_3 = 0 $, which maps to the bottom right configuration in figure \ref{fig:special twist}.  The lines connecting $1-1$, $2-2$ and $3-3$ are the lines along which the worldsheet is twisted (three lines map to three twist points on the target space). Note the simple-connectedness of the regions $M_1$ and $M_4$ 
    through the tube.}
    \label{fig:special_twist_WS}
\end{figure}

To get a mapping with $n_2=0,n_3=0$, we need to have 3 twist points. As above, each twist point should come from a degenerate line on the world sheet, that ends at two points on the boundary and maps to the twist point. Naively, one may think that one can put 3 such lines on the disk. However, when going around the Wilson loop (corresponding to going around the boundary of the worldsheet), we encounter these twist points in the order $1-2-3-1-2-3$, and there is no way to put 3 lines on the disk that do not intersect and whose end-points come in this ordering along the boundary. Thus, we must take the worldsheet to be a disk with a handle, and on this surface it is possible to have 3 twist lines that do not intersect and whose end-points on the boundary have this ordering, see figure \ref{fig:special_twist_WS}. 

\subsection{A Wilson loop on the torus}

Finally, let us discuss the Wilson loop on the torus, drawn in figure \ref{fig:Torus} and analyzed in the YM theory in section \ref{torus_YM}.

At leading order in $1/N$, the worldsheet is an annulus. We can take the target space to be a torus with modular parameter $\tau$, described by $Z = X + i Y \equiv Z + 1 \equiv Z + \tau$. We will take the two Wilson loops to sit at $Y=0$ with one orientation, and at $Y=Y_0$ with the opposite orientation. We can choose the worldsheet annulus to be described by a periodic coordinate $x$, $x \equiv x+1$, and another coordinate $0 < y < t$, with boundaries at $y=0$ and $y=t$. The parameter $t$ here is the real modulus of the annulus, which can take any values from $0$ to $\infty$.

The simplest mappings are linear functions, which can either have one orientation and involve covering numbers $n_1=n_2+1$, or have the opposite orientation and covering numbers ${\tilde n}_2={\tilde n}_1+1$. For simplicity, we discuss here just the lowest possible covering numbers, for which only $n_1=1$ or only ${\tilde n}_2=1$, with the other covering numbers equal to zero.

We can analyse the possible mappings either in the conformal gauge or in the induced gauge. Up to diffeomorphisms, the mappings are given by
\begin{equation} \label{first}
    X = x, \qquad Y = \frac{Y_0}{t} y
\end{equation}
with one orientation, and
\begin{equation} \label{second}
    X = x, \qquad Y = Y_0 + \frac{{\rm Im}(\tau)-Y_0}{t} (t-y)
\end{equation}
with the opposite orientation.
They are a sum of a holomorphic and an anti-holomorphic function, so they solve the EOM of $X^{\mu}$ both in the conformal gauge and in an induced gauge.

In the conformal gauge, naively the Virasoro condition does not hold, since for generic values of $t$ these mappings are not holomorphic or anti-holomorphic. However, since we have a modulus $t$ that we have not yet integrated over, we have not yet imposed its EOM, so we can allow configurations with constant $T_{zz}=T_{{\bar z}{\bar z}}$, which is what we get from these mappings. If we now compute the bottom component of the integral over $\theta$ in $S_0$, which is what we extremize in our path integral, we get for the mapping \eqref{first}
\begin{equation}
    s^0_0(t) = \frac{t}{2} + \frac{Y_0^2}{2t}.
\end{equation}
When we now do the integral over $t$, we find that it is extermized at $t=Y_0$, which is precisely when the mapping \eqref{first} is holomorphic (as expected, since the configurations that contribute in the conformal gauge should satisfy the Virasoro condition). Similarly, for the mapping \eqref{second} we see that the action is extremized for $t = {\rm Im}(\tau)-Y_0$, where the mapping is anti-holomorphic. Our path integral gives contributions just from these two mappings, and there is no moduli space, so each mapping just gives a factor of $N^{\chi({\rm annulus})} = 1$. More precisely, naively there is a moduli space corresponding to shifting $X$, but this is a conformal Killing vector so we need to identify the different configurations, which we can do just by fixing $Z(0,0)=0$ for the first case, and $Z(0,0)=i{\rm Im}(\tau)$ in the second case.

In the induced gauge, the mappings above are good solutions to the equations of motion for all $t$. So we still remain with an integration over $t$, but as discussed in \cite{Aharony:2023tam}, this cancels with the integration over the corresponding zero mode of the $B$ ghost, so we obtain the same result.

\appendix

\section*{Acknowledgments}
We would like to thank Netanel Barel, Rajesh Gopakumar, Shota Komatsu, Pranabesh Maity, Kiarash Naderi, and K. Sreeman Reddy for useful discussions. 
This work was supported in part by ISF grant no. 2159/22, by Simons Foundation grant 994296 (Simons Collaboration on Confinement and QCD Strings), by the Minerva foundation with funding from the Federal German Ministry for Education and Research, and by the German Research Foundation through a German-Israeli Project Cooperation (DIP) grant ``Holography and the Swampland''. OA is the Samuel Sebba Professorial Chair of Pure and Applied Physics.  The work of SK is supported by the ``Koshland Fellowship'' from the Weizmann Institute of Science, the ERC-COG grant NP-QFT No. 864583, ``Non-perturbative dynamics of quantum fields: from new deconfined phases of matter to quantum black holes'',
the MUR-FARE grant EmGrav No. R20E8NR3HX 
, and the MUR-PRIN2022 grant No. 2022NY2MXY. SK also gratefully acknowledges the hospitality of the TIFR String Theory group (during long-term visits) and the Simons Foundation while this work was in progress.

\section{Regularization of folds}
\label{appendix:folds}

In this appendix, we will show how we regulate mappings near a fold. This is necessary for the calculation of the extrinsic curvature of an orientation-reversing tube.

Let us consider a generic form of a fold on the planar target space $X^\mu \equiv (T(t,x), X(t,x))$ as
\begin{equation}
    \begin{split}
        T = t, \quad\quad X = f(t) - \frac{1}{2}x^2
    \end{split}
\end{equation}
where the fold is at $x=0$ on the world sheet and at $X=f(T)$ in the target space, and the strings are extended in the $X<f(T)$ direction. At a generic point, the extrinsic curvature is zero, but on the fold, it is not true. We will see a one-dimensional $\delta^{\perp}$-function source along the fold. Note that this non-zero extrinsic curvature occurs due to the sudden change in the world-sheet around the fold. To regulate this region, we will add an extra dimension $Y$ to space-time, and allow for a small change in this direction so that the world sheet no longer folds on top of itself by taking
\begin{equation}
    Y = \kappa  \tanh \left(\frac{x}{w}\right).
\end{equation}
Here, $\kappa$ is a very small positive number which helps regulate the fold, and $w$ is a scale which determines the shape of the regulated surface. We will see that in the $\kappa\rightarrow 0$ limit, the final expression will not depend on the value of $w$. \footnote{One can also check that the final answer does not depend on the choice of the function to regulate the fold ($w$ is a part of that choice).}

The target space metric that we take is
\begin{equation}
    ds^2 = G_{AB}dX^A dX^B = g_{\mu\nu}dX^\mu dX^\nu + dY^2 = dT^2+dX^2+dY^2.
\end{equation}
 
Note that the extra dimension $Y$ does not participate in the worldsheet dynamics. This is called a `Semi-Riemannian Foliation'. We will call this a `$3D$ target space' for simplicity.

In this $3D$ target space, we can calculate the induced metric, and accordingly the extrinsic curvature, using the normal direction ($n_A$) to the world-sheet, which is now available everywhere due to the presence of the extra dimension. Note that generically this normal vector is mostly along the extra dimension (in the $\kappa\rightarrow 0$ limit it is exactly along the extra dimension, except when we are on the fold, and then it lies in the $2D$ target space directions). 

The expression for the `second fundamental form' is
\begin{equation}
    \mathcal{K}^A_{ab}=\partial_a \partial_b X^{A }-h^{cd} \partial_c X^{A }\partial_d X_{B } \partial_a\partial_b X^{B }.
\end{equation}
Using this, we calculate the extrinsic curvature
\begin{equation}
    K^{AB} = (\mathcal{K}^A_{ac}\mathcal{K}^B_{bd} - \mathcal{K}^A_{ad}\mathcal{K}^B_{bc}) h^{ac} h^{bd}.
\end{equation}
One can easily check that on the regulated surface, the following relation is exactly true
\begin{equation}
    K_{AB} = K ~ n_A n_B,
\end{equation}
where $K = K_{CD}G^{CD}$ is the trace of the extrinsic curvature.

Note that on the fold $n_A \propto n_\mu$, where $n_\mu$ is the normal to the fold calculated in the $2D$ target space.  After scaling the $x$ direction around the fold $x\rightarrow \kappa \,\delta x$, and taking the $\kappa\rightarrow 0$ limit, we get
\begin{equation}
    \sqrt{h}K = -\frac{2 w f''(t)}{\kappa  \left(f'(t)^2 + 1+\delta x^2 w^2\right)^{3/2}}+ \cdots
\end{equation}
Note that this function is $\sim 1/\kappa$ near $x=0$, and when $\delta x\gg \kappa$ the function decays to zero. Hence, this is a $1D$ $\delta$ function situated along the fold $X = f(T)$. The final expression that we get by taking $\kappa \to 0$ is
\begin{equation}\label{fin_K}
    \sqrt{h}K_{\mu\nu} = \delta(x) \frac{4 f''(t)}{ f'(t)^2+1 } n_\mu n_\nu,
\end{equation}
where
\begin{equation}
    n_\mu = \left( \frac{f'(t)}{\sqrt{f'(t)^2+1}},-\frac{1}{\sqrt{f'(t)^2+1}} \right)
\end{equation}
is the vector normal to the fold. On the fold, we have
\begin{equation}
    \sqrt{h}K = -\frac{4 f''(t)}{f'(t)^2+1},
\end{equation}
which measures the local curvature of the shape of the fold. Note that the final answer doesn't depend on the regulator (in our case, the regulator is $\tanh (x/w)$). It only depends on the function $f(t)$, which is the position of the fold. 

We further claim that if the fold goes in a loop in the target space, then after integrating over the loop, we will get
\begin{equation}
    \oint \sqrt{h} K_{\mu\nu} = - 4\pi\, g_{\mu\nu}.
\end{equation}
To see this, let us
look at each component of $\sqrt{h} K_{\mu\nu}$ separately,
\begin{equation}
\begin{split}
        \oint \sqrt{h} K_{\mu\nu} dt &= \oint \left(
\begin{array}{cc}
 -\frac{4 f'(t)^2 f''(t)}{\left(f'(t)^2+1\right)^2} & \frac{4 f'(t) f''(t)}{\left(f'(t)^2+1\right)^2} \\
 \frac{4 f'(t) f''(t)}{\left(f'(t)^2+1\right)^2} & -\frac{4 f''(t)}{\left(f'(t)^2+1\right)^2} \\
\end{array}
\right) dt\\
&= \oint \left(
\begin{array}{cc}
 -\frac{4 q^2 dq}{\left(q^2+1\right)^2} & \frac{4 q dq}{\left(q^2+1\right)^2} \\
 \frac{4 q dq}{\left(q^2+1\right)^2} & -\frac{4 dq}{\left(q^2+1\right)^2}\\
\end{array}
\right) ,\quad\quad q = f'(t)\\
&= \oint \left(
\begin{array}{cc}
 d\left(-\frac{2q}{q^2+1}-2\tan ^{-1}q\right) & d\left(-\frac{2}{q^2+1}\right) \\
 d\left(-\frac{2}{q^2+1}\right) & d\left(\frac{2 q}{q^2+1}-2 \tan ^{-1}q\right)\\
\end{array}
\right)\\
&= -2\oint \left(
\begin{array}{cc}
 d\left(\tan ^{-1}q\right) & 0 \\
 0 & d\left(\tan ^{-1}q\right)\\
\end{array}
\right)\\
&= -4\pi\left(
\begin{array}{cc}
 1 & 0 \\
 0 & 1\\
\end{array}
\right)= -4\pi g_{\mu\nu}.
\end{split}
\end{equation}
This calculation was done on a planar target space; on a general curved space, the answer will be more complicated, but if
the loop is infinitesimally small in size, then the final answer would be proportional to $g_{\mu\nu}(X_0)$, where $X_0$ is the `center' or position of the infinitesimal loop.

\section{Collapsed handle}\label{appendix:coll_handle}

In the string duals of Yang-Mills theory suggested in \cite{Gross:1993hu,Horava:1995ic}, the contributions from collapsed handles vanish in the topological zero-coupling theory (in the case of $U(N)$ gauge groups, they vanish at finite coupling as well). Configurations with collapsed handles do appear in the sum over mappings. For instance, one can map a torus to a sphere just by squashing the torus onto a sphere, leading to a configuration which covers the worldsheet once in one region and three times in another region, that is bounded by four folds (with two sheets meeting at the first and third ones in cyclic order, and the first sheet meeting a third sheet at the second and fourth folds). The folds meet at four points that we will call `corners'. 

\begin{figure}[H]
    \centering

\tikzset{every picture/.style={line width=0.75pt}} 

\begin{tikzpicture}[x=0.75pt,y=0.75pt,yscale=-1,xscale=1]

\draw [color={rgb, 255:red, 74; green, 144; blue, 226 }  ,draw opacity=1 ][line width=2.25]    (244.5,219.46) -- (411.81,218.15) ;
\draw [color={rgb, 255:red, 74; green, 144; blue, 226 }  ,draw opacity=1 ][line width=2.25]    (245.5,347.46) -- (412.81,346.15) ;
\draw [line width=2.25]    (244.5,219.46) .. controls (255.5,279.46) and (253.5,293.46) .. (245.5,347.46) ;
\draw [line width=2.25]    (412.81,346.15) .. controls (402.5,300.46) and (403.5,269.46) .. (411.81,218.15) ;

\end{tikzpicture}

    \caption{A finite-size collapsed handle has four folds, with two sheets meeting on two of them, and one of these sheets meeting another sheet on the other two folds.}
    \label{four folds}
\end{figure}
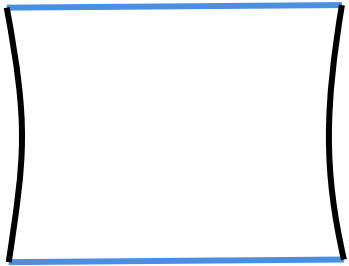

For finite-size folds, this is not a solution to the equations of motion, but, as in other configurations we discussed above, it becomes a solution when all the folds (and thus the handle) go to zero size (a ``collapsed handle''). Thus, in our formalism, when analyzing mappings whose topology allows for collapsed handles (which can be closed strings, as in mappings of the torus to the sphere, or open strings as for the disk with a handle that appeared in Section \ref{sec:special_twist}), a thorough study of the collapsed handle configuration and its contribution to the path integral is required, and we will perform it in this appendix. 

For simplicity, we focus on mapping a worldsheet torus to a space-time sphere (the analysis can easily be generalized to other cases).

Let us parameterize the world-sheet torus by periodic coordinates $\alpha,\beta \in [0,2\pi)$. A specific solution mapping the torus to a sphere (with a collapsed handle) is:
\begin{equation} \label{specific solution}
    X= \sin(\alpha) ~ \sin ^2(\frac{\beta }{2}),\quad Y = |\sin(\frac{\alpha}{2})|
\sin (\beta),\quad Z = 1-\frac{1}{2} (\cos (\alpha) -1) (\cos (\beta) -1),
\end{equation}
where we write the space-time sphere as $X^2+Y^2+Z^2=1$. The induced metric degenerates along the lines $\alpha=0$ and $\beta=0$ (that are non-trivial cycles). These lines map to the north pole $X=Y=0$, $Z=1$, and any non-trivial cycle of the torus necessarily goes through this point (which we identify with the ``collapsed handle''). This mapping with a mesh created by constant $\alpha$ and $\beta$ lines looks like:
\begin{figure}[H]
    \centering
    \includegraphics[width=0.5\linewidth]{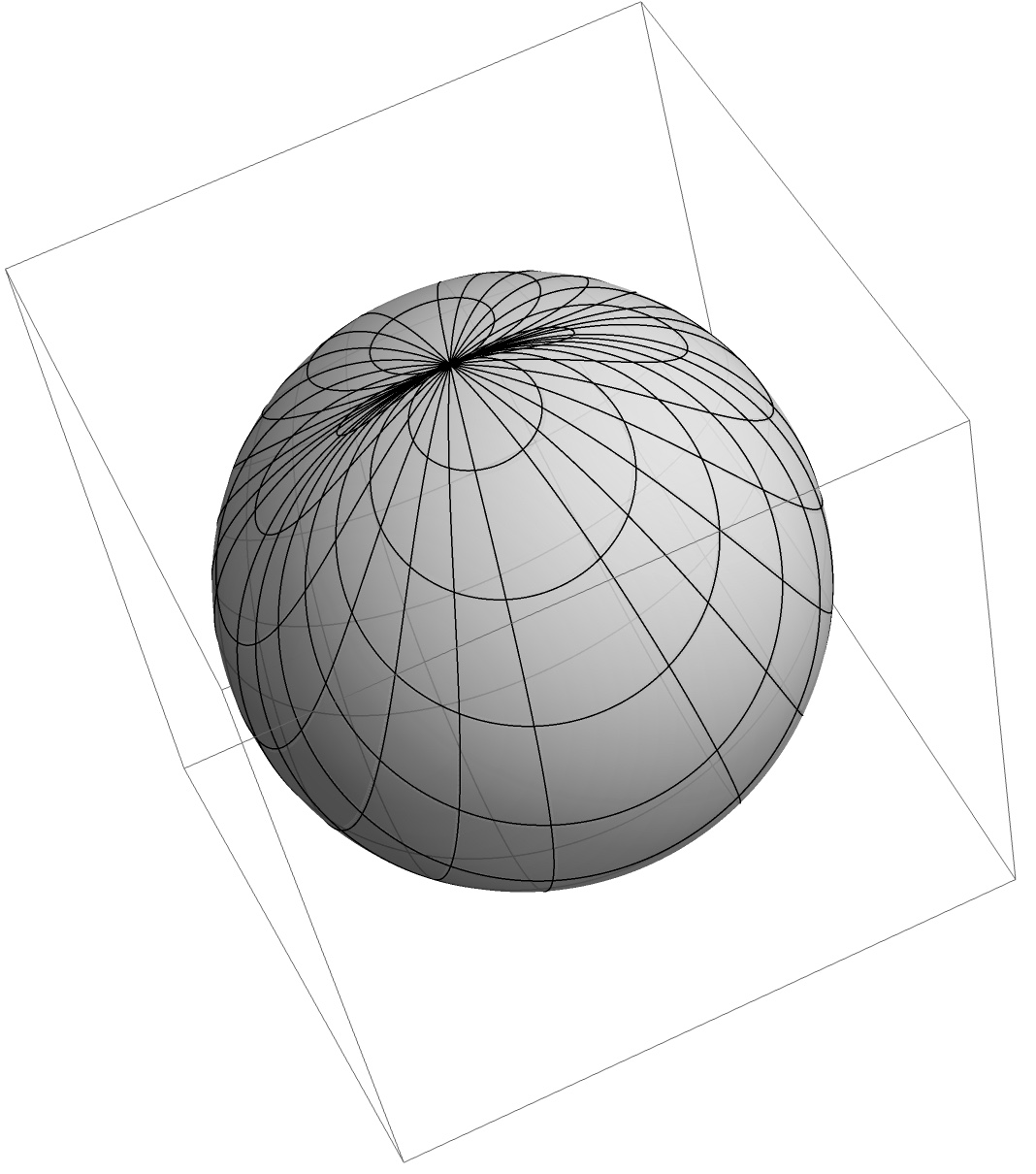}
    \caption{A singular mapping of a torus to a sphere with a ``collapsed handle''.}
    \label{fig:enter-label}
\end{figure}

\begin{figure}[h]
    \centering
    \includegraphics[width=0.5\linewidth]{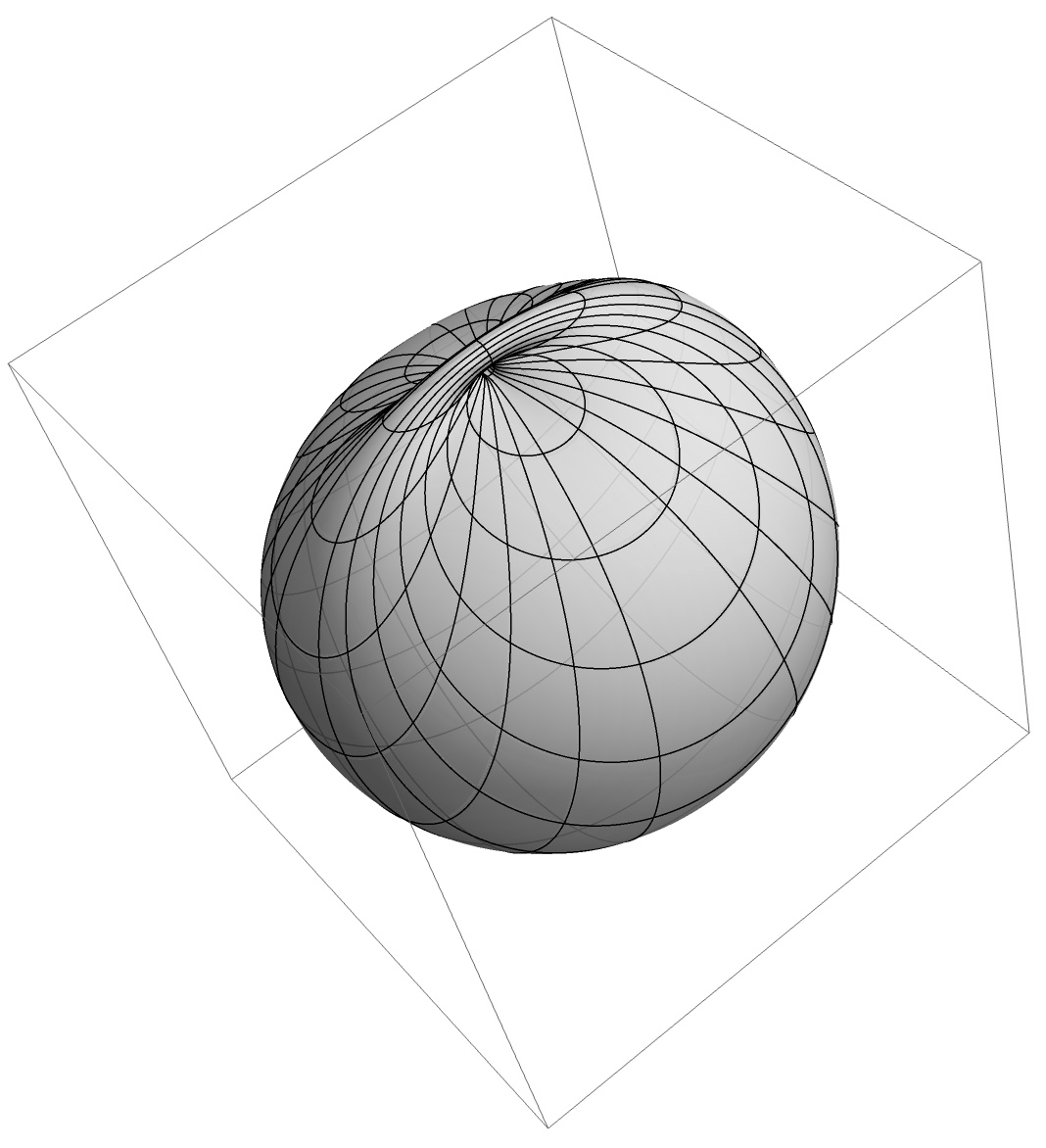}
    \caption{The regularized mapping with $\epsilon = 0.05$.}
    \label{fig:regularized handle}
\end{figure}

One way to regularize this configuration and to confirm that indeed it has a ``collapsed handle'' is by mapping it to a different surface embedded in three-dimensional space, by inflating the tube in the radial direction using the following mapping:
\begin{equation}
    \begin{split}
        X&= \frac{1}{2} \sin (\alpha)  ((\epsilon -1) \cos (\beta) +\epsilon +1),\\
        Y&=\frac{1}{2} \sin (\beta)  \sqrt{(\epsilon -1)^2 \sin ^2(\alpha) +(-(\epsilon +1) \cos (\alpha) +2 \epsilon +1)^2},\\
        Z&=\frac{1}{2} (-\cos (\alpha)  ((\epsilon +1) \cos (\beta) +\epsilon -1)+2 \epsilon  \cos (\beta) +\cos (\beta) +2 \epsilon +1).
    \end{split}
\end{equation}
For small values of $\epsilon$, the mapping exhibits a finite-size handle, as in figure \ref{fig:regularized handle}, and as $\epsilon\to 0$ it goes over to \eqref{specific solution}.

Note that for this specific mapping, the ``handle'' is elongated in the $Y$ direction. Near $\alpha,\beta\sim 0$ the mapping becomes
\begin{equation}
    X \sim \frac{\alpha \beta^2}{4};\quad Y\sim \frac{|\alpha|\beta}{2}; \quad Z\sim 1,
\end{equation}
such that $|Y| \gg |X|$. We can make this more precise by using spherical coordinates
\begin{equation}
X = \sin(\theta) \sin(\phi),\qquad Y = \sin(\theta) \cos(\phi), \qquad Z = \cos(\theta).
\end{equation}
We can view the torus as a square bounded by the lines $\alpha=0$, $\beta=0$, $\alpha=2\pi$ and $\beta=2\pi$ that all map to the north pole $\theta=0$. As we go slightly inside the square, $\theta$ increases to a non-zero small value, and then as we go around the square in the worldsheet we go around the north pole in the target space, such that $\phi$ increases monotonically from $\phi=0$ to $\phi=2\pi$ (the increase must be monotonic since we do not have any additional folds). We will argue below that the value of $\phi$ as we approach each of the corners of the square is a diffeomorphism-invariant quantity that characterizes the mapping. In the mapping described above, we have $\phi=0$ at the corners $\alpha=\beta=0$ and $\alpha=2\pi,\beta=0$ (where, as we go inside the square, $\phi$ increases very slowly along the $\alpha$ direction), and $\phi=\pi$ at the other two corners $\alpha=0,\beta=2\pi$ and $\alpha=\beta=2\pi$. 

As in other circumstances, we need to investigate the full moduli space of mappings that are not related by world sheet diffeomorphisms. There are two obvious moduli, which are just the position of the collapsed handle on the sphere. It is natural to expect that one modulus of the solution we found above would involve rotating it in the $X-Y$ plane (or shifting the $\phi$ coordinate), and one can check that this is indeed a modulus. However, we expect to also have non-trivial moduli representing the ``shape'' of the tube.
In particular, if we view the collapsed handle as coming from 4 folds with 4 corners as in figure \ref{four folds} collapsing to a point at the north pole, then there are four angles in the $X-Y$ plane (represented by $\phi_i$, $i=1,2,3,4$) representing the angular positions of the corners as they approach the north pole, and we will argue that these lead to four moduli of the mapping (including the rotation modulus mentioned above).

We begin in section \ref{local analysis} with a local analysis of the deformations near each corner of the square on its ``inside'', which shows that locally the only non-trivial possibilities are a translation or a rotation in $\phi$. Then, in section \ref{global} we argue that the full moduli space can be described in terms of the position of the handle and the four $\phi$ positions of the corners. Finally, in section \ref{metric collapsed handle} we argue that the metric for the latter four moduli vanishes, such that the integration over the fermionic zero modes gives zero for configurations with ``collapsed handles'' and they do not contribute to the path integral (at zero coupling), as required for agreement with the field theory.

\subsection{Deformations of the collapsed handle}
\label{local analysis}

A collapsed handle is a mapping from a torus to a sphere. An example of such a mapping is
\begin{equation} \label{specific solution_1}
    X_0= \sin(\alpha) ~ \sin ^2(\frac{\beta }{2}),\quad Y_0 = |\sin(\frac{\alpha}{2})|
\sin (\beta),\quad Z_0 = 1-\frac{1}{2} (\cos (\alpha) -1) (\cos (\beta) -1),
\end{equation}
where the target space is a sphere $X^2+Y^2+Z^2=1$; we will describe it using two coordinates $X^\mu\in \{X(\alpha,\beta), Y(\alpha,\beta)\}$ with the curved metric of the sphere\footnote{Cartesian coordinates are best suited for analyzing the deformations.}. We intend to find all possible deformations from this solution. It is hard to find all possible global deformations, which we will discuss in the next subsection. However, we can solve for the possible deformations perturbatively around the collapsed handle solution in one of the four quadrants (we will analyze them near $\alpha,\beta=0$, assuming we are in the quadrant where both are positive); in the next subsection, we will discuss how to put together the deformations in different quadrants. We assume that at least locally we can choose a diffeomorphism that takes the induced metric to be the one of the original solution $\{X_0,Y_0\}$, so that the analysis is similar to the analysis of deformations in the previous sections (by gauge-fixing the metric to be the induced metric of the original solution). We do not know if it is always possible globally to choose such a diffeomorphism, or if there are induced metrics that are not equivalent by diffeomorphisms (like the different values of the complex structure in the metric of a torus), but this issue will not play a role in our analysis.

As in section \ref{subsec:Induced Gauge}, the equations for infinitesimally deformed solutions ($X^\mu=X_0^\mu+\delta X^\mu$) are the equations of motion 
\begin{equation}\label{deformHandle}
\begin{split}
    &\frac{1}{\sqrt{H}}\partial_a \left( \sqrt{H} H^{ab} \partial_b~ \delta X^\mu \right) = 0
\end{split}
\end{equation}
and the linearised Virasoro constraint
\begin{equation}\label{deformHandle}
\begin{split}
    & \partial_a X^\mu\partial_b\delta X_\mu + \partial_a \delta X^\mu\partial_b X_\mu - H_{ab} \left(H^{cd} \partial_c X^\mu\partial_d\delta X_\mu\right)= 0,
\end{split}
\end{equation}
where $H_{ab} \left(= \partial_aX_0\cdot \partial_bX_0\right)$ is the induced metric calculated from the undeformed classical solution. We want to know all infinitesimal smooth deformations from $X_0^\mu$. Let us start with the following ansatz for the smooth deformation (expanded around $\alpha,\beta \sim \epsilon\ll 1$)
\begin{equation}\label{defor}
    \begin{split}
        \delta X = \sum_{m,n=0}^\infty x_{mn} \alpha^m \beta^n,\qquad \delta Y = \sum_{m,n=0}^\infty y_{mn} \alpha^m \beta^n.\\
    \end{split}
\end{equation}
Since we assume smooth functions (with no $|\alpha|$'s or $|\beta|$'s), we are solving here just in one quadrant, and in the next subsection we discuss how to glue these solutions along the lines $\alpha=0$ and $\beta=0$.
We solved perturbatively in $\epsilon$ (up to $\mathcal{O}(\epsilon^{10})$) and found the following solutions for linearized deformations of $(X,Y)$ (we denote them by either $\delta x_i^\mu$ or $\delta y_i^{\mu}$):
\begin{equation}\label{fullsol}
    \begin{split}
        \delta x^\mu_0=&\{1,0\}\\
        \delta y^\mu_0=&\{0,1\}\\
        \delta x^\mu_1=&\{\alpha  \beta -\frac{\alpha  \beta \left(\alpha ^2+4 \beta ^2\right)}{24}+\frac{\alpha  \beta \left(\alpha ^4+16 \beta ^4\right)}{1920}+\frac{-\alpha ^7 \beta +252 \alpha ^5 \beta ^3+2688 \alpha ^3 \beta ^5-64 \alpha  \beta ^7}{322560}\\
        &+\frac{\alpha  \beta \left(\alpha ^8-4320 \alpha ^6 \beta ^2-230400 \alpha ^2 \beta ^6+256 \beta ^8\right)}{92897280}+\cdots,-\frac{\alpha  \beta ^2}{2} + \frac{\alpha  \beta ^2 \left(2 \alpha ^2+\beta ^2\right)}{24} \\
        &+\frac{-6 \alpha ^5 \beta ^2+5 \alpha ^3 \beta ^4-2 \alpha  \beta ^6}{1440} + \frac{ \left(12 \alpha ^7 \beta ^2-273 \alpha ^5 \beta ^4-602 \alpha ^3 \beta ^6+3 \alpha  \beta ^8\right)}{120960}  + \cdots \}\\
        \delta y^\mu_1=&\{\frac{\alpha  \beta ^2}{2} - \frac{\alpha  \beta ^2 \left(2 \alpha ^2+\beta ^2\right)}{24} + \frac{6 \alpha ^5 \beta ^2 - 5 \alpha ^3 \beta ^4 + 2 \alpha  \beta ^6}{1440} - \frac{ \left(12 \alpha ^7 \beta ^2-273 \alpha ^5 \beta ^4-602 \alpha ^3 \beta ^6+3 \alpha  \beta ^8\right)}{120960} + \cdots,\\
        & \alpha  \beta -\frac{\alpha  \beta \left(\alpha ^2+4 \beta ^2\right)}{24}+\frac{\alpha  \beta \left(\alpha ^4+16 \beta ^4\right)}{1920}+\frac{-\alpha ^7 \beta +252 \alpha ^5 \beta ^3+2688 \alpha ^3 \beta ^5-64 \alpha  \beta ^7}{322560}\\
        &+\frac{\alpha  \beta \left(\alpha ^8-4320 \alpha ^6 \beta ^2-230400 \alpha ^2 \beta ^6+256 \beta ^8\right)}{92897280}+\cdots\}\\
        \delta x^\mu_2=&\{\alpha ^2 \beta ^2 -\frac{\alpha ^2 \beta ^2 \left(\alpha ^2+7 \beta ^2\right)}{12} +\frac{\alpha ^2 \beta ^2 \left(\alpha ^4+31 \beta ^4\right) }{360} + \frac{ -3 \alpha ^8 \beta ^2+392 \alpha ^6 \beta ^4+13496 \alpha ^4 \beta ^6-381 \alpha ^2 \beta ^8}{60480} + \cdots,\\
        & -\alpha ^2 \beta ^3 + \frac{\alpha ^2 \beta ^3 \left(5 \alpha ^2+6 \beta ^2\right)}{24}  +\frac{\left(-91 \alpha ^6 \beta ^3+980 \alpha ^4 \beta ^5-144 \alpha ^2 \beta ^7\right)}{5760} + \cdots \}\\
        \delta y^\mu_2=&\{\alpha ^2 \beta ^3 - \frac{\alpha ^2 \beta ^3 \left(5 \alpha ^2+6 \beta ^2\right)}{24}  -\frac{\left(-91 \alpha ^6 \beta ^3+980 \alpha ^4 \beta ^5-144 \alpha ^2 \beta ^7\right)}{5760} + \cdots,\\
        &\alpha ^2 \beta ^2 -\frac{\alpha ^2 \beta ^2 \left(\alpha ^2+7 \beta ^2\right)}{12} +\frac{\alpha ^2 \beta ^2 \left(\alpha ^4+31 \beta ^4\right) }{360} + \frac{ -3 \alpha ^8 \beta ^2+392 \alpha ^6 \beta ^4+13496 \alpha ^4 \beta ^6-381 \alpha ^2 \beta ^8}{60480} + \cdots\}\\
        \delta x^\mu_3=&\{\alpha ^3 \beta ^3 -\frac{\alpha ^3 \beta ^3 \left(\alpha ^2+10 \beta ^2\right)}{8} + \frac{13 \alpha ^7 \beta ^3+688 \alpha ^3 \beta ^7}{1920} + \cdots,-\frac{3 \alpha ^3 \beta ^4}{2} +\frac{3 \alpha ^3 \beta ^4 \left(\alpha ^2+2 \beta ^2\right)}{8} + \cdots \}\\
        \delta y^\mu_3=&\{\frac{3 \alpha ^3 \beta ^4}{2} -\frac{3 \alpha ^3 \beta ^4 \left(\alpha ^2+2 \beta ^2\right)}{8} + \cdots, \alpha ^3 \beta ^3 -\frac{\alpha ^3 \beta ^3 \left(\alpha ^2+10 \beta ^2\right)}{8} + \frac{13 \alpha ^7 \beta ^3+688 \alpha ^3 \beta ^7}{1920} + \cdots \}\\
        \delta x^\mu_4=&\{\alpha ^4 \beta ^4 -\frac{\alpha ^4 \beta ^4 \left(\alpha ^2+13 \beta ^2\right)}{6} + \cdots,-2\alpha ^4 \beta ^5+\cdots\}\\
        \delta y^\mu_4=&\{2\alpha ^4 \beta ^5+\cdots,\alpha ^4 \beta ^4 -\frac{\alpha ^4 \beta ^4 \left(\alpha ^2+13 \beta ^2\right)}{6} + \cdots\}\\
        \delta x^\mu_5=&\{\alpha ^5 \beta ^5+\cdots,\mathcal{O}(\epsilon^{11})\}\\
        \delta y^\mu_5=&\{\mathcal{O}(\epsilon^{11}),\alpha ^5 \beta ^5+\cdots\}\\
        \vdots
    \end{split}
\end{equation}
In two dimensions, the number of target space dimensions is the same as that of the world-sheet. Hence, all solutions can be written naively as a diffeomorphism of the world-sheet; this is true at regular points, but it may not be true around the singular collapsed handle. Next we write what are the diffeomorphisms that give rise to the solutions above (which may or may not become singular at the collapsed handle), namely we solve $X^\mu (\sigma^a) + \delta X^\mu_i (\sigma^a) = X^\mu (\sigma^a+\delta\sigma^a_i(\sigma)) $ for the corresponding infinitesimal $\delta\sigma_a(\alpha,\beta)$ functions. For these functions, we will just write the leading terms:
\begin{equation}\label{diffossol}
    \begin{split}
        \delta \sigma^a[\delta x^\mu_0] = \{-\frac{4}{\beta ^2} + \cdots,\frac{4}{\alpha  \beta }+\cdots\}\qquad&
        \delta \sigma^a[\delta y^\mu_0] = \{\frac{4}{\beta } + \cdots,-\frac{2}{\alpha }+\cdots\}\\
        \delta \sigma^a[\delta x^\mu_1] = \{-\frac{4 \alpha }{\beta } + \cdots, 4  +\cdots\}\qquad&
        \delta \sigma^a[\delta y^\mu_1] = \{2 \alpha + \cdots, \frac{\alpha ^2 \beta }{2} +\cdots\}\\
        \delta \sigma^a[\delta x^\mu_2] = \{ -4 \alpha ^2 + \cdots, 4 \alpha  \beta +\cdots\}\qquad&
        \delta \sigma^a[\delta y^\mu_2] = \{ 4 \alpha ^2 \beta + \cdots, 2 \alpha  \beta ^2 +\cdots\}\\
        \delta \sigma^a[\delta x^\mu_3] = \{ -4\alpha^3\beta+ \cdots, 4 \alpha ^2 \beta ^2 +\cdots\}\qquad&
        \delta \sigma^a[\delta y^\mu_3] = \{ - 2\alpha^3\beta^2 + \cdots, 4 \alpha ^2 \beta ^3 +\cdots\}\\
        \delta \sigma^a[\delta x^\mu_4] = \{ -4 \alpha ^4 \beta ^2 + \cdots, 4 \alpha ^3 \beta ^3 +\cdots\}\qquad&
        \delta \sigma^a[\delta y^\mu_4] = \{ -4\alpha ^4 \beta ^3 + \cdots, 6 \alpha ^3 \beta ^4 +\cdots\}\\
        \delta \sigma^a[\delta x^\mu_5] = \{-4\alpha ^5 \beta ^3 + \cdots, 4 \alpha ^4 \beta ^4 +\cdots\}\qquad&
        \delta \sigma^a[\delta y^\mu_5] = \{ -6\alpha ^5 \beta ^4 + \cdots, 8 \alpha ^4 \beta ^5 +\cdots\}\\
        \cdots
    \end{split}
\end{equation}

{\bf Comments:} The set of solutions written above in \eqref{fullsol} and \eqref{diffossol} has the following properties:
\begin{itemize}
    \item We can easily check that $\delta y^\mu_i = \epsilon^{\mu\nu} \delta x^\nu_i$ and $\delta x^\mu_i = -\epsilon^{\mu \nu} \delta y^\nu_i$ for all modes $i=0,1,2,\cdots$, namely $\delta y_i^X = - \delta x_i^Y$ and $\delta y_i^Y = \delta x_i^X$. It can be shown from the equation of motion and Virasoro constraint that if there is a generic solution $\{\delta X(\alpha,\beta),\delta Y(\alpha,\beta)\}$ then $\{-\delta Y(\alpha,\beta),\delta X(\alpha,\beta)\}$ is also a valid solution to the same equations.
    \item In equation \eqref{diffossol}, only three modes, $\delta x_0^\mu$, $\delta y_0^\mu$, and $\delta x_1^\mu$, correspond to a singular diffeomorphism near $\alpha=\beta=0$. They constitute the non-trivial deformations from the solution. All other deformations, with non-singular diffeomorphisms, can be undone by a diffeomorphism.
    \item The modes $\delta x^\mu_0$ and $\delta y^\mu_0$ represent translations of the collapsed handle on the sphere. 
    \item At leading order $\delta y^\mu_1$ is proportional to $\delta \lambda^\mu=\{X_0,Y_0\}$, so this is an overall dilatation of the solution. Similarly, $\delta x^\mu_1$ is proportional to $\delta r^\mu=\{-Y_0,X_0\}$, so this mode represents a rotation. Although these identifications don't continue in sub-leading orders, and we get the following relations (correct only up to relevant orders),
    \begin{equation}
    \begin{split}
        \delta y^\mu_1 &= 2\delta\lambda^\mu + \frac{1}{144} \delta y^\mu_3 + \frac{1}{230400} \delta y_5^\mu + \cdots,\\
        \delta x^\mu_1 &= 2\delta r^\mu + \frac{1}{144} \delta x^\mu_3 + \frac{1}{230400} \delta x_5^\mu + \cdots,\\
        \end{split}
    \end{equation}
    Note that the deviation from $\delta \lambda^\mu$ (dilatation mode) or $\delta r^\mu$ (rotation mode) is a non-singular diffeomorphism, which can be removed. So we identify these modes with a rotation (which is non-trivial) and a dilatation (which can be undone by a diffeomorphism).
    \item In equation \eqref{defor} we started with any arbitrary function of $(\alpha,\beta)$ represented by the Taylor series coefficients $x_{mn}$ and $y_{mn}$. After solving the equations, we got modes only with free $x_{mm}$ (represented by $\delta x^\mu_m$) and $y_{mm}$ (represented by $\delta y^\mu_m$), shown in \eqref{fullsol}. These free coefficients $x_{mm}$ and $y_{mm}$ are equivalent to two functions with a single-variable Taylor expansion. Hence, we can say that there are two one-variable functions worth of modes allowed. The reader can convince herself that this counting is reasonable from the following argument. In two dimensions, for an arbitrary metric in complex coordinates ($ds^2 = g_{z\bar z}(z,\bar z)\,dz\,d\bar z$), we can solve the conformal Killing vector (CKV) equations exactly. The solutions are written as $\xi=f(z)\partial_z+\bar f(\bar z) \partial_{\bar z}$. Here $f(z)$ and $\bar f(\bar z)$ are two one-variable arbitrary analytic functions. On the other hand, we can show that if an infinitesimal diffeomorphism of the solution also satisfies the Virasoro constraint, then the change in the diffeomorphism function has to be a CKV on the world-sheet. Hence, the number of modes we got is the same as the number of CKVs on the world-sheet.
\end{itemize}

Locally, we thus find three non-trivial deformations in each quadrant (two translations and a rotation), and in the next subsection, we will discuss how to join them together into global deformations.

\subsection{Global analysis} \label{global}

In the previous subsection, we found that locally inside each corner, new solutions that are not related by diffeomorphisms to our original one can be either translations on the sphere, or a rotation in the angular $\phi$ direction.

Clearly, since we need to have a collapsed handle that maps to a single point on the sphere, the translation modes in the four quadrants need to take the same value, so that the whole collapsed handle moves together to a different point. Thus, the translation modes lead to two moduli altogether, corresponding to the position of the collapsed handle on the sphere.

The other modes correspond to general $\phi$ rotations of each quadrant, so that as we approach (say) $\alpha=\beta=0$ from the four different quadrants we obtain values $\phi_1,\phi_2,\phi_3,\phi_4$; or, alternatively, when we go around the inside of the square close to its boundary, we pass near each of these values when we pass near the four corners. Obviously, this requires that the four values should be ordered in a cyclic order on the $\phi$ circle (with $\phi \equiv \phi + 2\pi$).

The analysis of the previous subsection shows that different values of the $\phi_i$ are not related by non-singular diffeomorphisms, so we just need to show that smooth solutions exist with any values of the $\phi_i$. It is enough to show this near one of the boundaries of the square, since the analysis near the other boundary is analogous, and since in the interior it is easy to smoothly fill any configuration in which $\phi$ goes around the circle once. 

Let us analyze the behavior near $\alpha=0$, on both sides of this line. We know that $\theta$ should go to zero at $\alpha=0$ from both sides, so the generic behavior close to $\alpha=0$ is $\theta(\alpha,\beta) = |\alpha| f(\beta) + O(\alpha^2)$ with $f(\beta) > 0$. A priori we could have had different functions $f(\beta)$ for positive and negative $\alpha$'s, but since the induced metric at $\alpha=0$ is given by $h_{\alpha \alpha} = f^2(\beta) + O(\alpha)$ and it should be smooth, we must choose the same $f$ on both sides (the other components of the induced metric vanish at $\alpha=0$).

For the $\phi$ coordinate, we claim that we can choose any values with a finite limit as $\alpha\to 0$, in particular we can choose that for $\alpha \to 0^+$ we have $\phi(\alpha,\beta) = g_1(\beta) + O(\alpha)$, and for $\alpha \to 0^-$ we have $\phi(\alpha,\beta) = g_2(\beta) + O(\alpha)$, with appropriate values of $g_{1,2}$ at $\beta=0,2\pi$ so that we pass near the $\phi_i$ at the four corners. The values of $g_i$ for intermediate values of $\beta$ are not diffeomorphism-invariant, but their values at $\beta\to 0,2\pi$ are diffeomorphism-invariant, as we showed in the local analysis of the previous subsection. The claim is that these solutions are continuous (as mappings to the sphere) near $\alpha=0$ for any $g_{1,2}$, so that we can complete them to continuous solutions on the whole torus for any value of the $\phi_i$. Note that in order to match to the other boundary (which has some analogous functions ${\tilde f}(\alpha)$, ${\tilde g}_1(\alpha)$ and ${\tilde g}_2(\alpha)$) we need to have $f(\beta) \propto \beta$ as $\beta\to 0^+$, and $f(\beta) \propto (2\pi - \beta)$ as $\beta$ approaches $2\pi$ from below, but otherwise we can make any choice of $f$ that we want.

The specific solution \eqref{specific solution} is a special case of this setup, with $f(\beta) = \sqrt{\sin^4(\beta/2) + \sin^2(\beta)/4}$, and ${\tilde f}(\alpha) = |\sin(\alpha/2)|$. In this special solution the function ${\tilde g}_1(\alpha)$ happens to vanish (and the function ${\tilde g}_2(\alpha)$ happens to equal $\pi$), while $g_1(\beta)$ and $g_2(\beta)$ are non-trivial, but for generic solutions all the functions will be non-constant and interpolate monotonically between the appropriate values of the $\phi_i$.

The conclusion is that we can construct continuous mappings for any cyclically-ordered $\phi_i$ $(i=1,2,3,4)$, and that the solutions with different $\phi_i$ are inequivalent, so that they give four coordinates on the moduli space. In the next subsection, we will compute the metric on the full moduli space, including these four coordinates and the two coordinates from the position of the collapsed handle.

\subsection{Moduli space metric}
\label{metric collapsed handle}

In the previous two sub-sections, we found that there are a total of six moduli. In this sub-section, we will calculate the metric on this six-dimensional moduli space. We will show that the metric vanishes in all four $\phi_i$ directions. Recall \cite{Aharony:2023tam} that the general form of the metric on the moduli space is
\begin{equation} \label{metric moduli}
    m_{ij} = \int d^2\sigma ~ \partial_i X^\mu K_{\mu\nu}(X) \partial_j X^\nu,
\end{equation}
where $i$ and $j$ go over the 6 possible deformations, which can modify the solution $X^{\mu}(\alpha,\beta)$.

Let us expand in the deformations around a collapsed handle solution with some $\phi_i$, and assume that a smooth deformation that changes $\phi_j$ and not any of the other $\phi_i$ is given by $\delta r_j$. The curvature $K_{\mu\nu}$ is localized only near the collapsed handle, and in fact it has delta function contributions there (that can be explicitly computed from the form of our solutions). So, to compute the metric, it is enough to know the form of the deformations close to $\alpha=\beta=0$. The translations have $\partial_i X^{\mu} = \delta_i^{\mu}$ so in these directions we have
\begin{equation}
        m_{ab} = \int d^2\sigma K_{ab}, \qquad a,b\in (x_0,y_0).
\end{equation}
On the other hand, the rotations behave locally as $\delta r^\mu =\{-Y_0,X_0\}$ near $\alpha=\beta=0$, such that they and their derivative with respect to $\phi_i$ vanish there, so if one or both indices in \eqref{metric moduli} is a rotation, its contribution vanishes at $\alpha=\beta=0$ and we get zero.

Thus, the metric is non-zero only in the two translation directions (where it is the metric of the sphere), and vanishes in all other directions. This means that we will have fermionic zero modes that will cause the path integral to vanish whenever we have a collapsed handle. This is consistent with the expected results in the Yang-Mills theory with zero coupling.

\printbibliography
\end{document}